\documentclass[iop]{emulateapj}
\usepackage{txfonts}
\usepackage{mathrsfs}
\usepackage{natbib} 
\usepackage{graphicx} \usepackage{color}
\usepackage{epstopdf}
\newcommand{\Lower}[1]{\smash{\lower 1.5ex \hbox{#1}}}
\newcommand\T{\rule{0pt}{2.6ex}}
\newcommand\B{\rule[-1.2ex]{0pt}{0pt}}

\begin{document}

\title{
The Radio Continuum--Star Formation Rate Relation in WSRT SINGS Galaxies}

\author{Volker Heesen\altaffilmark{1,2}, Elias Brinks\altaffilmark{1}, Adam
  K.\ Leroy\altaffilmark{3}, George Heald\altaffilmark{4}, Robert
  Braun\altaffilmark{5}, Frank Bigiel\altaffilmark{6}, \\ and  Rainer Beck\altaffilmark{7}}

\altaffiltext{1}{Centre for Astrophysics Research, University of
  Hertfordshire, Hatfield AL10 9AB, UK; e.brinks@herts.ac.uk
}
\altaffiltext{2}{School of Physics \& Astronomy, University of
  Southampton, Southampton SO17 1BJ, UK; v.heesen@soton.ac.uk
}
 \altaffiltext{3}{National Radio
  Astronomy Observatory, 520 Edgemont Road, Charlottesville, VA 22903-2475,
  USA; aleroy@nrao.edu}
\altaffiltext{4}{Netherlands Institute for Radio Astronomy (ASTRON), Postbus
  2, 7990 AA Dwingeloo, The Netherlands; heald@astron.nl
}
\altaffiltext{5}{CSIRO Astronomy and Space Science, P.O. Box 76, Epping, NSW
  1710, Australia; Robert.Braun@csiro.au}
\altaffiltext{6}{Institut f\"ur theoretische Astrophysik, Zentrum f\"ur
  Astronomie der Universit\"at Heidelberg,
  Albert-Ueberle-Str.\ 2, D-69120 Heidelberg, Germany; bigiel@uni-heidelberg.de}
\altaffiltext{7}{Max-Planck-Institut f\"ur Radioastronomie, Auf dem H\"ugel 69,
  D-53121 Bonn, Germany; rbeck@mpifr-bonn.mpg.de}

\date{Received 2013 August 1 / accepted 2014 February 4} 

\begin{abstract}
 We present a study of the spatially resolved radio
  continuum--star
 formation rate (RC--SFR) relation using state-of-the-art 
  star formation tracers in a sample of 17 THINGS galaxies. We use
  SFR surface density ($\Sigma_{\rm SFR}$) maps created by a
  linear combination of
 \emph{GALEX} far-UV (FUV) and \emph{Spitzer} $24~\mu\rm m$
  maps. We use RC maps at
 $\lambda\lambda$ 22 and 18~cm from the WSRT SINGS
  survey and H$\alpha$
 emission maps to correct for thermal RC emission. We
  compare azimuthally
 averaged radial profiles of the RC and FUV/mid-IR (MIR) based
  $\Sigma_{\rm SFR}$ maps and study pixel-by-pixel correlations at fixed
  linear scales of $1.2$ and
 $\rm 0.7~kpc$. The ratio of the integrated SFRs
  from the RC emission to that of
 the FUV/MIR-based SF tracers is $\mathscr{R}_{\rm
    int} = 0.78\pm
 0.38$, consistent with the
 relation by Condon. We find a
  tight correlation between the radial profiles
 of the radio and FUV/MIR-based
  $\Sigma_{\rm SFR}$ for the entire extent of the disk. The ratio
 $\mathscr{R}$ of the
  azimuthally averaged radio to FUV/MIR-based $\Sigma_{\rm SFR}$ agrees with
  the integrated ratio and has
 only quasi-random fluctuations with galactocentric radius that
  are relatively small
 (25\%). Pixel-by-pixel plots show a tight correlation
  in log-log
 diagrams of radio to FUV/MIR-based $\Sigma_{\rm SFR}$, with a
  typical standard deviation of a
 factor of two. Averaged over our sample we
  find $\rm (\Sigma_{SFR})_{RC}\propto
 (\Sigma_{SFR})_{hyb}^{0.63\pm0.25}$,
  implying that data points with high $\Sigma_{\rm SFR}$ are
 relatively radio
  dim, whereas the reverse is true for low $\Sigma_{\rm SFR}$. We
 interpret
  this as a result of spectral aging of cosmic-ray electrons (CREs), which
 are diffusing away from
  the star formation sites where they are injected
 into the interstellar
  medium. This is supported by our finding that the
 radio spectral index is a
  second parameter in pixel-by-pixel plots: those data
 points dominated by
  young CREs are relatively radio dim,
 while those dominated by old CREs are slightly
  more RC bright than
 what would be expected from a linear extrapolation. We
  studied the ratio $\mathscr{R}$ of radio
 to FUV/MIR-based integrated SFR as a function
  of global galaxy parameters and
 found no clear correlation. This suggests
  that we can use RC emission as a
 universal star formation tracer for
  galaxies with a similar degree of
 accuracy as other tracers, if we restrict
  ourselves to global or azimuthally averaged measurements. We can reconcile
  our finding of an almost linear RC--SFR relation and sub-linear resolved (on
  1~kpc scale) RC--$\Sigma_{\rm SFR}$ relation by proposing a non-linear
  magnetic field--SFR relation, $B\propto \rm SFR_{hyb}^{0.30\pm 0.02}$, which
  holds both globally and locally.
\end{abstract}


\section{Introduction}
\label{sec:introduction}
Radio continuum (RC) emission in galaxies stems from massive stars, allowing
us to use it as a star formation (SF) tracer. Ionizing UV radiation from massive stars creates \ion{H}{2}
regions
 where thermal electrons give rise to RC radiation through
the process of 
 {\em bremsstrahlung} or thermal (free--free) 
 emission. These
same stars, upon reaching the end of their lives, are the source of
cosmic-ray electrons (CREs) that are accelerated in supernova shock
waves. When these CREs 
encounter a magnetic field, they generate non-thermal
(synchrotron) emission. If all CREs lose their energy exclusively
{\em within} the galaxy, it can be considered a {\em calorimeter}
\citep{voelk_89a,lisenfeld_96a}. More realistically, a galaxy is a leaky box,
and models become far more involved \citep{bell_03a,lacki_10a}. The relative
success of theoretical models lends
confidence that massive SF and RC emission are indeed closely tied
together.

RC can act as tracers for the star formation rate (SFR) with the added advantage over other tracers---such as the optical Balmer H$\alpha$
line emission, far-UV (FUV) emission as measured by the \emph{GALEX} satellite, and
hybrid combination of these \citep[e.g.,][]{leroy_12a}---that they are unaffected by dust attenuation. The advantage offered by probing SF via radio astronomical means is that this does not rely on the availability of (cryogenically cooled) IR satellites. The motivation of this paper is to deepen our understanding of the relation between the SFR and the RC luminosity of normal star-forming galaxies,
hereafter the RC--SFR relation, as originally explored by \citet{condon_92a}.

RC emission observed at 33~GHz is an almost ideal SF tracer
\citep{murphy_11a}. The reason is that thermal free--free emission is the
dominating emission mechanism in the 20--40~GHz range. However, galaxies
can be a factor of 10 brighter at lower frequencies such as between 1 and 2~GHz
than at 33~GHz, where the non-thermal emission dominates. This frequency range
is hence potentially more valuable, especially for cosmological studies. For
this to work, the non-thermal component needs to be calibrated as successfully
as the thermal RC against other SF indicators \citep[see][for a recent review
of SF tracers]{kennicutt_12a}. A properly calibrated RC--SFR for non-thermal
continuum emission would offer a powerful method to probe the cosmic SFR,
initially with the Very Large Array (VLA) and with Square Kilometre Array (SKA) precursors, and eventually with the SKA
\citep[e.g.,][]{murphy_09a,lacki_10b}.
 At the highest redshifts inverse
Compton losses could dominate to such an extent that only the (weaker) thermal
RC is expected to survive, although opinion on this remains divided \citep[e.g.,][]{murphy_09a,lacki_10b,schleicher_13a}.

A semi-empirical model was introduced by \citet{condon_92a} to quantify the
RC--SFR relation. It assumes that the RC luminosity is proportional to the CRE
production rate, which itself is proportional to the supernova rate and hence
to the SFR. The Condon relation is a \emph{calorimetric model}, which requires
at a minimum the following two assumptions to hold
\citep{voelk_89a,lisenfeld_96a}: (1)
 galaxies act as a calorimeter both for the
CREs and for the ionizing UV-photons: and (2) although CREs have competing
 energy
losses to synchrotron radiation such as inverse Compton (IC) radiation,
ionization, bremsstrahlung, and adiabatic losses, the fraction of these
additional losses to the synchrotron
 radiation should remain constant. The
assumption of a CRE calorimeter does hold in starburst galaxies
\citep[e.g.,][]{paglione_12a,yoast-hull_13a}, but probably not in normal
spiral galaxies. Therefore, as alternatives, there are \emph{non-calorimetric}
models that do account for a possible CRE escape
\citep{helou_93a,niklas_97a,lacki_10a,dumas_11a,tabatabaei_13a}. The models of
\citet{helou_93a} and \citet{lacki_10a} were proposed to explain the
RC--far-IR (FIR)
correlation, allowing for a possible escape of dust heating FUV photons to
compensate the escape of CREs. These models are hence not strictly applicable
to the RC--SFR relation, because, as we will see, we are using SF tracers that
do account for the escape of FUV photons. Hence, we do not consider them any
further. The model by \citet{niklas_97a} and their spatially resolved
derivatives \citep{dumas_11a,tabatabaei_13a}, however, although again posed to
explain the RC--FIR correlation, do assume a direct proportionality between
the FIR luminosity and the SFR. Hence, they are essentially models for the
RC--SFR relation and thus relevant for our study.
We will use Condon's relation as the basis of our study
and investigate whether it holds for our sample for integrated SFRs, as well as
for SFR surface densities ($\Sigma_{\rm SFR}$). Two emission processes contribute to the RC emission:
massive stars contribute to both the thermal and the non-thermal
emission. The thermal emission originates from free--free emission of electrons
in the \ion{H}{2} region surrounding these stars. The H$\alpha$ emission can
be used as a proxy for the thermal emission; the radio spectral index of the
thermal emission is flat with $\alpha=-0.1$ ($S_\nu\propto \nu^\alpha$).  The
non-thermal emission is generated by young CREs accelerated in
supernova shockwaves. They emit synchrotron emission in the magnetized interstellar
medium with a non-thermal spectral index of typically $\alpha_{\rm nt}=
-0.83$ \citep{niklas_97b}. \citet{tabatabaei_07c,tabatabaei_07b} separated the
thermal and non-thermal RC emission in M33 based on a novel method to
correct Balmer H$\alpha$ emission for internal extinction by dust. They
found that the non-thermal emission is distributed more smoothly than the thermal
emission, as expected from cosmic-ray transport \citep{tabatabaei_07a}.
Young supernova remnants are visible as strong non-thermal sources but
contribute only about $10\%$ to the instantaneous non-thermal RC emission at
wavelengths around $\lambda$20~cm \citep{condon_92a,lisenfeld_00a}.  The main
 contribution to the RC emission
comes from diffuse emission in the interstellar
 medium. Cosmic rays diffuse
along magnetic field lines or are advected in
 galactic outflows. The electrons
lose their energy and remain visible
 typically for as long as $10^8~\rm yr$ after
the SF event that
 generated them, depending on the magnetic field
strength (for a strong field,
 this time can be shorter). The non-thermal RC
distribution can therefore be
 understood as a smeared-out version of the SFR
distribution
 \citep{bicay_89a,murphy_06a,murphy_06b,murphy_08a}. As we
  can see, the thermal RC and non-thermal RC behave in different ways, motivating
  a separation of the RC emission in order to study them separately. Even
  though the non-thermal RC dominates at wavelengths around $\lambda$20~cm,
  the thermal RC can become more important for local measurements (1~kpc
  scale). In the following, RC refers to the \emph{total} RC, i.e.,
  the sum of the thermal and non-thermal RC, except when we specifically refer
  to one of the components.
Studies of the RC--SFR relation started out based on integrated properties of
the galaxies under study. Often, no allowance could be made for contaminating
background sources within the fairly large beams of single-dish
telescopes. Likewise, single-dish radio flux densities can be contaminated by
the presence of an active galactic nucleus (AGN.
\citet{condon_02a} tried to separate AGNs from their sample by taking
out the outliers with too high a radio luminosity compared to the SFR
observed, but that selection is obviously not able to detect some of the
weaker AGNs. Spatially resolved studies circumvent this problem as an
AGN can be readily identified as an excess of radio brightness in the center
of a galaxy. Similarly, background radio sources can be identified and their
contribution removed.

Recent observational
 efforts have concentrated on studies of the RC--FIR or
RC--mid-infrared (RC--MIR) correlation rather than the RC--SFR relation at
ever
 higher spatial resolutions. Using wavelet correlation spectra,
  \citet{dumas_11a} show that the correlation
 between $\lambda 20~\rm cm$ RC
  and $24~\rm\mu m$ MIR emission in M51 (NGC~5194) holds
 well down to a scale
  of $1.8~\rm kpc$ (the typical width of a spiral arm) and to a less extent
  to a scale of 500~pc (the CRE diffusion length), but breaks down at smaller
  scales. A study of the spatially resolved RC--FIR correlation in the
 Small
  Magellanic Cloud by \citet{leverenz_13a} found a good correlation
 down to a
  scale of 15~pc; however, this galaxy has a high thermal RC fraction, meaning
  that cosmic-ray diffusion is of less importance.  Other papers dealing with spatially resolved studies of the
RC--FIR
 correlation are those by \citet{tabatabaei_07a} on M33 and
\citet{tabatabaei_13a} on NGC~6946, among others.

\begin{deluxetable*}{lrrrrrrrrrrrr}
\tabletypesize{\scriptsize}
\tablecaption{Basic Data of the Galaxies in Our Sample.\label{tab:data_sample}}
\tablewidth{0pt}
\tablehead{
\colhead{Galaxy\B} & \colhead{FWHM} & \colhead{$i$} & \colhead{$D$} & \colhead{$D_{25}$} & \colhead{$M_{\rm B}$} & \colhead{$M_{\rm HI}$}  & \colhead{$M_{\rm H2}$} & \colhead{$\log_{10}(M_{\rm s})$} & \colhead{$PI$} & \colhead{Type} & \colhead{$z$} & \colhead{$Z$} \\ & & & \colhead{(Mpc)} & & & \colhead{($10^8M_\odot$)} & \colhead{($10^8M_\odot$)} & \colhead{($M_\odot$)} & \colhead{(\%)} & & $(10^{-2})$ &  \\
\colhead{(1\T)} & \colhead{(2)} & \colhead{(3)} & \colhead{(4)} & \colhead{(5)} & \colhead{(6)} & \colhead{(7)} & \colhead{(8)} & \colhead{(9)} & \colhead{(10)} & \colhead{(11)} & \colhead{(12)} & \colhead{(13)}}
\startdata
Ho~II     & $13\farcs 5$ & $41\degr$ & $3.4$ & $6\farcm 61$  & $-16.87$ & $6.0$   & $<0.4$ & $8.3$  & $<1.8$ & 10 & $0.05$ & $7.68$ \\
IC~2574   & $13\farcs 5$ & $53\degr$ & $4.0$ & $12\farcm 88$ & $-18.11$ & $14.8$  & $<0.8$ & $8.7$  & $<9  $ & 9  & $0.02$ & $7.94$ \\
NGC~628   & $46\farcs 0$ &  $7\degr$ & $7.3$ & $9\farcm 77$  & $-19.97$ & $38.0$  & $10.0$ & $10.1$ & $12  $ & 5  & $0.22$ & $8.33$ \\
NGC~925   & $23\farcs 0$ & $66\degr$ & $9.2$ & $10\farcm 72$ & $-20.04$ & $45.8$  & $2.5 $ & $9.9$  & $<0.8$ & 7  & $0.18$ & $8.24$ \\
NGC~2403  & $13\farcs 6$ & $63\degr$ & $3.2$ & $15\farcm 85$ & $-19.43$ & $25.8$  & $0.2 $ & $9.7$  & $7.8 $ & 6  & $0.04$ & $8.31$ \\
NGC~2841  & $16\farcs 1$ & $73\degr$ & $14.1$& $6\farcm 92$  & $-21.21$ & $85.8$  & $3.2 $ & $10.9$ & $6.2 $ & 3  & $0.21$ & $8.52$ \\
NGC~2903  & $34\farcs 1$ & $65\degr$ & $8.9$ & $11\farcm 75$ & $-20.93$ & $43.5$  & $21.8$ & $10.0$ & $4.0 $ & 4  & $0.19$ & $9.12$ \\
NGC~2976  & $13\farcs 3$ & $65\degr$ & $3.6$ & $7\farcm 24$  & $-17.78$ & $1.4$   & $0.6 $ & $9.1$  & $4.4 $ & 5  & $0.00$ & $8.30$ \\
NGC~3184  & $18\farcs 6$ & $16\degr$ & $11.1$& $7\farcm 41$  & $-19.92$ & $30.7$  & $15.9$ & $10.3$ & $6.4 $ & 6  & $0.20$ & $8.48$ \\
NGC~3198  & $17\farcs 6$ & $72\degr$ & $13.8$& $6\farcm 46$  & $-20.75$ & $101.7$ & $6.3 $ & $10.1$ & $<1.4$ & 5  & $0.22$ & $8.32$ \\
NGC~3627  & $55\farcs 6$ & $62\degr$ & $9.3$ & $10\farcm 23$ & $-20.74$ & $8.2$   & $12.6$ & $10.6$ & $2.5 $ & 3  & $0.24$ & $8.43$ \\
NGC~4736  & $19\farcs 1$ & $41\degr$ & $4.7$ & $7\farcm 76$  & $-19.80 $& $4.0$   & $4.0 $ & $10.3$ & $4.3 $ & 2  & $0.10$ & $8.31$ \\
NGC~4826  & $33\farcs 6$ & $65\degr$ & $7.5$ & $10\farcm 47$ & $-20.63$ & $5.5$   & $18.1$ & $10.4$ & $1.0 $ & 2  & $0.14$ & $8.59$ \\
NGC~5055  & $18\farcs 6$ & $59\degr$ & $10.1$& $11\farcm 75$ & $-21.12$ & $91.0$  & $50.1$ & $10.8$ & $4.7 $ & 4  & $0.17$ & $8.42$ \\
NGC~5194  & $17\farcs 1$ & $42\degr$ & $8.0$ & $7\farcm 76$  & $-21.04$ & $25.4$  & $25.1$ & $10.6$ & $6.9 $ & 4  & $0.15$ & $8.54$ \\
NGC~6946  & $14\farcs 1$ & $33\degr$ & $5.9$ & $11\farcm 48$ & $-20.61$ & $41.5$  & $39.8$ & $10.5$ & $10  $ & 6  & $0.02$ & $8.40$ \\
NGC~7331  & $22\farcs 1$ & $76\degr$ & $14.7$& $9\farcm 12$  & $-21.67$ & $91.3$  & $50.1$ & $10.9$ & $3.8 $ & 3  & $0.27$ & $8.36$
\enddata                                                                                                                       
\tablecomments{Column 1: galaxy name; Column 2: resolution of the radio map as full-width-half-mean in arcsec; Column 3: inclination angle in degree; Column 4: distances in Mpc; Column 5: optical size in arcmin; Column 6: $B$-band absolute magnitude from \citet{walter_08a}; Column 7: hydrogen mass from \citet{walter_08a}; Column 8: molecular hydrogen mass from \citet{leroy_08a}, NGC~2903 is from \citet{leroy_09a}, NGC~4826 from \citet{wilson_12a}; Column 9: decadal logarithm of stellar mass from \citet{leroy_08a}, NGC~2841 and NGC~2903 are from \citet{deblok_08a}, NGC~4826 was determined from \emph{IRAS} data as in \citet{leroy_08a}; Column 10: fractional polarization at $\lambda$22~cm from \citet{heald_09a} with respect to our integrated flux density; Column 11: galaxy type from \citet{walter_08a}; Column 12: redshift; Column 13: metallicity defined by $\log_{10}({\rm O/H})+12$ from \citet{moustakas_06a}.}
\end{deluxetable*}

But until recently, no studies of the RC--SFR have been forthcoming. Obviously, such
studies should have sufficient angular resolution in order to be able to isolate any AGN activity or unrelated
background sources from the area under study. Moreover, spatially resolved studies allow one
to probe the RC--SFR relation as a function of location within a galaxy, such as arm versus inter-arm.
Thanks to the THINGS \citep[The \ion{H}{1} Nearby Galaxy Survey,][]{walter_08a} 
collaboration---making use of SINGS \citep[SIRTF Nearby Galaxy Survey;][]{kennicutt_03a} and the ``GALEX Ultraviolet Atlas of
  Nearby Galaxies'' \citep{gildepaz_07a}---spatially resolved SFR maps have become available of a significant sample of nearby
galaxies. In this paper, we present a sample of 17
galaxies where we compare the RC brightness and the SFR surface
density ($\Sigma_{\rm SFR}$). We use $\Sigma_{\rm SFR}$ maps
based on a combination of \emph{Spitzer} MIR and \emph{GALEX} FUV maps \citep[see][for
details]{leroy_08a}. The RC data are from the Westerbork Synthesis Radio
Telescope (WSRT) SINGS sample \citep{braun_07a}, taken at $\lambda\lambda$ 22 and 18~cm.

The paper is organized as follows: Section~\ref{sec:observations} explains the
selection of the sample and briefly summarizes the data processing applied to the
maps that we used. Section~\ref{sec:methodology} deals among other things with 
the azimuthally (Section~\ref{subsec:radial})
averaged plots where we compare the RC brightness with the $\Sigma_{\rm SFR}$ as a function of
galactocentric radius. In Section~\ref{subsec:pixel} we produce pixel-by-pixel plots of
our sample at a fixed resolution. The results are analyzed in Section~\ref{sec:results} 
and discussed in Section~\ref{sec:discussion} where we explore the
dependence of the RC--SFR relation on various galaxy
parameters. We present our conclusions in Section~\ref{sec:conclusions}.
\section{Observations}
\label{sec:observations}
\subsection{The Galaxy Sample}
\label{sec:the_sample}
\begin{figure*}[tbhp]
\centering
\resizebox{1.0\hsize}{!}{ \includegraphics{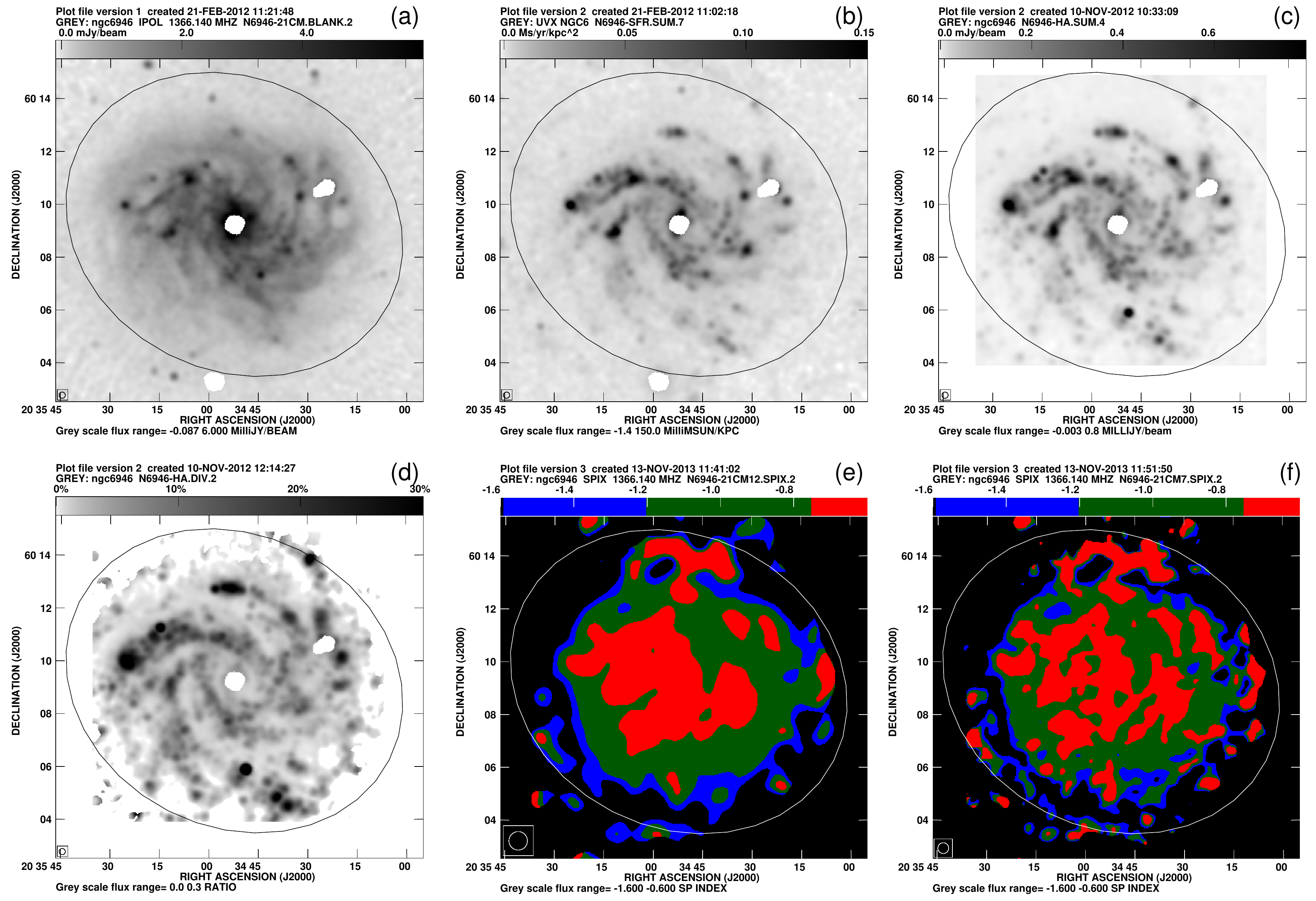}}
\caption{NGC~6946. (a) RC emission at $\lambda 22~\rm
  cm$. (b) Hybrid SFR surface density $(\Sigma_{\rm SFR})_{\rm hyb}$. (c) Thermal RC emission
  as derived from H$\alpha$. (d) Predicted thermal RC fraction. (e) Radio
  spectral index at $\rm 1.2~kpc$ resolution. (f) Radio spectral index at $\rm 0.7~kpc$
  resolution. The total RC intensity maps were clipped at 3 $\times$ the
  rms noise level prior to combination resulting in a spectral
  index error of $0.6$, which is lower in areas of higher intensities.}
\label{fig:n6946_maps}
\end{figure*}

The overlap between the THINGS and the WSRT SINGS sample consists of 17
galaxies that we included in our analysis. All maps were transformed to the
same coordinate system and convolved to the same angular resolutions before
any comparison. Table~\ref{tab:data_sample} summarizes the properties of the
objects used in this study. In Figure~\ref{fig:n6946_maps}(a) we show the radio
brightness at $\lambda$22~cm in NGC~6946 and in Figure~\ref{fig:n6946_maps}(b)
the SFR surface density in the same galaxy.

\subsection{Radio Continuum Maps}
 
The RC maps come from the WSRT SINGS survey
\citep{braun_07a}. This survey contains 28 galaxies observed at
$\lambda\lambda$ 22 and 18~cm in radio continuum.  Observations were obtained
in the ``maxi-short'' configuration, which is optimized for imaging extended
emission. All data were calibrated in AIPS,\footnote{AIPS is free software
  released by the National Radio Astronomy Observatory (NRAO).}  with
subsequent imaging and self-calibration in Miriad. All maps were primary
 beam
corrected using the task {\sc linmos} in Miriad \citep{sault_95a}. Because the WSRT is
 an
east--west interferometer (with 2.7~km extent), the resolution in the
north--south direction is declination dependent. This results in an
elliptical
 Gaussian kernel for the restoration after the maps were cleaned
\citep{hoegbom_74a}. We convolved the maps with a further Gaussian kernel to
obtain circular beam shapes. In Table~\ref{tab:data_sample} we present the
resolution of the maps expressed by the full width at half maximum (FWHM). The
resolution can
 be as high as $13\farcs 1$ and is better than $20\arcsec$ for
most of our
 sample galaxies.
The RC maps were observed with an interferometer, which is not sensitive for
emission on angular scales larger than about half the diameter of the primary
beam
 ($\approx 30\arcmin$ for the WSRT at $\lambda 20~\rm cm$). This figure
is
, however, dependent on the $(u,v)$-coverage and also on the source
structure. In
 particular, some of our RC maps, notably at $\lambda$18~cm, are
affected by a lack of observed spacings shorter than 27~m, which results in
 the
emission sitting within a negative ``bowl''.\footnote{The shortest physical
  baseline length at the WSRT is 36~m. However, the shortest projected
  baseline lengths are typically 27~m before the data are flagged due to
  antenna shadowing at low elevations.} The depth of the bowl can be
 up to
$0.5$ $\times$ the rms noise level in some cases. As the
 $\lambda$18~cm maps
are more severely affected than the $\lambda$22~cm maps,
 the effect on the
measured radio spectral index is thus that it becomes too
 steep, particularly
in the outskirts of galaxies where the emission levels drop
 to the rms
noise level. As a first-order correction we applied a constant
 offset prior to
calculating integrated flux densities, integrated spectral indices, and the
radio spectral index maps.

\subsection{Hybrid SFR Surface Density Maps}
\citet{leroy_08a,leroy_12a} motivated the use of combined \emph{Spitzer} $24~\mu\rm m$
and \emph{GALEX} FUV maps as a tracer for SF and the resolved
SFR surface density. Their study was based on the THINGS sample, which
consists of 34 galaxies observed with the NRAO\footnote{The National Radio
  Astronomy Observatory is a facility of the National Science Foundation
  operated under cooperative agreement by Associated Universities, Inc.} VLA in B-, C-, and D-configuration in the $\lambda$21~cm line of neutral hydrogen
(\ion{H}{1}). These galaxies were part of the SINGS survey by the \emph{Spitzer}
satellite consisting of 75 nearby galaxies \citep{kennicutt_03a}. They were
also part of the \emph{GALEX} FUV survey \citep{gildepaz_07a}. The FUV traces
the young SF via the UV radiation of massive stars. The dust
absorbs the UV light efficiently, which gets re-radiated in the
infrared. Therefore, the Multiband Imaging Photometer for Spitzer (MIPS) $24~\mu\rm m$ maps trace the embedded SF. In massive spiral galaxies the inner region shows little FUV but
strong $24~\rm\mu m$ emission due to the locally higher dust content. The two tracers
for SF are complementary and can be linearly combined to give the
total SFR. The MIPS maps were convolved with a custom kernel
to $13\farcs 5$ resolution referred to as the half-power beam width of the
Gaussian beam. The FUV maps were convolved to the same Gaussian beam before
the combination with the MIPS maps; the $\Sigma_{\rm SFR}$ maps therefore have a resolution
of $13\farcs 5$. A detailed description of the motivation for the construction
of the $\Sigma_{\rm SFR}$ maps can be found in \citet{leroy_08a}.
We refer to these maps as \emph{hybrid} $\Sigma_{\rm SFR}$ maps in order to
distinguish them from the radio maps, which can be converted to \emph{radio}
$\Sigma_{\rm SFR}$ maps using Condon's relation (see
Section~\ref{subsec:radio_sfrd}). We use
 the hybrid $\Sigma_{\rm SFR}$ maps
by \citet{leroy_08a} as an absolute reference of the
 SFR
against which we compare our radio-derived $\Sigma_{\rm SFR}$ maps.
 We note
that \citet{leroy_12a} suggested that the hybrid combination of \emph{Spitzer}
$24~\rm \mu m$ MIR emission and Balmer H$\alpha$ emission is a
more accurate SF tracer than the hybrid FUV/MIR combination. FUV maps
  provided by \emph{GALEX} are much more sensitive, though, and have revealed current
  SF extending throughout the outskirts of galaxies as traced by
  \ion{H}{1} \citep[e.g., see][]{thilker_05a,bigiel_10a,kennicutt_12a}. It is for that
  reason that we prefer to use the FUV/MIR hybrid $\Sigma_{\rm SFR}$ maps.

\subsection{H$\alpha$ Maps}
\label{subsec:halpha}
We are using H$\alpha$ emission as a tracer for thermal RC
emission (see Section~\ref{subsec:thermal}). Maps for the galaxies in our sample
were obtained from the Local Volume Legacy (LVL) survey, which covers all galaxies
within a distance of 11~Mpc \citep{kennicutt_08a}. The remaining maps were
obtained from the SINGS legacy program \citep{kennicutt_03a} with the
exception of NGC~2903 \citep[from][]{hoopes_01a}, NGC~4736 \citep[from][]
{sanchez_12a}, and NGC~6946 \citep[from][]{ferguson_98a}.
All maps were continuum subtracted using a broad $R$-band continuum image. LVL
maps were found to be in agreement to within 5\%--10\% of the flux densities
published by \citet{kennicutt_08a}. For the SINGS legacy maps, however,
background levels fluctuated and were not always properly
adjusted. \citet{sanchez_12a} reprocessed some of the SINGS maps, which
resulted in an improved background subtraction. Where possible, we used their
maps.  We typically got flux densities within $30\%$ of the published
values. This is of sufficient accuracy, because the H$\alpha$ maps are only used
to correct for the thermal component to the radio continuum emission and the
typical contribution to the radio continuum flux density by the thermal
emission is less than $30\%$. We used $E(B-V)$ values as published by
\citet{schlegel_98a} to correct for foreground absorption by the Galaxy. 
  We note that we correct  for foreground absorption only and not for
  \emph{internal} absorption by dust within the observed galaxies (see also
  Section~\ref{subsec:thermal}).
 With the exception of NGC~6946 the
    emission maps include the contribution from the [\ion{N}{2}] line, which lies
    within the filter coverage (see Section~\ref{subsec:thermal} for
    further details).

\subsection{Masking of AGNs and Background Sources}
\label{subsec:masking}
Our resolved study allows us to identify regions of radio emission in our maps
that is coming from an AGN or from background radio
galaxies. AGNs are visible as compact nuclear sources. Background radio
galaxies are significant sources in radio emission but weak in the $\Sigma_{\rm SFR}$
maps. The preferred way of dealing with the above-mentioned objects, is to
mask them in the maps, so they do not influence our results.
This can be done by ``blinking'' the radio and hybrid $\Sigma_{\rm SFR}$ maps, which
clearly highlights those objects. Sources are blanked within AIPS with the
task {\sc blank}, which basically creates a mask by selecting regions on a
display. This mask is then applied to all maps (RC, $\Sigma_{\rm SFR}$, H$\alpha$) in
order to remain consistent. In general, we masked all significant
sources that are in the center of any given galaxy. Only galaxies that do not
have a significant central source are not masked. Secondly, we masked
background radio galaxies that are showing up as strong radio sources while
no counterpart is visible in the $\Sigma_{\rm SFR}$ maps. Thirdly, we masked sources in the $\Sigma_{\rm SFR}$
maps that do not show up in the radio. These sources are relatively rare and
could be stars or possibly submillimeter galaxies.
Nuclear sources are easy to identify, but background sources in either the
radio or hybrid $\Sigma_{\rm SFR}$ maps can be somewhat ambiguous. We stress that the
number of sources that are removed in each galaxy is small, typically less
than five. Therefore, we expect the accuracy of the masking to be
sufficient. As an example we quote NGC~628, which is the galaxy most affected
by background radio sources. The difference between integrated luminosity with
or without this correction is
smaller than 15\%. In all other galaxies the difference is yet smaller. For
spurious sources in the hybrid $\Sigma_{\rm SFR}$ maps, the difference is considerably
smaller than in the radio maps. Therefore again, we expect errors or omissions
in the masking procedure not to influence our analysis or conclusions.
\section{Methodology}
\label{sec:methodology}
\subsection{Resolved RC--SFR Relation}
\label{subsec:radio_sfrd}
As discussed in Section~\ref{sec:introduction}, Condon's relation states that the
integrated SFR and the RC luminosity of a galaxy are proportional to each
other. In this section we explain how we can derive resolved radio $\Sigma_{\rm SFR}$
maps. According to Equation~(27) in \citet{condon_02a}, the SFR
can be expressed as
\begin{equation}
\left .\frac{\rm SFR_{RC}}{M_\odot\, {\rm yr^{-1}}}\right |_{>0.1\,M_\odot} =
0.75\times10^{-21} \left ( \frac{L_{1.4\,\rm GHz}}{\rm
    W\,Hz^{-1}}\right ),
\label{eq:sfr_total}
\end{equation}
where $L_{1.4\,\rm GHz}$ is the RC luminosity at $1.4$~GHz of an
unresolved galaxy. We note that Condon used a Salpeter initial mass function (IMF) to extrapolate
  from
 the massive stars ($M>5M_\odot$) that show up in the RC to that of all
  stars
 formed ($0.1 < M/M_\odot<100$). The hybrid $\Sigma_{\rm SFR}$ maps of
\citet{leroy_08a} are based on a broken
 power-law IMF as described in
\citet{calzetti_07a}, so their derived SFRs are
 a factor of $1.59$ lower than
using the Salpeter IMF. We have thus scaled
 Condon's relation in this paper
accordingly.
\citet{condon_92a} derived the above relation by adding the thermal and
non-thermal RC contribution. For the non-thermal emission it was assumed that
one can scale the non-thermal luminosity of the Milky Way as modeled by
\citet{beuermann_85a} based on the 408~MHz observations by \citet{haslam_82a} to the observed Type~II supernova rate of
\citet{tammann_82a}. The thermal RC contribution was derived from the Balmer
H$\alpha$ line emission following the SFR relation as presented in
\citet{kennicutt_83a}. As Condon's relation is widely
used in the literature, we will use it as the basis of our analysis.
If we use $L_{1.4\,\rm GHz} =
4\pi A I_{1.4\,\rm GHz}$, where $A$ is the area observed, we obtain for resolved
emission the relation converted to the SFR surface density measured in the plane of the sky, $\Sigma_{\rm SFR}^\prime$ as
\begin{equation}
\left .\frac{(\Sigma_{\rm SFR})_{\rm RC}'}{M_\odot\, {\rm yr^{-1}\, kpc^{-2}}}\right |_{>0.1\,M_\odot} =
8.8\times10^{-8} \left (\frac{I_{1.4\,\rm GHz}}{\rm
    Jy\,ster^{-1}}\right ),
\label{eq:sfrd_ster}
\end{equation}
where $I_{1.4\,\rm GHz}$ is the RC intensity at $1.4$~GHz. The RC intensity is
measured in units of $\rm Jy~beam^{-1}$, where the area of the synthesized
beam is
prescribed by the FWHM, effectively the resolution of
the radio map. The solid angle a Gaussian beam subtends is $\Omega =
\rm 1.133\times (FWHM)^2~ ster$ or $\Omega=\rm 2.66\times 10^{-11} \times
(FWHM [arcsec])^2~ster$, where the FWHM is measured in arcsec. Thus, the
$\Sigma_{\rm SFR}$ can be expressed as
\begin{eqnarray}
\nonumber \left . \frac{(\Sigma_{\rm SFR})_{\rm RC}'}{M_\odot\, {\rm yr^{-1}\, kpc^{-2}}}\right
|_{>0.1\,M_\odot} & =&
3.31\times10^{3} \cdot \left ( \frac{\rm FWHM}{\rm arcsec}\right )^{-2} \\
 &&  \cdot \left ( \frac{I_{1.4\,\rm GHz}}{\rm
    Jy\,beam^{-1}}\right ).
\label{eq:sfrd}
\end{eqnarray}
Because the hybrid $\Sigma_{\rm SFR}$ maps are inclination corrected, we need
to correct the
 radio $\Sigma_{\rm SFR}$ for inclination angle as well and as
follows: ${\rm
 (\Sigma_{\rm SFR})_{RC}} = \cos(i)\times \rm {(\Sigma_{\rm
    SFR})_{RC}'}$, where $0\degr \leq i \leq90\degr$ is the
 inclination angle
of the galaxy in question, and $i=0\degr$ is defined as
 face-on.
\begin{figure*}[tbhp]
  \centering
  \resizebox{0.9\hsize}{!}{\includegraphics{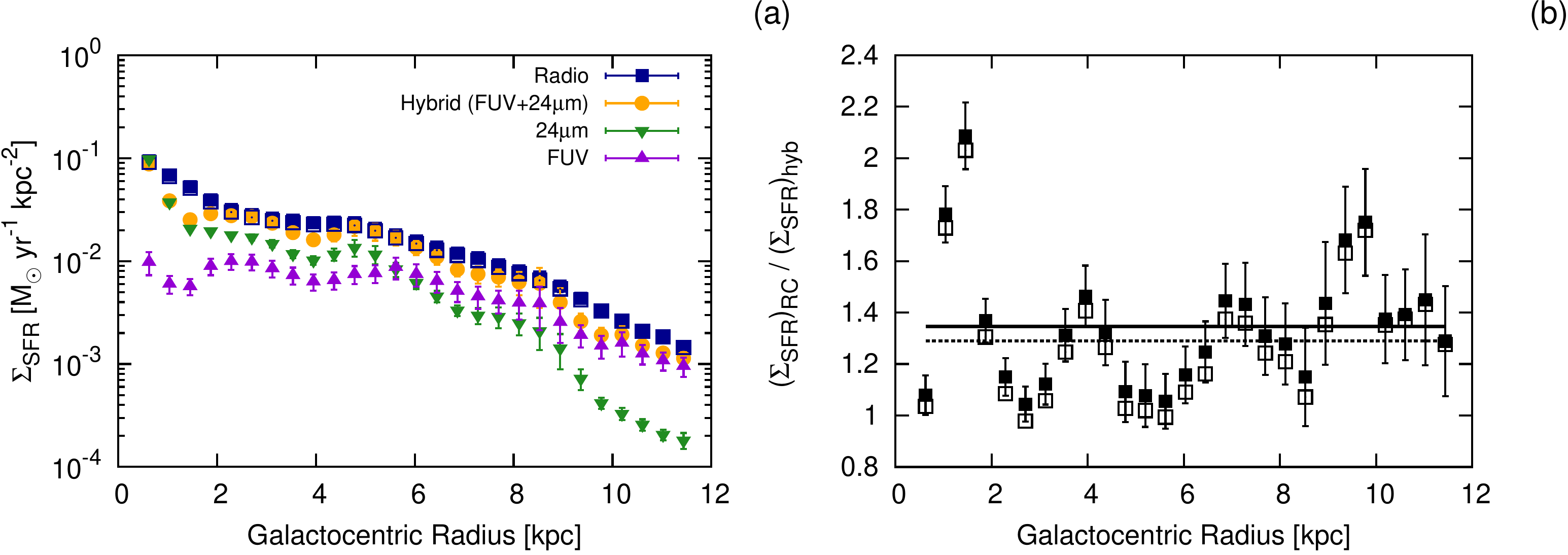}}
  \caption{NGC~6946. (a) Radial distribution of the $(\Sigma_{\rm SFR})_{\rm
      RC}$ from $\lambda 22~\rm cm$ RC emission
 (dark blue), hybrid (FUV +
    $24~\rm\mu m$) $(\Sigma_{\rm SFR})_{\rm hyb}$ (orange), $24~\rm\mu m$ alone
    (green), and FUV alone (violet). (b) Ratio $\mathscr{R}$ of RC to hybrid $\Sigma_{\rm
      SFR}$ as a
 function of galactocentric radius. Open symbols represent the non-thermal RC
    emission
 alone. The solid line shows a least-squares fit to the data points
    assuming a
 constant ratio (dashed line for the non-thermal RC emission
    alone).}
  \label{fig:n6946_rad}
\end{figure*}
\subsection{Thermal Radio Continuum Emission}
\label{subsec:thermal}
The RC emission consists of non-thermal synchrotron and thermal
 free--free
emission. As eventually we wish to also study the resolved non-thermal
  RC--SFR relation, we need to subtract a map of the thermal RC emission
  alone. A standard technique is to use H$\alpha$ emission for
 estimating the
free--free RC emission. We use the H$\alpha$ maps described in
Section~\ref{subsec:halpha} to create spatially resolved distributions of
the
 thermal RC emission. According to \citet{deeg_97a},
\begin{eqnarray}
\frac{S_{\rm th}(\nu)}{\rm erg\, cm^{-2}\, s^{-1}\, Hz^{-1}} & = &1.14 \times
 10^{-14} \left ( \frac{\nu}{\rm GHz} \right )^{-0.1}\\\nonumber
&&\cdot \left (\frac{T_{\rm
 e}}{10^4~\rm K}\right )^{0.34}\left ( \frac{F_{\rm H\alpha}}{\rm
   erg\,cm^{-2}\,s^{-1}}\right ),
\end{eqnarray}
where $S_{\rm th}$ is the thermal RC flux density and $F_{\rm H\alpha}$ is
the Balmer H$\alpha$ emission line flux.  We use an electron
temperature of $T_{\rm e} =10^4~\rm K$, which is a good approximation for
most cases. The thermal contribution was then subtracted from the radio data
before further analysis (Sections~\ref{subsec:radial} and
\ref{subsec:pixel}). We did not correct for the contribution by [\ion{N}{2}],
which according to \citet{kennicutt_08a} is typically 40\% in our
sample. \citet{kennicutt_08a} provided only global estimates of the
[\ion{N}{2}]/H$\alpha$ ratio, so we do not have any spatial information of the [\ion{N}{2}]
contribution. On the other hand, as mentioned in Section~\ref{subsec:halpha}, we did not correct the 
H$\alpha$ emission for internal extinction, which in normal galaxies is typically in the 
0--1~mag range at H$\alpha$. In Figures~\ref{fig:n6946_maps}(c) and (d) we present the distribution of the
thermal RC emission and the fraction of thermal emission in NGC~6946, respectively.

These maps can be compared with the those by \citet{beck_07a}, who separated thermal and
non-thermal emission  based on observations at $\lambda\lambda$ 20 and 3~cm, assuming a constant non-thermal
spectral index of $\alpha_{\rm nt}=-1.0$ in NGC~6946. Another approach to derive maps of the thermal emission 
was taken by \citet{tabatabaei_13a}, who de-redden the H$\alpha$ maps by first modeling the
dust spectral energy distribution. Qualitatively these maps of the thermal RC are all quite similar. Given that the
integrated thermal fraction is on average less than 10\% across the angular scales
discussed here, we consider our estimate for the thermal RC contribution at $\lambda$20~cm adequate.  In
Appendix~\ref{app:sample} we present relevant maps for the entire sample.

 We
notice that we did correct for thermal RC emission to measure the integrated
non-thermal spectral index, but not for the spectral index on a 1~kpc
scale. This is because in the pixel measurements, the uncertainty in the
H$\alpha$ maps would contribute to the already relatively high error of the
radio spectral index. The integrated non-thermal radio spectral index is typically
only $0.1$ steeper than the integrated radio spectral index.
\subsection{Radial Profiles}
\label{subsec:radial}
As the first step we analyzed the data using azimuthal averaging.  We
  convolved the radio and hybrid $\Sigma_{\rm SFR}$ maps with a Gaussian
  kernel to bring them to the same resolution and regridded the maps
 to a
  common coordinate system. The resolution is either
 $13\farcs 5$, set by the
hybrid $\Sigma_{\rm SFR}$ maps, or limited by the resolution of the radio maps. 
 We
used inclination and position angles for the elliptical annuli as given in
Table~\ref{tab:data_sample}.
 We created azimuthal averages by using elliptical
annuli spaced by one FWHM using the task {\sc iring} of
AIPS. This results in plots of the $\Sigma_{\rm SFR}$ based on the RC emission
at $\lambda 22~\rm cm$ or the hybrid $\Sigma_{\rm SFR}$ maps as a function of
galactocentric radius.
The two dwarf irregular galaxies, Holmberg~II
and IC~2574, are only barely detected in  RC emission, even though they are
quite prominent in SFR maps based on UV or optical/IR wavelengths. For these two galaxies we chose the
center of the elliptical annuli to be centered on the brightest emission
visible in the RC maps. For all other galaxies we chose the center of the elliptical annuli to coincide with the center of the
galaxies.
We converted the averaged RC brightnesses into averaged $\Sigma_{\rm SFR}$ in each annulus
using Equation~(\ref{eq:sfrd}). For the hybrid $\Sigma_{\rm SFR}$ we integrated them in the same
annuli and calculated again the averaged $\Sigma_{\rm SFR}$. We note that the $\Sigma_{\rm SFR}$ are
inclination corrected. In Figure~\ref{fig:n6946_rad} we show the results for
NGC~6946 as an example. We present both the radial distributions of the radio
and hybrid $\Sigma_{\rm SFR}$ and their ratio. The results for the entire sample are shown
in Appendix~\ref{app:sample}.
We also use {\sc iring} to measure integrated values of the RC flux densities,
hybrid SFRs, and thermal RC flux densities. We use an ellipse with the major
axis of the galaxy and a minor axis corresponding to an aspect ratio for an
inclination angle of $40\degr$. This ensures that we include the RC emission
in the halos of the galaxies. These integrated values are listed in
Table~\ref{tab:sample}. We note that these values are usually lower than the
values given in \citet{braun_07a}, because we masked the maps to exclude AGNs
and background galaxies (see Section~\ref{subsec:masking}). The values we
present should represent the best estimate of the RC emission in each galaxy
that originates from SF.
\begin{deluxetable*}{lrrrrrrrrr}
\tabletypesize{\scriptsize}
\tablecaption{Measured Properties of the Galaxies in Our Sample.\label{tab:sample}}
\tablewidth{0pt}
\tablehead{
\colhead{Galaxy\B} & \colhead{$S_{\rm 22\,cm}$} & \colhead{$f_{\rm th}$} & \colhead{$\rm SFR_{RC}$} & \colhead{$\rm SFR_{hyb}$} & \colhead{$\Re_{\rm int}$} & \colhead{$<\Re>$} & \colhead{$\Re^{\rm nth}_{\rm int}$} & \colhead{$<\Re^{\rm nth}>$} & \colhead{$\delta$}
\\& \colhead{(Jy)} & \colhead{(\%)} & \colhead{($M_\odot \,\rm yr^{-1}$)} & \colhead{($M_\odot \,\rm yr^{-1}$)} & & & \\
\colhead{(1\T)} & \colhead{(2)} & \colhead{(3)} & \colhead{(4)} & \colhead{(5)} & \colhead{(6)} & \colhead{(7)} & \colhead{(8)} & \colhead{(9)} & \colhead{(10)}}
\startdata
Ho~II    & $0.027\pm0.001$ & $16$ & $0.028\pm0.002$ & $0.033 \pm 0.002$ & $0.87\pm0.09$ & $0.80\pm0.38$ & $0.68\pm0.07$ & $0.67\pm0.40$ & $0.26\pm0.06$\\
IC~2574  & $0.007\pm0.001$ & $41$ & $0.010\pm0.001$ & $0.033 \pm 0.002$ & $0.31\pm0.03$ & $0.29\pm0.09$ & $0.02\pm0.00$ & $0.17\pm0.07$ & $1.21\pm0.27$\\
NGC~628  & $0.192\pm0.010$ & $9 $ & $0.929\pm0.047$ & $0.848 \pm 0.042$ & $1.10\pm0.11$ & $1.56\pm0.81$ & $0.99\pm0.10$ & $1.50\pm0.86$ & $\ldots\qquad$\\
NGC~925  & $0.079\pm0.004$ & $12$ & $0.608\pm0.031$ & $1.417 \pm 0.071$ & $0.43\pm0.04$ & $0.40\pm0.06$ & $0.37\pm0.04$ & $0.33\pm0.05$ & $0.62\pm0.04$\\
NGC~2403 & $0.357\pm0.018$ & $22$ & $0.331\pm0.017$ & $0.868 \pm 0.043$ & $0.38\pm0.04$ & $0.35\pm0.09$ & $0.31\pm0.03$ & $0.26\pm0.09$ & $0.64\pm0.03$\\
NGC~2841 & $0.093\pm0.005$ & $11$ & $1.677\pm0.086$ & $2.464 \pm 0.123$ & $0.68\pm0.07$ & $0.64\pm0.07$ & $0.57\pm0.06$ & $0.51\pm0.10$ & $0.42\pm0.03$\\
NGC~2903 & $0.351\pm0.018$ & $6 $ & $2.514\pm0.126$ & $2.999 \pm 0.150$ & $0.84\pm0.08$ & $0.83\pm0.19$ & $0.79\pm0.08$ & $0.87\pm0.19$ & $\ldots\qquad$\\
NGC~2976 & $0.068\pm0.003$ & $11$ & $0.080\pm0.004$ & $0.205 \pm 0.010$ & $0.39\pm0.04$ & $0.34\pm0.09$ & $0.34\pm0.03$ & $0.29\pm0.10$ & $0.50\pm0.03$\\
NGC~3184 & $0.078\pm0.004$ & $10$ & $0.868\pm0.044$ & $0.835 \pm 0.042$ & $1.04\pm0.10$ & $1.06\pm0.24$ & $1.04\pm0.10$ & $0.95\pm0.21$ & $0.56\pm0.03$\\
NGC~3198 & $0.036\pm0.002$ & $6 $ & $0.614\pm0.033$ & $2.138 \pm 0.107$ & $0.29\pm0.03$ & $0.25\pm0.02$ & $0.29\pm0.03$ & $0.22\pm0.02$ & $0.27\pm0.04$\\
NGC~3627 & $0.501\pm0.025$ & $3 $ & $3.928\pm0.197$ & $4.916 \pm 0.246$ & $0.80\pm0.08$ & $0.80\pm0.06$ & $0.77\pm0.08$ & $0.77\pm0.06$ & $\ldots\qquad$\\
NGC~4736 & $0.265\pm0.013$ & $12$ & $0.530\pm0.027$ & $0.503 \pm 0.025$ & $1.05\pm0.11$ & $1.38\pm0.56$ & $0.97\pm0.10$ & $1.23\pm0.52$ & $0.75\pm0.03$\\
NGC~4826 & $0.072\pm0.004$ & $9 $ & $0.368\pm0.019$ & $0.753 \pm 0.038$ & $0.49\pm0.05$ & $0.53\pm0.08$ & $0.41\pm0.04$ & $0.47\pm0.06$ & $\ldots\qquad$\\
NGC~5055 & $0.372\pm0.019$ & $4 $ & $3.437\pm0.172$ & $3.778 \pm 0.189$ & $0.91\pm0.09$ & $0.87\pm0.05$ & $0.86\pm0.09$ & $0.83\pm0.05$ & $0.79\pm0.02$\\
NGC~5194 & $1.185\pm0.059$ & $5 $ & $6.870\pm0.344$ & $3.929 \pm 0.196$ & $1.75\pm0.17$ & $1.71\pm0.89$ & $1.68\pm0.17$ & $1.63\pm0.89$ & $0.67\pm0.02$\\
NGC~6946 & $1.471\pm0.074$ & $5 $ & $4.637\pm0.232$ & $3.497 \pm 0.175$ & $1.33\pm0.13$ & $1.35\pm0.25$ & $1.22\pm0.12$ & $1.29\pm0.25$ & $0.80\pm0.02$\\
NGC~7331 & $0.381\pm0.019$ & $4 $ & $7.461\pm0.373$ & $11.765\pm 0.588$ & $0.63\pm0.06$ & $0.66\pm0.14$ & $0.63\pm0.06$ & $0.66\pm0.14$ & $\ldots\qquad$
\enddata                                                                                                                       
\tablecomments{Column 1: galaxy name; Column 2: RC flux density at $\lambda$22~cm; Column 3: thermal fraction of the RC emission at $\lambda 22~\rm cm$; Column 4: radio derived star-formation rate at $\lambda$22~cm; Column 5: star-formation rate from hybrid FUV + $24~\mu\rm m$ tracer; Column 6: ratio of integrated radio to hybrid SFR; Column 7: radially averaged ratio of radio to hybrid $\Sigma_{\rm SFR}$; Column 8: ratio of integrated non-thermal radio to hybrid SFR; Column 9: radially averaged ratio of non-thermal radio to hybrid $\Sigma_{\rm SFR}$; Column 10: slope of pixel-by-pixel plot at $1.2$~kpc resolution for the total RC emission (see text).}
\end{deluxetable*}

\subsection{Pixel-by-pixel Analysis}
\label{subsec:pixel}
\begin{figure*}[tbhp]
\centering
\resizebox{0.9\hsize}{!}{ \includegraphics{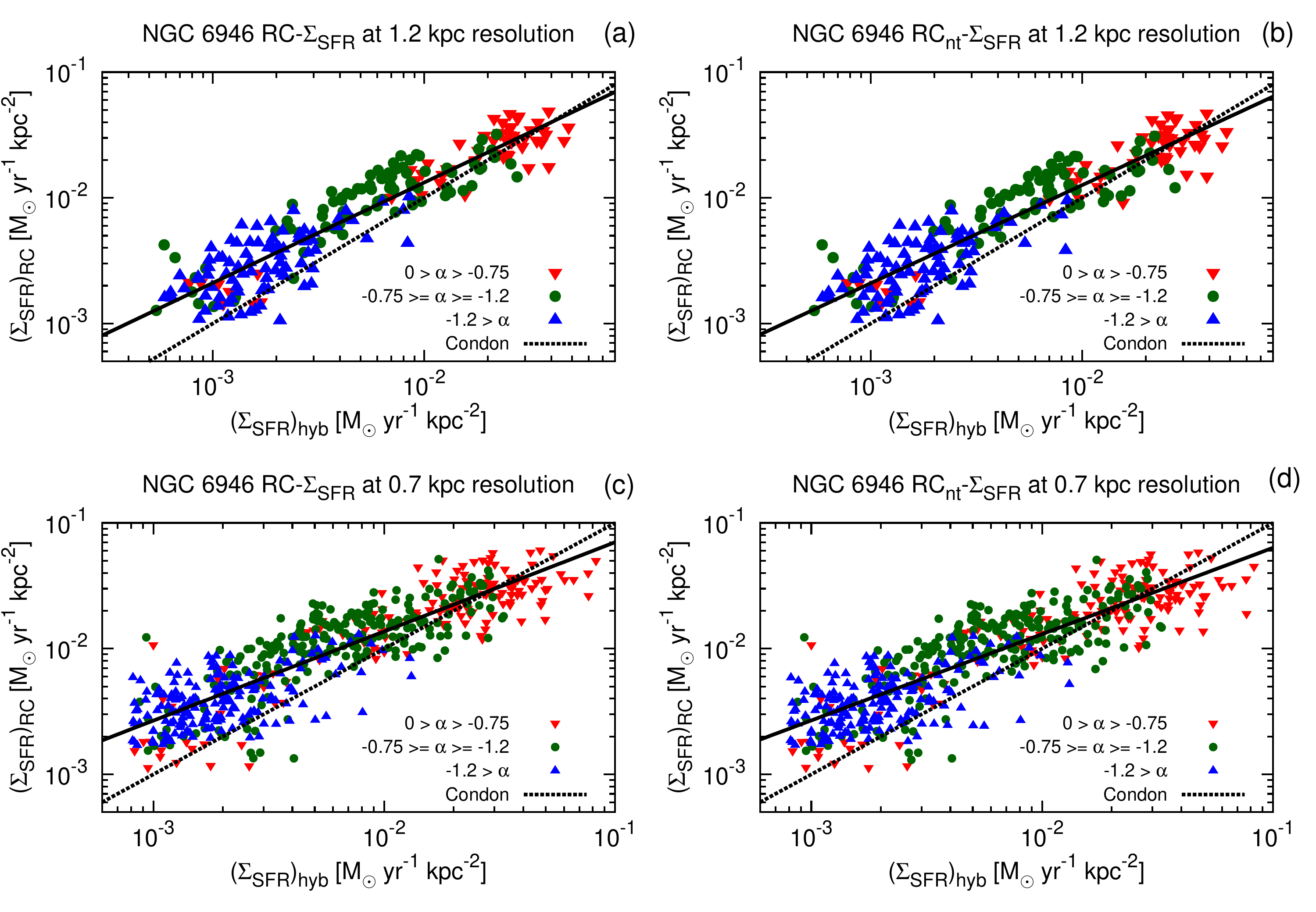}}
\caption{NGC~6946. (a) Radio $\Sigma_{\rm SFR}$ plotted against hybrid
  $\Sigma_{\rm SFR}$ at $\rm 1.2~kpc$
 resolution. (b) Same as (a) but for the
  non-thermal RC emission only based radio $\Sigma_{\rm SFR}$. (c) and (d)
  Same as (a)
  and
 (b) but at $\rm 0.7~kpc$ resolution. The dashed line shows the one-to-one
  Condon
 relation, and the solid line is a linear least-squares fit to the data
  points. Red data points indicate a ``young'' non-thermal spectral index ($\alpha
  > -0.75$), blue data points an ``old'' non-thermal spectral index
  ($\alpha<-1.2$), and green data points a non-thermal spectral index of
  intermediate age ($-0.75\geq \alpha \geq -1.2$). The typical uncertainties
  for the radio spectral index are $0.4$, $0.2$
, and $0.1$ for the blue, green,
  and red data points, respectively.}
\label{fig:n6946_pix}
\end{figure*}

We subsequently investigated the relation between the radio and hybrid $\Sigma_{\rm SFR}$ by
averaging in ``pixels'', for which we chose squares with a constant linear
scale for all galaxies. Taking into account the various angular resolutions
and distances across our sample, we chose two linear scales of $\rm 0.7~kpc$ and
$\rm 1.2~kpc$, respectively. These scales are chosen in order to be able to analyze most
galaxies at a set resolution. We can analyze six galaxies at $\rm 0.7~kpc$, and
this number increases to 12 for a scale of $\rm 1.2~kpc$.  These linear scales are
similar to recent studies of resolved $\Sigma_{\rm SFR}$ in THINGS galaxies
\citep{leroy_08a,bigiel_08a}.
The maps were convolved with a circular Gaussian kernel to the 
angular resolution corresponding to the linear scale studied. A grid with the corresponding
resolution was overlaid, and the pixel values were determined in Miriad using
{\sc imstat}. We used a cutoff level of three times the rms noise level
for the radio and hybrid $\Sigma_{\rm SFR}$ data. We calculated the radio spectral index
between $\lambda\lambda$ 22 and 18~cm using the convention that $S_\nu\propto
\nu^\alpha$, where $\alpha$ is the radio spectral index (of the total RC emission, i.e., not corrected for
thermal emission). For this, we again
used a cutoff level of three times the rms noise level for the
$\lambda$18~cm RC maps.
$\Sigma_{\rm SFR}$ values are inclination corrected, as for the azimuthally
averaged
 analysis (Section~\ref{subsec:radial}). We plot pixel-by-pixel
values of the
 $\Sigma_{\rm SFR}$ in NGC~6946 in Figure~\ref{fig:n6946_pix},
where we used a log-log
 representation of the $\Sigma_{\rm SFR}$ of the radio
and FUV/MIR data. We show the radio $\Sigma_{\rm SFR}$ at $\lambda 22~\rm
  cm$ both for the total RC emission and for the non-thermal RC emission only.
Data points
 are color coded according to radio spectral index between
$\lambda\lambda$
 22 and 18~cm, where red represents a spectral index
corresponding to young
 CREs, green an intermediate spectral index, and blue
data
 points with a spectral index dominated by old CREs. At
 wavelengths around
$\lambda 20\rm ~cm$ the RC emission is dominated by
 non-thermal synchrotron
emission. Young CREs have a non-thermal spectral index of
$\alpha_{\rm nt}\approx -0.6$
 \citep{green_09a}, which steepens as they age to
$\alpha_{\rm nt}< -1$, referred to as
 cosmic-ray ``aging''. We will refer
to the non-thermal spectral index
 as either ``young'' or ``old'' depending on
whether it is dominated
 by young or old CREs.  We also present the
 distribution of
the radio spectral index as maps with resolutions of $\rm 0.7~kpc$
 and $\rm
1.2~kpc$ matching those of the pixel-by-pixel plots. The colors used for
these maps are identical to those of the pixel-by-pixel plots. Results for
the
 entire sample are again shown in Appendix~\ref{app:sample}.
\subsection{Estimate of the Uncertainties}
\label{met:unc}
We calculated the error in the integrated flux densities from two error
contributions. The first is from an absolute calibration uncertainty, which is
assumed to be 5\% \citep{braun_07a}. We note that the relative calibration
error between WSRT
 $\lambda\lambda$ 22 and 18~cm is much smaller with $0.5\%$,
allowing us to
 reliably measure the radio spectral index. Secondly, the
rms noise of a
 map needs to be taken into account by $\sigma_{\rm N} =
\sigma \sqrt{N}$, where 
 $N$ is the number of independent beams and $\sigma$
the rms noise in the map. This effect becomes
 important the larger the
integrated area and thus the number of independent
 beams within an annulus
is. This error is, however, small in comparison to the
 intrinsic fluctuations
within one individual annulus. We therefore used the
 standard deviation of the
measured intensities within one annulus as our
 second error contribution. The
two error contributions were quadratically
 added within each annulus.
 Because the non-thermal RC flux densities are derived by subtracting the
  thermal RC emission from the total RC emission, we would need to include an additional
  error term stemming from the H$\alpha$ flux uncertainty. However, because
  the thermal RC contributes on average only 11\% to the global RC flux
  density, we neglect this error contribution.
For surface brightness measurements, and hence for extended emission, the
  estimated uncertainty for the \emph{GALEX} FUV maps is $\rm 0.15~mag$
  \citep{gildepaz_07a}, thus a 15\% calibration error. For the \emph{Spitzer}
  $24~\mu\rm m$ maps the photometric uncertainty is 2\% for both unresolved sources
  and extended emission \citep{engelbracht_07a}. As the
 hybrid $\Sigma_{\rm SFR}$ maps are a combination of FUV and
$24~\rm\mu m$ emission with
 the latter dominating, we use an average formal
calibration error of $5\%$ for our
 hybrid $\Sigma_{\rm SFR}$ maps.
We note that these are of course only the formal uncertainties to measure the
MIR and FUV flux densities. Converting them into measurements of the SFR
introduces further calibration errors. \citet{leroy_12a} compares the hybrid
SF tracer we use with various other tracers in order to determine the
calibration error. Their best estimate of the calibration error is $\pm 50\%$
for galaxy integrated SFRs and a factor of 2--3 for locally (1~kpc scale)
derived hybrid $\Sigma_{\rm SFR}$. In the following analysis, we would like to investigate to what
degree the RC emission can be used as an SF tracer for both global and local
(1~kpc) measurements. In order to do this, we use the formal errors only, for
both the radio and hybrid $\Sigma_{\rm SFR}$ maps, so that we can see how much
the radio-derived SFRs are deviating from the hybrid SFRs in comparison with the
formally expected uncertainties.

\citet{leroy_12a} proposed a correction for the so-called cirrus emission
in the MIR, which is not related to any recent SF but comes from the background
radiation field of the old stellar population. Since we used the hybrid $\Sigma_{\rm SFR}$
as originally proposed by \citet{leroy_08a}, no cirrus correction was
applied. We do not expect this to affect our analysis, because the cirrus
contribution to the $24~\mu\rm m$ MIR emission is only $19\%$ for global
measurements. Locally, the cirrus contribution becomes only important in the
outskirts of galaxies, where the FUV emission is anyway dominating the
$\Sigma_{\rm SFR}$. The difference between locally (1~kpc scale) derived
hybrid $\Sigma_{\rm SFR}$ values with and without the correction for cirrus
emission is typically less than $\approx 25\%$.
\section{Results}
\label{sec:results}
\subsection{Integrated RC--SFR Relation}
\label{res:int}
\begin{figure}[tbhp]
\centering
\resizebox{1.0\hsize}{!}{ \includegraphics{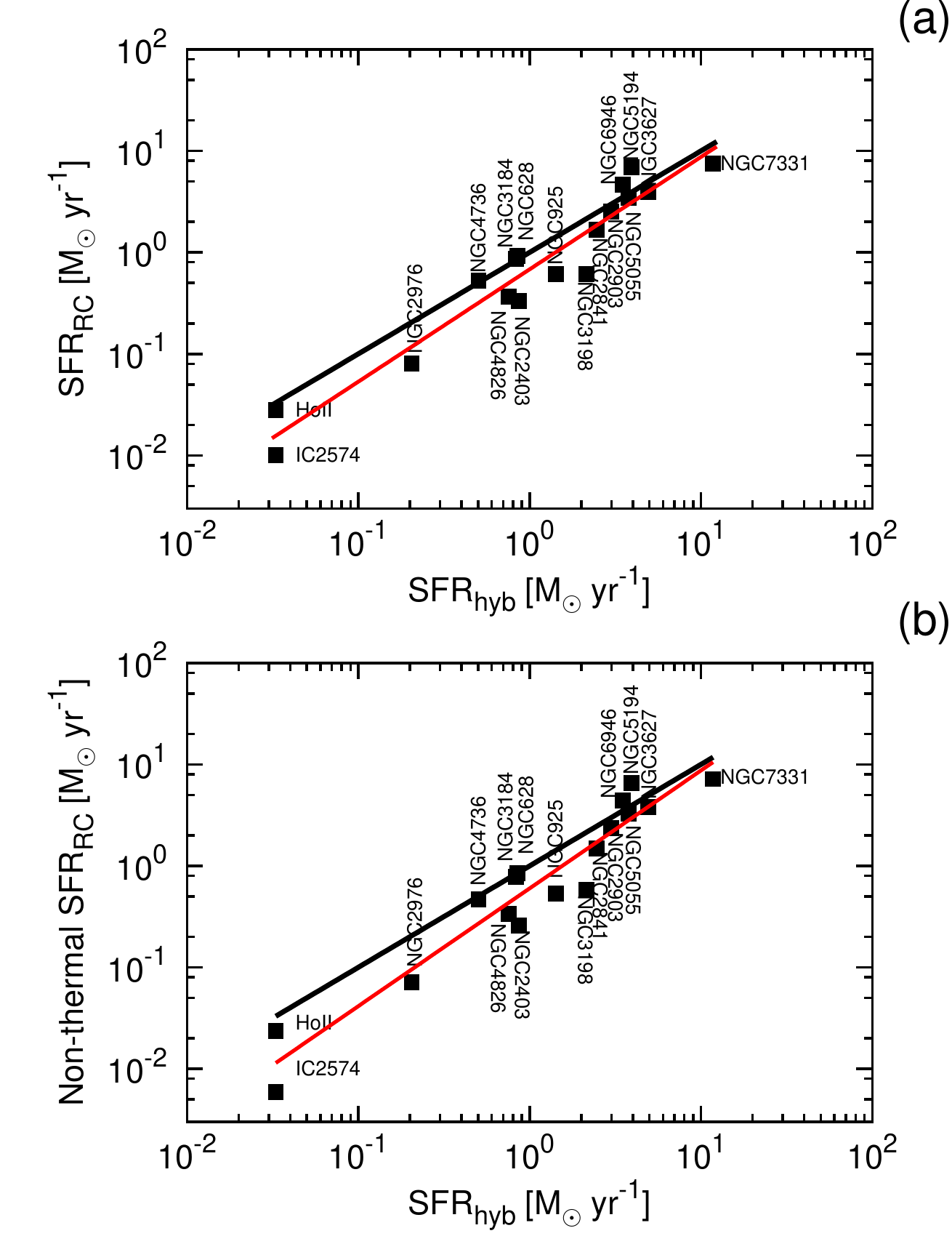}}
\caption{Comparison between the integrated SFRs from the total RC emission
  and
 hybrid SF tracers. The black line shows the relation expected from the
  Condon relation, i.e., would correspond to a linear (one-to-one)
  relation. The red line is a fit to the data points. (a) For the total RC
  emission. (b) For the non-thermal RC emission alone. The error bars, in both  x- and y-direction, are
 smaller than the size of the symbols.}
\label{fig:par_int_comb}
\end{figure}

As a first exercise, we calculate integrated radio luminosities and
SFRs for each galaxy and thus test the integrated RC--SFR
relation. For this we integrated both the radio intensity and the hybrid $\Sigma_{\rm SFR}$ in
elliptical annuli as explained in Section~\ref{subsec:radial}. The integrated
values are presented in Table~\ref{tab:sample} for each galaxy, where $\rm
SFR_{RC}$ is the radio SFR using the Condon relation from
Equation~(\ref{eq:sfr_total}) and $\rm SFR_{hyb}$ is the value as defined by the
hybrid $\Sigma_{\rm SFR}$ maps.
We present our result in Figure~\ref{fig:par_int_comb}(a) for the total RC emission and
in Figure~\ref{fig:par_int_comb}(b) for the non-thermal RC emission alone. Clearly,
the SFRs as derived from the RC luminosity and the hybrid SF tracers agree
well. Our sample covers 2--3 orders of magnitude in SFR. This confirms
the validity of the assumption that we can use the RC as an SF tracer. It also
confirms that Condon's relation holds for the integrated SFRs in our galaxy
sample. In the plot we show as a black line the relation that we would expect
from the Condon relation and in red a line derived from a fit to the data, where we use the following representation:
\begin{equation}
\log_{10}({\rm SFR_{RC}}) = \zeta \cdot \log_{10}({\rm SFR_{hyb}}) + \xi.
\label{eq:int_corr}
\end{equation}
We find $\zeta = 1.11\pm 0.08$ and $\xi=-0.17\pm0.05$ using a standard
bivariate least-squares fit for the total RC emission including thermal RC
emission. For the non-thermal RC emission we find $\zeta_{\rm nt}=1.16\pm0.08$
and $\xi_{\rm nt}=-0.22\pm0.06$, very similar to that for the total emission.
Strictly speaking, we cannot apply the Condon relation to the
  non-thermal RC emission only. But we would like to study the
  relation between the non-thermal RC emission and the hybrid SFR in order to
  better understand the origin of the former. This relation has important
  consequences, for instance, for the generation of magnetic fields in galaxies,
as we will discuss in Section~\ref{dis:mf-gas}.
Both fits depend quite heavily on the inclusion of IC~2574, a dwarf
  galaxy with a weak detection. Excluding IC~2574, the fit to the data
  becomes linear within the error margin ($\zeta=1.05\pm 0.09$).

In Table~\ref{tab:sample} we also show the ratio of integrated SFRs as calculated from
the RC and hybrid SF maps in each galaxy. We will use this ratio in various
places throughout the paper. Therefore, we define it here as
\begin{equation}
\mathscr{R}_{\rm int} \equiv \frac{\rm SFR_{RC}}{\rm SFR_{hyb}},
\label{eq:rat_int}
\end{equation}
for the integrated SFRs. Another way of looking at our data
is to take the average of this ratio. For the average ratio in our sample we
find for the total RC emission:
%
\[\left \langle \frac{\rm SFR_{RC}}{\rm SFR_{hyb}} \right\rangle = 0.78 \pm
0.38,\]
%
where the ratio is the arithmetic average and the quoted error is the standard
deviation of our measurement set. Similarly, we
find for the non-thermal RC emission only:
%
\[\left \langle \frac{\rm SFR_{RC}}{\rm SFR_{hyb}} \right\rangle_{\rm nt} =
0.70 \pm 0.39.\]
%
The galaxies in our sample lie systematically slightly below, although
individually they fall well within one standard deviation of the Condon
relation. This is a remarkable success of Condon's
 relation, particularly if we
consider the systematic effects involved in the way the
 relation was derived
(Section~\ref{subsec:radio_sfrd}). The spread in the
 ratio $\mathscr{R}_{\rm int}$ ranges from
$0.31$ (IC~2574) to $1.74$ (NGC~5194). The first
 galaxy is a dwarf irregular
Galaxy and the second one is a galaxy that is
 interacting with its nearby
neighbor. The standard deviation of $0.38$ is
 about half of the ratio
$\mathscr{R}_{\rm int}$;
hence, an individual galaxy can be either 
 radio dim or radio bright by a
factor of $\pm 50\%$. This is identical to the lower limit for the
uncertainty
 that is inherent to the hybrid SF tracers as well when used for
integrated SFRs \citep{leroy_12a}. We can
 thus conclude that for integrated
SFRs the RC luminosity measurements are of
 similar accuracy as SF values
derived on the basis of other tracers. In the next sections we
 investigate
whether this holds also locally in galaxies.
\subsection{Radial Profiles of Radio and Hybrid $\Sigma_{\rm SFR}$}
\label{res:rad}
As we have seen in Section~\ref{res:int}, the integrated RC luminosity and the
SFR from the hybrid SF tracers are directly proportional, as
expected from the Condon relation. We next proceed to investigate the
 behavior of the two SF tracers averaged in elliptical annuli (Figure~\ref{fig:n6946_rad}). 
We  see that the two $\Sigma_{\rm SFR}$ agree
quite well, with the radio $\Sigma_{\rm SFR}$ offset by an almost constant distance above the
hybrid $\Sigma_{\rm SFR}$. This means that the ratio of the radio to the hybrid $\Sigma_{\rm SFR}$ is
almost constant with galactocentric radius; this is shown in a separate plot. We define
  the local ratio of the radio $\Sigma_{\rm SFR}$ to the hybrid $\Sigma_{\rm SFR}$ as
\begin{equation}
\mathscr{R} \equiv \frac{\rm (\Sigma_{SFR})_{RC}}{\rm (\Sigma_{SFR})_{hyb}},
\end{equation}
equivalent to the integrated SFRs in Equation~(\ref{eq:rat_int}). If the Condon
relation were fulfilled, then the ratio should be equal to one. In the sample
galaxy shown, NGC~6946, the ratio is $\mathscr{R}\approx 1.4$ and is remarkably
constant as a function of galactocentric radius. There are some sinusoidal fluctuations
 with
a period of $\approx 3~\rm kpc$ in radius. These may be
 caused by the
prominent spiral arms that are located at $2.4$ and $\rm 5.4~kpc$ 
 in the hybrid $\Sigma_{\rm SFR}$
map. The radii of the spiral arms agree with the
 minima in $\mathscr{R}$, because
locally the hybrid SF tracer is proportionally higher than the more
 diffuse
RC-based tracer. There is a further minimum at 8~kpc where no obvious
 spiral
arm is located, although there are some SF regions in a northwestern direction
($\rm
 P.A.=-45\degr$) at that distance from the nucleus.  For NGC~6946 we
find a radially arithmetic average ratio of
 $\langle \mathscr{R}\rangle =1.35\pm0.25$, where the
error is the standard deviation of all data
 points. For this galaxy the
relative deviation of the ratio from its average
 is thus only $18\%$. The
integrated ratio as determined in
 Sect.~\ref{res:int} is $\mathscr{R}_{\rm int}=1.32$,
so these two are
 identical. The non-thermal radially averaged ratio is
$\langle \mathscr{R}
_{\rm nt}\rangle =1.28\pm0.26$, 
 identical within the error to the value
 for the
total RC emission, with a relative variation around the mean of 20\%.

\begin{figure}[tbhp]
\centering
\resizebox{1.0\hsize}{!}{ \includegraphics{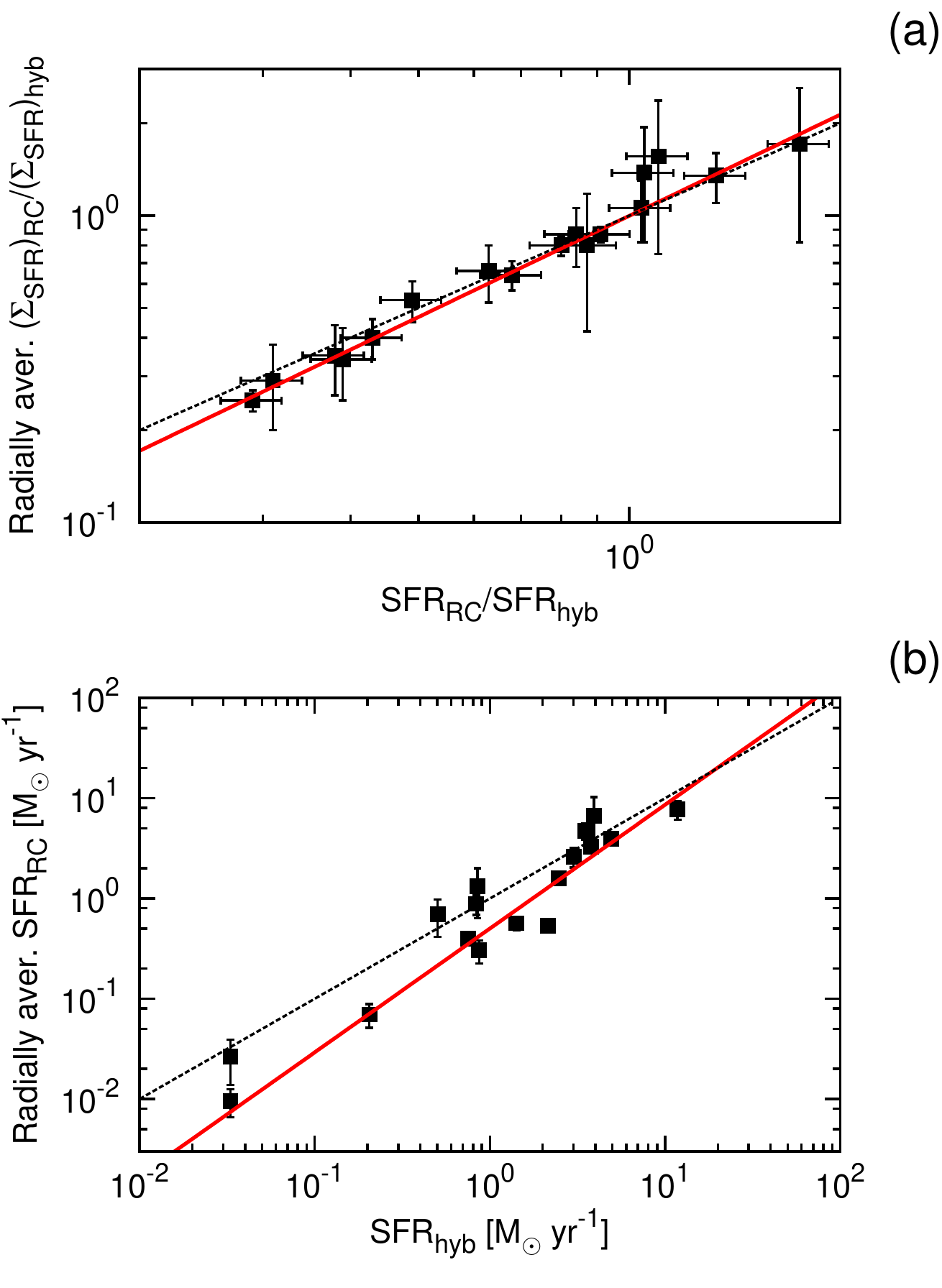}}
\caption{(a) Comparison between the radially averaged ratio $\langle \mathscr{R}\rangle$ and the
  integrated ratio $\mathscr{R}_{\rm int}$. The black dashed line shows the one-to-one
 relation,
  and the red line is a fit to the data points. (b) Comparison between the
  radially averaged radio $\rm \langle SFR_{RC}\rangle$ (which is defined by
  $\rm \langle SFR_{RC}\rangle \equiv \langle \mathscr{R}\rangle \cdot SFR_{hyb}$ ) and the integrated hybrid $\rm SFR_{hyb}$. The
 black
  dashed line indicates a one-to-one relation, and the red line is a fit to the
  data points. The shown error bars are derived from the
 standard deviation of
  the radial data points from the arithmetic
 average.}
\label{fig:par_av_comb}
\end{figure}

This trend of a constant ratio $\mathscr{R}$ as a function of galactocentric radius is also found in
the other galaxies. In Table~\ref{tab:sample} we present the integrated
values of the SFR both from the radio ($\rm SFR_{RC}$) and from the hybrid
data ($\rm SFR_{hyb}$) and their ratio ($\mathscr{R}_{\rm int}$). We also list the
radially averaged ratio of the local ratio $\langle \mathscr{R}\rangle$. It is instructive to
compare the radially averaged ratio with the integrated ratio for all of our
sample galaxies (Figure~\ref{fig:par_av_comb}(a)). Both ratios agree remarkably
well, and the radial ratios agree within the errors with the
integrated ones. We can also see from that figure  that the radial variation
compared to the average is small for most of the galaxies. The mean standard
deviation is $25\%$ of the radially averaged ratio. NGC~6946 is thus a
typical example with respect to the amplitude of its radial variation of the
ratio. In Figure~\ref{fig:par_av_comb}(b) we show the relation between the
radially averaged radio SFR and the SFR as measured from the hybrid SF tracers,
which can be compared with Figure~\ref{fig:par_int_comb}(a). The agreement
between the two figures is striking, underlining our conclusion that the
azimuthally averaged ratios behave very similarly to the integrated ones.
The radial variation is mostly changing quasi-periodic, which means it cannot be
described by a simple common function like a constant slope. Generally
speaking, the ratio tends to increase with galactocentric radius rather than
decrease. The best example for this is NGC~628, where the ratio increases almost
monotonically with galactocentric radius. A more step-like increase is seen in
NGC~2403, where the ratio jumps at a radius of 3~kpc from $0.3$ to $0.45$. There
are no prominent cases where the ratio drops as a function of galactocentric radius, although
NGC~2976 shows a monotonic decrease beyond a maximum at $\rm 1.5~kpc$.  Three
galaxies have a relative ratio variation close to 50\%, which is the largest
variation in our sample. These are Holmberg~II, NGC~628, and NGC~5194. At the
other end of the scale are NGC~2841, NGC~3198, and NGC~3627, where the
relative radial ratio variation is close to 10\%. 
In conclusion, we find that the radial variation appears to be quasi-periodic
and cannot be easily described by a linear fit, for instance. We do not find
any obvious candidate for a second parameter that could explain this
behavior. Relative variations are small, however, and the arithmetic average
of the radial data points agrees, within the standard deviation, with
integrated ratios of RC to hybrid SFRs.
\subsection{Local RC--SFR Relation}
\label{res:pix}
Next we turn to the local correlation between the $\Sigma_{\rm SFR}$ as measured from the
RC emission and hybrid SF tracers. For this we use the pixel-by-pixel data
measured as described in Section~\ref{subsec:pixel}. In
Figure~\ref{fig:n6946_pix} we
 show the result for NGC~6946. We plot the $\Sigma_{\rm SFR}$
of the RC emission as a
 function of the hybrid $\Sigma_{\rm SFR}$ at scales of
both $\rm 1.2~kpc$
and $\rm 0.7~kpc$. We chose these scales to maximize the numbers of galaxies for
each scale, depending on their distance and angular resolution, so that we can
test whether our results are sensitive to the actual spatial resolution
used. Both
 $\Sigma_{\rm SFR}$ are shown on a logarithmic scale. In this double logarithmic
plot a
 clear correlation between the radio and hybrid $\Sigma_{\rm SFR}$ is visible
at
 both scales. The correlation as expected from the Condon relation is shown
as
 a black dashed line. For NGC~6946 most data points lie above the Condon
relation. This is expected as this galaxy is radio bright, i.e., has a ratio
of radio to hybrid $\Sigma_{\rm SFR}$ of $\mathscr{R}_{\rm int}=1.3$ as measured in
Section~\ref{res:int}.
Motivated by the observation of a correlation between the radio and hybrid
$\Sigma_{\rm SFR}$, we use the following representation of the radio
$\Sigma_{\rm SFR}$ as
 a function of the hybrid $\Sigma_{\rm SFR}$
:
\begin{equation}
  \log_{10}({\rm (\Sigma_{SFR})_{RC}}) = \delta \cdot \log_{10}({\rm (\Sigma_{SFR})_{hyb}}) + \epsilon,
\label{eq:pix_corr}
\end{equation}
where $\delta$ and $\epsilon$ are constants. This definition is equivalent to the one
given in Section~\ref{res:rad} (Equation~(\ref{eq:int_corr})).
In Figure~\ref{fig:n6946_pix} we show the result  of a fit to the pixel-by-pixel
data for NGC~6946 as a red solid line in both the $1.2$ and $\rm 0.7~kpc$ scale
plots. The line is a fit to the data: the deviation of the data points
represents typically a factor of two from the fit. The measured slope for the
data in NGC~6946 is $\delta=0.79\pm0.03$ at a scale of $\rm 1.2~kpc$ and
$\delta=0.71\pm0.01$ at $\rm 0.7~kpc$.  The fit has a slope significantly smaller
than one, the value expected from the Condon relation.
The results for the rest of our sample are similar to NGC~6946 in the sense
that they all have sub-linear slopes. The average of all galaxies at
$\rm 1.2~kpc$ scale is $\langle\delta\rangle=0.63\pm0.25$ for the total emission and
$\langle\delta_{\rm nt}\rangle=0.61\pm0.29$ for the non-thermal emission alone. Errors are
standard deviation and give thus the typical deviation of the slope for a
particular galaxy from the mean. At a scale of $\rm 0.7~kpc$ we find
$\langle\delta\rangle=0.63\pm 0.22$ for the total emission and $\langle\delta_{\rm
  nt}\rangle=0.58 \pm 0.22$ for the non-thermal emission alone. The results for the
two scales thus virtually agree. We also note that the results for the
non-thermal emission alone show only marginally more shallow slopes than for
the total emission.

\begin{figure}[tbhp]
\centering
\resizebox{1.0\hsize}{!}{ \includegraphics{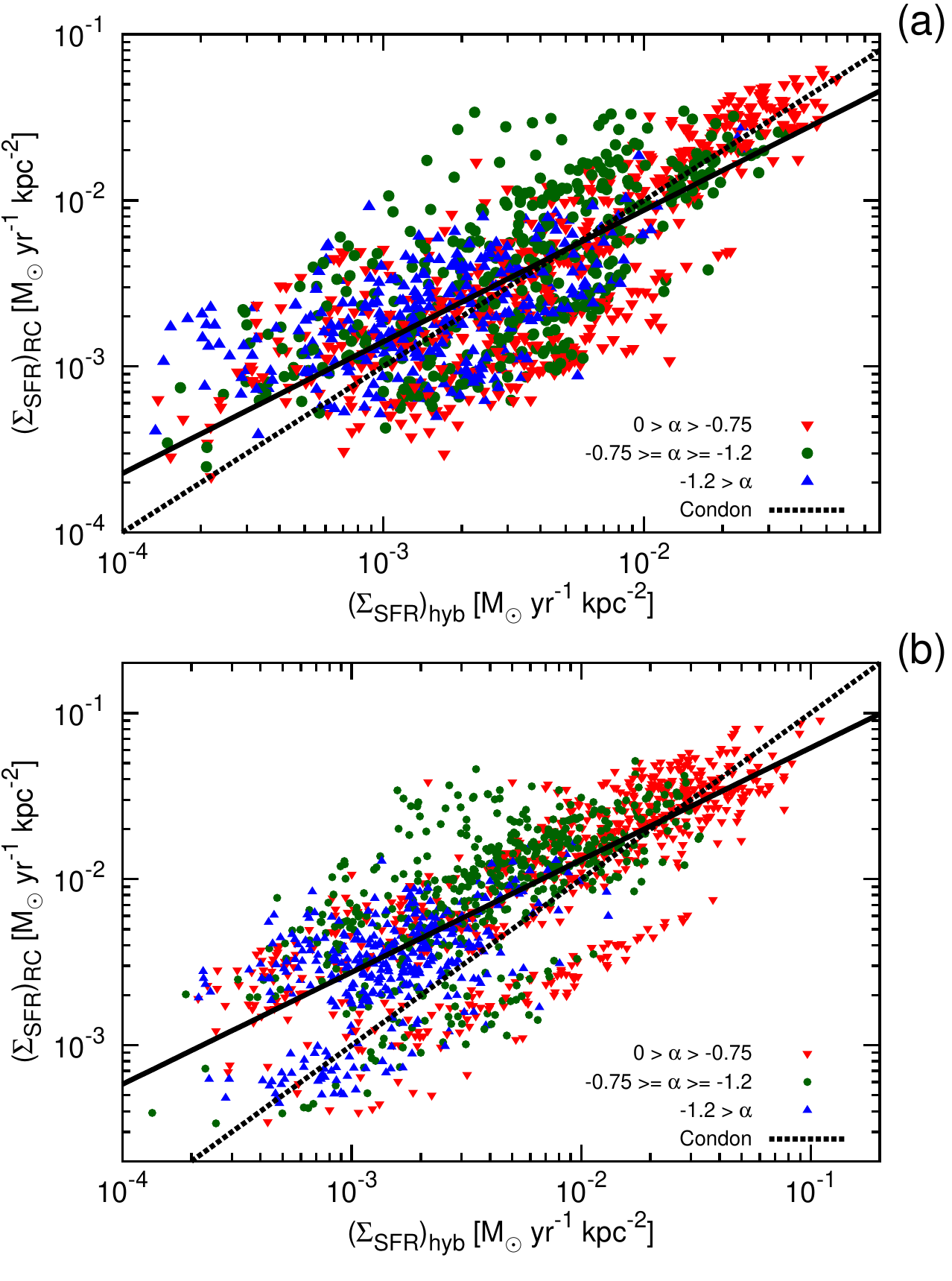}}
\caption{Cumulative pixel-by-pixel plot of all galaxies at $\rm 1.2~kpc$
  (upper panel a) and $\rm 0.7~kpc$ (lower panel b) resolution, where we
  present the radio $\Sigma_{\rm SFR}$ as function of hybrid $\Sigma_{\rm
    SFR}$. The dashed line shows the one-to-one Condon
 relation and the solid
  line is a linear least-square fit to the data
 points. Red data points
  indicate a ``young'' non-thermal spectral index ($\alpha > -0.75$), blue data
  points an ``old'' non-thermal spectral index ($\alpha<-1.2$) and green data
  points a non-thermal spectral index of intermediate age ($-0.75\geq \alpha
  \geq -1.2$). The typical uncertainties for the radio spectral index are 0.4,
  0.2
 and 0.1 for the blue, green, and red data points respectively.}
\label{fig:grand}
\end{figure}

We found a slope $\delta$ between $0.22$ and $1.21$ for the 12 galaxies in our
sample for which we have $\rm 1.2~kpc$ spatial resolution. The flattest slopes are
found in NGC~3198 ($0.22$) and Holmberg~II ($0.26$), whereas the steepest (and the
only super-linear slope) is found in IC~2574 ($1.21$).
The sub-linear slopes mean that within an individual galaxy data points
 at low
hybrid $\Sigma_{\rm SFR}$ are slightly more radio bright than expected from a linear
correlation. In Figure~\ref{fig:n6946_maps} we color coded all data points in
NGC~6946 depending on their radio spectral index between $\lambda\lambda$ 22
and 18~cm. Data points with ``young'' non-thermal spectral indices
($\alpha\geq -0.75$), color coded in red, are lying predominantly to the right
-hand side at high $\Sigma_{\rm SFR}$; data points with ``old'' non-thermal spectral indices
($\alpha<-1.2$), color coded in blue, are lying almost exclusively at the left
-hand side at low $\Sigma_{\rm SFR}$. Green data points with intermediate spectral indices
($-0.75>\alpha \geq -1.2$) lie in between at medium $\Sigma_{\rm SFR}$. It is remarkable how well
the data points separate depending on their radio spectral index, at both $1.2$
and $\rm 0.7~kpc$ scales. Combined with our result of sub-linear slopes, for
NGC~6946 we have $\delta=0.79\pm0.03$, and we find that data points dominated by
old CREs have large ratios $\mathscr{R}$ compared to the average,
whereas data points dominated by young CREs have low ratios
$\mathscr{R}$. This result should be taken only in a relative sense, as NGC~6946 is
radio bright overall with respect to integrated RC emission ($\mathscr{R}_{\rm
  int}=1.3$), most data points lying above the Condon relation.
\begin{figure}[tbhp]
\centering
\resizebox{1.0\hsize}{!}{ \includegraphics{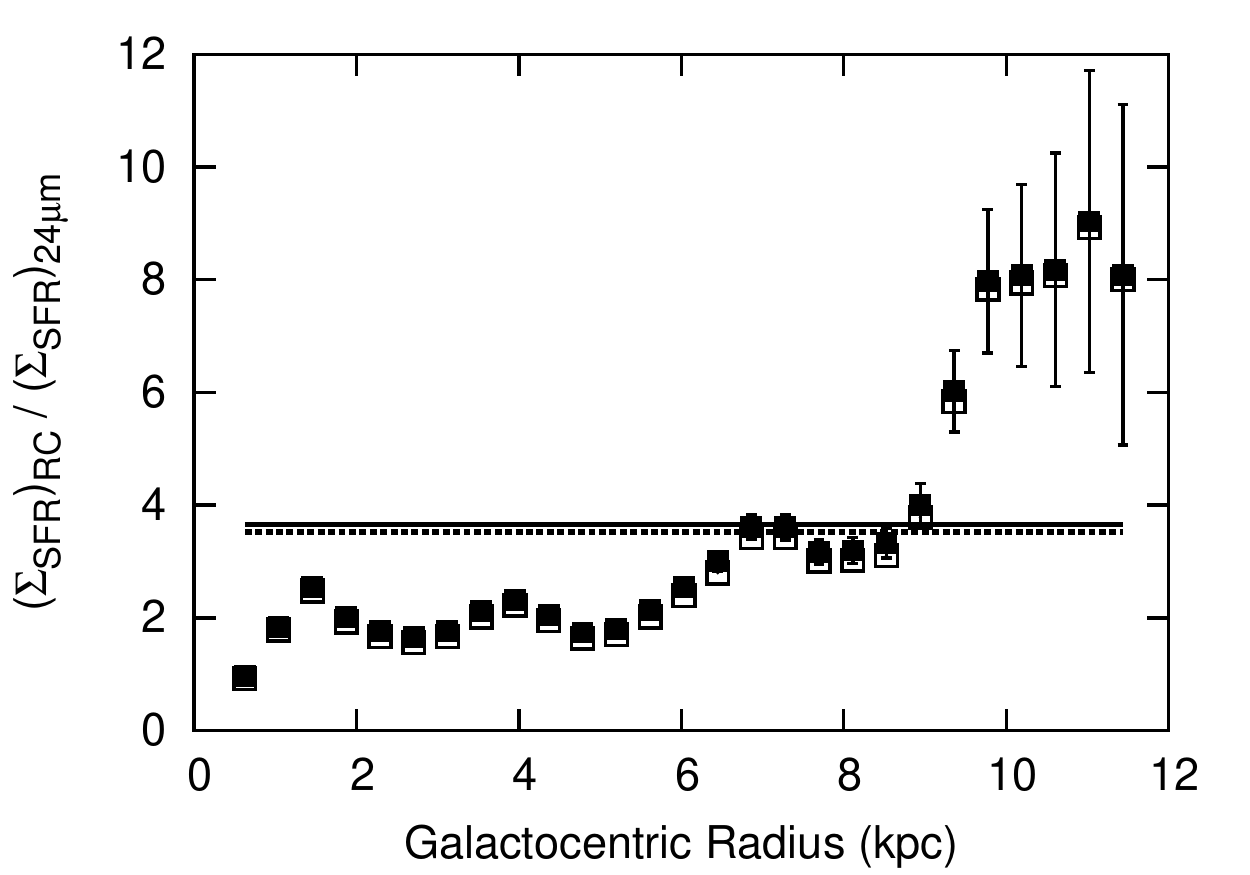}}
\caption{NGC~6946. Ratio $\mathscr{R}_{\rm 24~\mu m}$ of radio $\Sigma_{\rm SFR}$ to
  the $\Sigma_{\rm SFR}$ as derived
 from \emph{Spitzer} $\rm 24~\mu m$
  alone. Open symbols represent the
 non-thermal RC emission alone.  The solid
  (dashed) line shows the
 least-squares fit assuming a constant ratio (for the
  non-thermal RC alone). To be compared with
 Figure~\ref{fig:n6946_rad}(b).}
\label{fig:n6946-fit-24mu}
\end{figure}

In our sample at $\rm 1.2~kpc$ scale, with 12 galaxies, four are showing this clear
separation of the data points in the pixel-by-pixel plots depending on the
radio spectral index. These are NGC~2403, NGC~4736, NGC~5194, and
NGC~6946. A fifth galaxy, NGC~2976, has data points that may be separated a
bit less clearly than in the other galaxies. In all cases, data points with
young non-thermal spectral indices are relatively radio dim, whereas data points
with ``old'' non-thermal spectral indices are radio bright. CREs
diffusing away from SF sites undergo spectral aging, resulting in a
steeper observed radio spectral index. This could explain our observed
results, which we will discuss in more detail in Section~\ref{dis:crs}.
In Figure~\ref{fig:grand}(a) we show the cumulative pixel-by-pixel plot for all
galaxies combined at a scale of $\rm 1.2~kpc$. We have color coded the
radio spectral index as before, with red data points indicating a ``young''
non-thermal radio spectral index, green data indicating an intermediate age,
and blue data points indicating an ``old'' non-thermal spectral index. We show the
relation as expected from Condon's relation as a solid black line. We fitted
the data shown as a dashed line where we found $\delta=0.79\pm 0.02$ and
$\epsilon = -0.47\pm 0.04$. Obviously, the spread around the fit is
significantly larger than within individual galaxies. This is due to galaxies
having integrated ratios $\mathscr{R}_{\rm int}$ that vary by a factor of 2. For hybrid $\Sigma_{\rm SFR}$
of $\gtrapprox 10^{-2}~M_{\odot}~\rm  yr^{-1}~kpc^{-2}$, the spread becomes
smaller, though, and the data points mostly agree within a factor of 2--3 with
the Condon relation. For these high $\Sigma_{\rm SFR}$ the radio spectral index is
predominantly of a ``young'' non-thermal radio spectral index with $\alpha \geq
-0.75$. We note that the radio spectral index of most young supernova remnants
  is between $-0.5$ and $-0.6$ \citep{green_09a}, so that even in these areas
  we have some contribution from an aged population of CREs.
Figure~\ref{fig:grand}(b) shows the cumulative pixel-by-pixel plot for all
galaxies at a scale of $\rm 0.7~kpc$, which is otherwise identical to
Figure~\ref{fig:grand}(a). Due to the smaller number of galaxies, there is a
gap where one galaxy (NGC~2403) is clearly radio dim in comparison to the other
galaxies. A least-squares fit to the data gives $\delta=0.68\pm 0.01$ and
$\epsilon=-0.53\pm 0.03$. Hence, the slope is less steep than at $\rm 1.2~kpc$
resolution. This can possibly be explained by cosmic-ray transport within the
disk, which we discuss in Section\ref{dis:crs}.
\subsection{The Importance of FUV as an SF Tracer}
\label{res:fuv}
The $\rm 24~\mu m$ MIR emission can be used as an SF indicator in regions
dominated by molecular gas where the associated dust absorbs the UV photons
from young stars. This is predominantly the case in the inner $\rm H_2$
dominated parts of galaxies. In the outer parts, where the dust column density
is lower, young stars can be better traced directly by their UV emission as
traced by the {\em GALEX} FUV maps. This is shown in \citet{leroy_08a}, where
the $\Sigma_{\rm SFR}$ maps of NGC~6946 based on MIPS $24~\mu\rm m$ only and on {\em
  GALEX} FUV emission only are presented (their Appendix F).
To quantify the importance of FUV as an SF tracer in addition to MIR emission, we
carried out the same analysis as explained in Sections.~\ref{res:rad} and
\ref{res:pix}, namely, that of the radial and pixel-by-pixel behavior
of the SF based on the MIPS $\rm 24~\mu m$ emission alone. This is in principle
equivalent to measuring the local ratio of RC to $\rm 24~\mu m$ MIR intensities,
$q_{24}\equiv\log_{10}(I_{\rm 22\, cm}/I_{\rm 24\, \mu m}$). 
Figure~\ref{fig:n6946_rad} shows that the hybrid $\Sigma_{\rm SFR}$ based on a linear
combination of FUV and $\rm 24~\mu m$ follows much better the radio $\Sigma_{\rm SFR}$ than
the $\rm 24~\mu m$ emission alone. The MIR starts
to drop considerably at larger radii beyond 8~kpc where the FUV emission
becomes important out to a galactocentric radius of 12~kpc. At small radii, within 5~kpc,
$\rm 24~\mu m$ emission dominates over FUV, and at intermediate radii between
5 and 8~kpc both contribute similar amounts to the hybrid $\Sigma_{\rm SFR}$. The ratio of
RC to $\rm 24~\mu m$ emission shown in Figure~\ref{fig:n6946-fit-24mu} is
thus not constant to the same degree with galactocentric radius as is the ratio of radio to
hybrid $\Sigma_{\rm SFR}$. The relative radial variation is only 18\% for the RC to hybrid
ratio, whereas it is 66\% for the RC to $\rm 24~\mu m$ ratio. This difference
is also found in the galaxies in the rest of the sample. The averaged relative
radial variation of the RC to $\rm 24~\mu m$ ratio is 35\% for the total RC
emission and reaches 40\% for the non-thermal emission alone. This compares with
25\% (28\%) for the radio (non-thermal) to hybrid $\Sigma_{\rm SFR}$ ratio.
MIR $\rm 24~\mu m$  emission on its own is not sufficient for use as an SF tracer.
The ratio of RC to $\rm 24~\mu m$ MIR emission, $q_{24}$, is not
constant and, if used at all, is only a valid tracer within the $\rm H_2$
dominated region in a galaxy.

\section{Discussion}
\label{sec:discussion}
\subsection{Global RC--SFR Relation}
\label{dis:global}
The RC--SFR relation holds globally, and the RC emission derived SFR is
to a good approximation linearly proportional to the SFR as measured by the hybrid SF tracers
(Section~\ref{res:int}). This essentially implies that the ratio $\mathscr{R}_{\rm int}$
of SFRs measured by the RC to hybrid SF tracers is constant within our entire
sample of galaxies. We do not find any obvious correlation with any of the
fundamental global galaxy parameters as shown in Appendix~\ref{app:par}. We
note that we only find a factor of $\pm 50\%$ variation in the integrated ratio
$\mathscr{R}_{\rm int}$ across our galaxy sample. This factor could be easily
accounted for by the uncertainty in the hybrid SF tracer as discussed in
detail in \citet{leroy_12a}.  

As has been mentioned in Section~\ref{sec:introduction}, the Condon relation
is based on a calorimetric model. In this case the RC emission is proportional
to the CRE production rate, which is proportional to the frequency of Type~II
supernovae and hence also proportional to the SFR. Cosmic rays are assumed to be
trapped long enough within a galaxy so that 
 their radiation timescale (set by
synchrotron and
 IC losses) is shorter than their escape
time. Also, it is assumed
 that CRE emit a constant fraction of synchrotron
relative 
 to IC radiation. Furthermore, one has to account for
the various timescales that can influence the validity of the RC--SFR
relation. The fastest timescale is that of O stars with a few $10^6~\rm yr$
that emit the ionizing UV radiation and are responsible for creating and
sustaining the \ion{H}{2} regions. The thermal RC emission has hence the same
short timescale. Similarly, the $24~\mu\rm m$ MIR emission stems from heating
by O stars and has the same short timescale. The \emph{GALEX} emission traces
FUV emission as emitted by O and B stars with a timescale of 10--100~Myr. The RC emission has the longest timescale, assuming that stars above a
mass of $\approx 8~M_\odot$ evolve into a Type~II supernova \citep{heger_03a},
which do have a lifetime of a few $10^7~\rm yr$. This is the time the RC
emission needs to ramp up after a single starburst. Furthermore, in contrast
to the other tracers that vanish once the O and B stars have gone off as a
Type~II supernova, the RC emission decays slowly. For a typical magnetic field
strength of $B=8~\rm\mu G$, the CRE lifetime due to synchrotron radiation is
$3\times 10^7~\rm yr$. In order for the RC--SFR relation to work, there has to
be a quasi-steady state resulting in a constant or slowly varying SFR over
time. In most galaxies this condition is fulfilled when averaging over areas
that are $\approx 1$~kpc or larger, which ensures an ensemble of SF regions at
any one time. This assumption of course definitely holds in spiral galaxies,
when investigated globally. It becomes, however, an issue with dwarf irregular
galaxies, which have a more burst-like SF history.
Galaxies can differ from the
 idealized calorimetric model in several
ways. Firstly, cosmic rays can escape in a
 galactic wind or by diffusion along
vertical magnetic field lines
\citep[e.g.,][]{heesen_09a,heesen_09b}. Secondly, the ratio of the magnetic
 to
photon energy density may vary, resulting in a change of
 synchrotron to
IC radiation losses. The non-calorimetric model by
\citet{niklas_97a} addresses the escape of CREs in an elegant way. They assume
that the strength of the magnetic field is related to the gas mass and hence
to the SFR, assuming a Kennicutt--Schmidt SF law. The magnetic field regulates
the energy content that is stored in the CREs. This makes sense, as it is known
that the cosmic-ray energy density (including CREs and the dominating proton
component) and the magnetic field energy density are approximately in energy
equipartition \citep[e.g.,][]{beck_96a,beck_05a}. This is motivated by the
fact that too large a
 cosmic-ray energy density would give rise to a Parker
instability where the
 relativistic gas with its buoyancy escapes from the
disk. Also, the total energy stored in the magnetic field and cosmic rays
together is close to the minimum value in case of energy equipartition.

Energy equipartition is a hotly debated subject. The best evidence in support
of it comes from gamma-ray observations, which allow us to directly derive the
CRE spectrum and hence measure the CRE energy density. \citet{strong_11a} used
\emph{Fermi}--LAT observations to determine the magnetic field strength at the
position of the Sun as $B_{\rm ran}=7.5~\mu\rm G$ (for the dominating random
component). The non-thermal RC model of the Milky Way by \citet{beuermann_85a}
can be used to derive the magnetic field strength assuming equipartition as
$6\pm 2~\mu\rm G$ at the solar radius \citep[Berkhuijsen, in][]{beck_01a}. \citet{sun_08a} found
$4~\mu\rm G$ using rotation measures of polarized background sources. For
external galaxies there are only a handful of detections in gamma-ray
emission, which do suggest energy equipartition in the Large Magellanic Cloud
\citep{mao_12a}, a dwarf irregular galaxy, and in the starburst galaxies
NGC~253 \citep{rephaeli_10a} and M82 \citep{yoast-hull_13a}. For starburst
galaxies the assumption of energy equipartition may give the true magnetic
field strength only by sheer coincidence \citep{lacki_13a}, but staying away
from those, for our sample of normal star-forming galaxies the
 assumption of
equipartition is a reasonable one. In the following we will discuss aspects of
the calorimetric model by \citet{voelk_89a} and the non-calorimetric model by
\citet{niklas_97a}.
\begin{figure*}[tbhp]
  \centering
  \resizebox{0.75\hsize}{!}{ \includegraphics{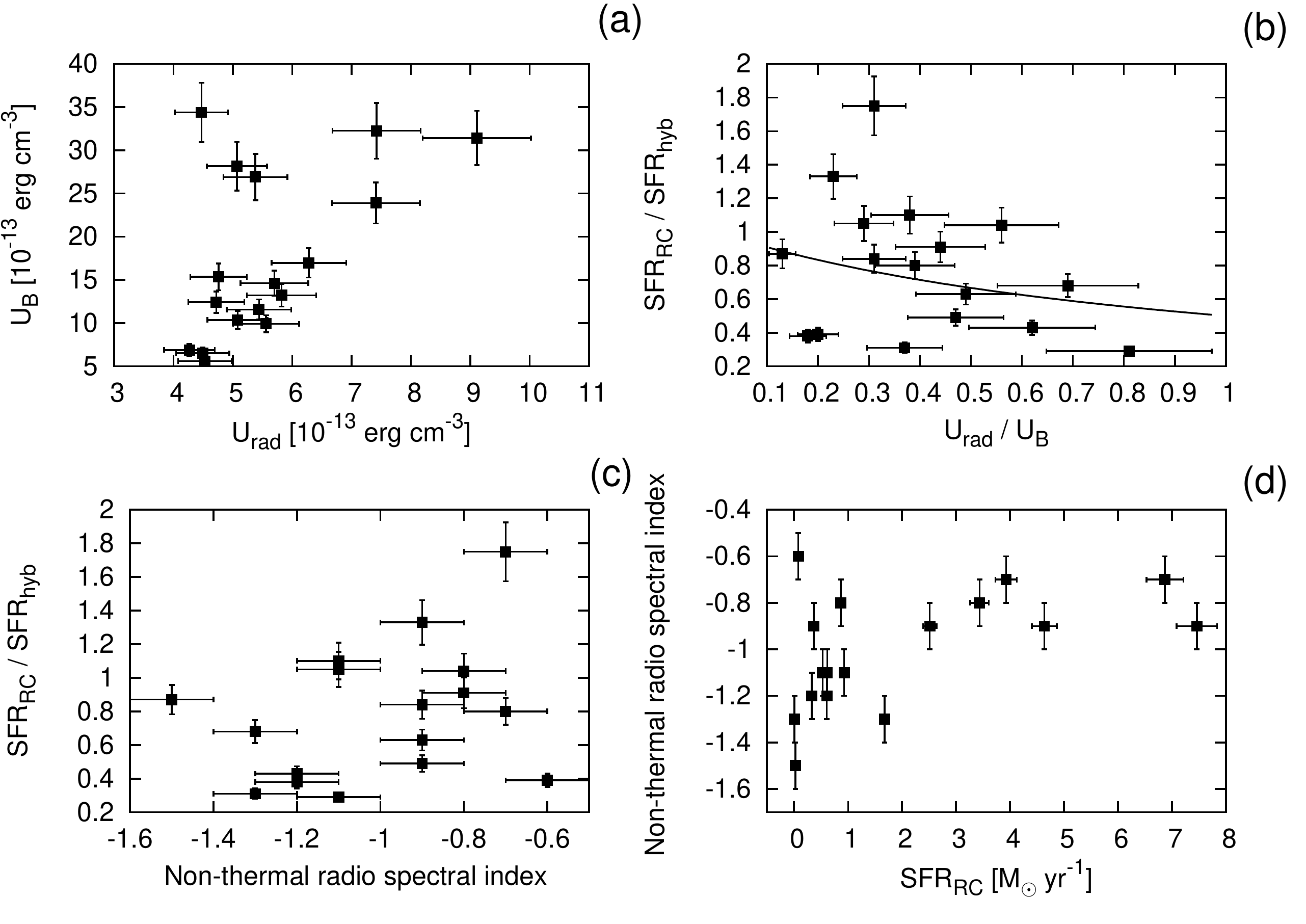}}
  \caption{Test of the calorimetric model. (a) Magnetic field energy density
    $U_{\rm B}$ as a function of radiation energy density $U_{\rm rad}$. (b)
    Ratio $\Re_{\rm int}$ of integrated radio to hybrid SFR as a function of
    ratio of radiation energy density to magnetic field energy density. The
    solid line shows the theoretical expectation. (c) Ratio $\Re_{\rm int}$
    of integrated radio to hybrid SFR as a function of the integrated
    non-thermal radio spectral index. (d) Non-thermal radio spectral index
    as a function of the radio SFR.}
  \label{fig:calor}
  \end{figure*}
\begin{deluxetable}{lccc}
\tabletypesize{\scriptsize}
\tablecaption{Data to Test the Calorimetric Model.\label{tab:calor}}
\tablewidth{0pt}
\tablehead{
\colhead{Galaxy\B} & \colhead{$U_{\rm rad}$} & \colhead{$U_{\rm rad}/U_{\rm B}$} & \colhead{$\alpha_{\rm nt}$} \\
& \colhead{$(10^{-13}~\rm erg\,cm^{-3})$} & & \\
\colhead{(1\T)} & \colhead{(2)} & \colhead{(3)} & \colhead{(4)}}
\startdata
Ho~II     & $5.2$ & $0.11$ & $-1.5\pm0.1$  \\
IC~2574   & $1.0$ & $0.29$ & $-1.3\pm0.1$  \\
NGC~628   & $5.9$ & $0.35$ & $-1.1\pm0.1$  \\
NGC~925   & $4.7$ & $0.58$ & $-1.2\pm0.1$  \\
NGC~2403  & $6.9$ & $0.16$ & $-1.2\pm0.1$  \\
NGC~2841  & $5.3$ & $0.65$ & $-1.3\pm0.1$  \\
NGC~2903  & $6.0$ & $0.30$ & $-0.9\pm0.1$  \\
NGC~2976  & $7.7$ & $0.19$ & $-0.6\pm0.1$  \\
NGC~3184  & $8.2$ & $0.53$ & $-0.8\pm0.1$  \\
NGC~3198  & $5.4$ & $0.79$ & $-1.1\pm0.1$  \\
NGC~3627  & $8.6$ & $0.38$ & $-0.7\pm0.1$  \\
NGC~4736  & $1.8$ & $0.27$ & $-1.1\pm0.1$  \\
NGC~4826  & $7.9$ & $0.44$ & $-0.9\pm0.1$  \\
NGC~5055  & $8.9$ & $0.43$ & $-0.8\pm0.1$  \\
NGC~5194  & $1.3$ & $0.31$ & $-0.7\pm0.1$  \\
NGC~6946  & $1.3$ & $0.22$ & $-0.9\pm0.1$  \\
NGC~7331  & $6.9$ & $0.48$ & $-0.9\pm0.1$  
\enddata                                                                                                                       
\tablecomments{Column 1: galaxy name; Column 2: radiation energy density; Column 3: ratio of radiation energy density to magnetic field energy density; Column 4: non-thermal radio spectral index between $\lambda\lambda$ 22 and 18 cm.}
\end{deluxetable}

\subsubsection{Inverse Compton Losses}
The calorimetric model predicts that the
ratio $\mathscr{R}_{\rm int}$ of radio to hybrid SFR is proportional to the inverse
of $1+U_{\rm rad} /
 U_{\rm B}$, where $U_{\rm rad}$ is the radiation energy
density and $U_{\rm
 B}$ the magnetic field energy density. This is motivated
by the fact that
 the ratio of IC to synchrotron radiation of
CREs is proportional to the ratio of radiation to magnetic
field energy density,
 $U_{\rm rad} / U_{\rm B}$. We calculated the radiation
energy density
 $U_{\rm rad}$ using the FIR luminosity $L_{\rm FIR}$ from
\citet{galametz_13a}. For NGC~2403 and NGC~5194 we obtained the FIR
luminosity from \citet{draine_07a} and for NGC~2903 from
\citet{sanders_03a}. After correcting for our
 distances and excluding nuclear
starbursts and background galaxies, the
 FIR radiation energy density can be
calculated as $ U_{\rm rad,FIR} = L_{\rm
 FIR}/(2\pi r_{\rm int}^2 c)$. Here,
$r_{\rm int}$ is the galactocentric radius of the galaxy
 to which we integrated the radio
flux densities, and $c$ is the speed of light. We
 have to add $U_{\rm CMB} =
4.1\times 10^{-13}~\rm erg\,cm^{-3}$, the
 radiation energy density of the
cosmic microwave background (CMB) at redshift zero,
 and the stellar radiation energy
density $U_{\rm rad,stars}$.  Following \citet{tabatabaei_13a}, we scaled
$U_{\rm rad,stars}$ to
 the FIR radiation energy density via $U_{\rm
  rad,stars}=1.73\times U_{\rm rad,FIR}$, which is obtained from the values
for the Milky Way in the solar neighborhood
\citep{mathis_83a,draine_11a}. Resulting radiation energy densities are
tabulated in Table~\ref{tab:calor}. We calculated the magnetic field
energy density using the equipartition approximation between the cosmic-ray
and
 magnetic field energy density \citep{beck_05a}. We used the usual
assumption for the proton-to-electron ratio, $k =100$,
 which is the observed
value in the solar system as direct measurements of the cosmic-ray proton and
electron fluxes show \citep{adriani_11a,shikaze_07a}.  For the integration
length (vertical to the galactic plane) we chose 1~kpc and for the non-thermal spectral index $\alpha_{\rm
  nt}=-0.85$. Resulting ratios $U_{\rm rad} / U_{\rm B}$ of radiation energy
density to the magnetic field energy density are tabulated in
Table~\ref{tab:calor}.
The average magnetic field energy density is
 $\langle U_{\rm B}\rangle  = 2.8\times
10^{-12}~\rm erg~ cm^{-3}$ (corresponding to $\langle B\rangle =8.4~\mu\rm G$), the
average radiation
 energy density is $\langle U_{\rm
  rad}\rangle  = 8.4\times 10^{-13}~\rm erg~ cm^{-3}$, and
 the average ratio is
$\langle U_{\rm rad}/U_{\rm B}\rangle =0.38$. This is a similar result to that of
\citet{niklas_95a}, who found an average magnetic field energy density of
$\langle U_{\rm B}\rangle =3.2\times 10^{-12}~\rm erg\,cm^{-3}$ in his sample.
 Firstly, we
can test whether the ratio $U_{\rm rad}/
 U_{\rm B}$ is constant within our
sample. For this we plot in Figure~\ref{fig:calor}(a) the dependence of $U_{\rm
  B}$ as a function of $U_{\rm rad}$. There is no clear correlation between them,
resulting in a spread of the ratio $U_{\rm rad}/
U_{\rm B}$, which ranges
between $0.11$ (Holmberg~II) and $0.79$ (NGC~3198). Because of the rather small
ratio $\langle U_{\rm rad}/U_{\rm B}\rangle =0.38$, we expect only a spread of $\pm
40\%$ in the RC luminosity. This is comparable to the uncertainty of $\pm50\%$
in the integrated SFRs from the hybrid SF tracer. Hence, we cannot expect to
detect a clear influence of IC radiation on the RC in comparison
to the hybrid SFR. In Figure~\ref{fig:calor}(b) we present the ratio
$\mathscr{R}_{\rm int}$ of
radio to hybrid SFR as a function of the ratio $U_{\rm rad}/
U_{\rm B}$ and as a
line the theoretical expectation from the inverse of $1+U_{\rm rad}/
U_{\rm B}$
normalized to our observed averaged ratio $\langle \mathscr{R}_{\rm int}\rangle =0.8$. We can see that the scatter
of the ratio $\mathscr{R}_{\rm int}$ is indeed too high to be solely explained by a variation of
the IC radiation in comparison to synchrotron radiation.
\subsubsection {Cosmic-ray Escape} 
The other possibility to explain a variation in the RC luminosity in
comparison to the SFR is the escape of CREs in a galactic wind. This can be
tested by using the non-thermal radio spectral index as a tracer for the CRE
escape fraction \citep{lisenfeld_00a}. Radiation losses (synchrotron and
IC) steepen the
 spectral index with time, because electrons with
the highest energies lose their energy fastest. In contrast, adiabatic losses
that occur in an accelerated outflow where the relativistic cosmic-ray gas
loses energy by $P{\rm d}V$ work on the interstellar medium leave the radio
spectral index unchanged as these losses are directly proportional to the
cosmic-ray energy. In our sample of galaxies the gas densities are not high
enough to cause cosmic-ray ionization losses that are important in
starburst galaxies \citep{thompson_06a}. Hence, if CREs are contained within a
galaxy, radiation losses should be important and the non-thermal radio
spectral index should be steep, $\alpha_{\rm nt}<-1.2$.  On the other hand, if
adiabatic losses are important or CREs can escape freely, the observed electron
spectral index should be identical to the injection spectral index
\citep{lisenfeld_00a}. In this case
 we expect a ``young'' non-thermal radio
spectral index of $\alpha_{\rm nt}\approx
 -0.6$.
Our galaxies have an averaged radio spectral index of $\alpha=-0.88\pm 0.22$,
and the averaged non-thermal radio spectral index is $\alpha_{\rm
  nt}=-0.99\pm 0.22$. This is steeper than what \citet{niklas_97b} found in
their sample with $\alpha_{\rm nt}=-0.83$, who used integrated un-resolved
flux densities. However, their measurements are between $\lambda\lambda$
20 and 3~cm, so that their non-thermal radio spectral index is more
appropriate for intermediate wavelengths such as $\lambda$6~cm.
A non-thermal radio spectral index of $\alpha_{\rm nt}=-1.0$
would be indicative of almost no escape of CREs, so that the galaxy is an
effective electron calorimeter. However, some of our galaxies have spectral
indices that are in agreement with a ``young'' non-thermal spectral index
($\alpha\geq -0.7$), possibly indicative of an almost free CRE escape, whereas
other galaxies have a fairly steep spectral index $\alpha\leq -1.1$ possibly
caused by high CRE radiation losses. Hence, we can test whether this spread
has an influence on RC--SFR correlation. This is shown in
Figure~\ref{fig:calor}(c), where we present the ratio $\mathscr{R}_{\rm int}$ as a function of the
non-thermal radio
 spectral index. It can be seen that there is no
correlation between them. The calorimetric model predicts that for higher SFRs
the radio spectral index should be steeper. This is tested in
Figure~\ref{fig:calor}(d), where no such correlation is detected. In summary,
the non-thermal spectral indices suggest that CREs can indeed escape from
galaxies in a galactic wind. This argues against the calorimetric model.

\begin{figure}[tbhp]
\centering
\resizebox{1.0\hsize}{!}{ \includegraphics{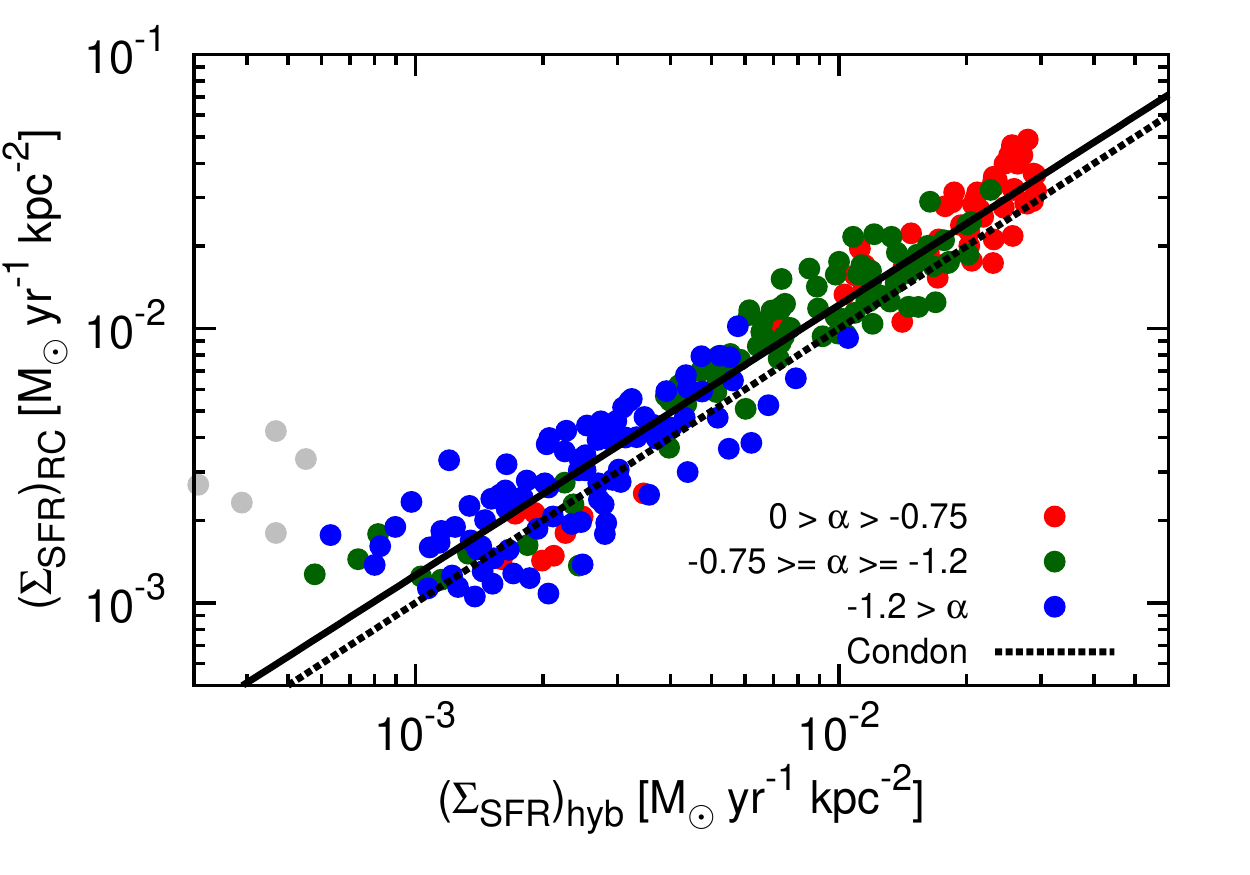}}
\caption{Resolved RC--SFR relation in NGC~6946 with a pixel size of
  $\rm 1.2~kpc$. The hybrid $\Sigma_{\rm SFR}$ map was convolved with a Gaussian kernel of
  $\rm 3.4~kpc$ to simulate cosmic-ray diffusion. The dashed line shows the
  one-to-one Condon
 relation, and the solid line is a linear least-squares fit
  to the data
 points. Red data points indicate a ``young'' non-thermal spectral
  index ($\alpha > -0.75$), blue data points an ``old'' non-thermal spectral index
  ($\alpha<-1.2$), and green data points a non-thermal spectral index of
  intermediate age ($-0.75\geq \alpha \geq -1.2$). The typical uncertainties
  for the radio spectral index are $0.4$, $0.2,$
 and $0.1$ for the blue, green, and
  red data points, respectively. Gray data points were excluded from the
  least-squares fit.}
\label{fig:n6946-24}
\end{figure}
\subsubsection{Magnetic Field--Gas Relation}
\citet{niklas_97a} found
 a correlation between the non-thermal RC luminosity and the
integrated mass of atomic
 hydrogen $M_{\rm HI}$. Assuming equipartition, they
concluded that the magnetic field correlates
 with the mass surface density as
$B \propto \Sigma_{\rm HI}^{0.48\pm 0.05}$. Using $I_{\nu}\propto
B^{3-\alpha_{\rm nt}}$, we expect a relation between the RC
luminosity (and thus the radio SFR) and the atomic hydrogen mass as ${\rm
  SFR_{RC}}\propto M_{\rm HI}^{1.8}$, a significantly super-linear
relation. This is in agreement with the
 Kennicutt--Schmidt relation, which predicts
a super-linear relation between the integrated SFR and the atomic hydrogen
(and possibly also dense
 molecular hydrogen) mass as ${\rm SFR}\propto
M_{\rm HI,H2}^{1.5}$ \citep{kennicutt_12a}.
We tested for our sample the relation between
 the radio SFR and the
  masses of atomic and
 molecular hydrogen (separately and for the sum of both;
  see Appendix~\ref{app:par}), but we find no clear correlation with a
  dispersion much larger than what we observe for the
 RC--SFR relation. Our
sample is probably too small to test the integrated SFR--gas relation, where
the spread in the data points is dominating the estimate of the slope. However, since we find a tight integrated RC--SFR relation, we can derive a
  magnetic field--gas relation, if we assume a local Kennicutt--Schmidt relation
  between the gas mass surface density $\Sigma_{\rm gas}$ and $\Sigma_{\rm
    SFR}$ (Section~\ref{dis:mf-gas}).
\begin{figure*}[tbhp]
  \centering
  \resizebox{0.75\hsize}{!}{ \includegraphics{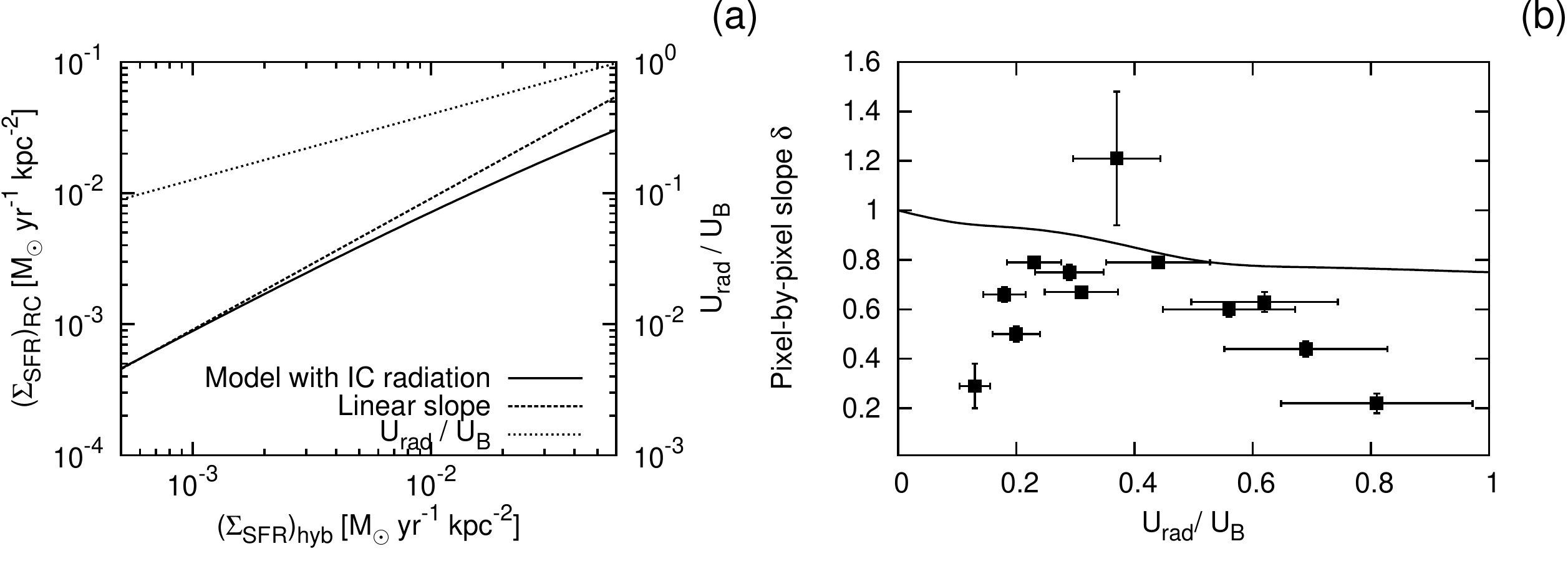}}
  \caption{(a) Model for the resolved RC--SFR relation, where the ratio
    $U_{\rm rad}/U_{\rm B}$ of the radiation energy density to the magnetic
    field energy density is a function of the hybrid $\Sigma_{\rm SFR}$. At
    high $(\Sigma_{\rm SFR})_{\rm hyb}$ the RC emission is suppressed, because
    cosmic-ray electrons emit predominantly inverse Compton (IC) radiation. (b)
    Slope of the pixel-by-pixel plots as a function of the ratio of radiation
    energy density to magnetic field energy density. The line shows the
    predicted slope when including IC radiation.}
  \label{fig:ic-comb}
\end{figure*}
\subsection{RC--SFR Relation on a 1 kpc Scale}
\label{dis:crs}
The RC--SFR relation when integrating over entire galaxies is almost linear: however, on a 1~kpc scale
we find that the RC--SFR slopes are significantly sub-linear. This is explored in the following.

\subsubsection{Cosmic-ray Diffusion}
Cosmic rays diffuse away during their
Lifetime, while massive stars will be hardly moving from
their birth place. We thus expect the RC image to be smeared out with respect
to the hybrid $\Sigma_{\rm SFR}$ map \citep{murphy_06a,murphy_06b,murphy_08a},
explaining the sub-linear slopes of our pixel-by-pixel plots: the
RC emission is suppressed at the peaks of the SF distribution, because the
CREs have diffused away from the SF sites. In contrast, away from the
SF sites we see, relatively speaking, more RC emission than we would expect
from the hybrid $\Sigma_{\rm SFR}$. The former situation we find in spiral arms, whereas the latter
situation we find in 
the inter-arm regions. \citet{berkhuijsen_13a}
  studied the RC--SFR correlation in M31 and M33 and found also sub-linear
  slopes. They convolved the thermal RC map---assuming it to be a representation of
  $\Sigma_{\rm SFR}$---with a Gaussian kernel, so that the
  RC--SFR relation becomes linear, and use this as an estimate for the
  cosmic-ray diffusion length. We tested this by convolving the
hybrid $\Sigma_{\rm SFR}$ map with a Gaussian $\rm 3.4~kpc$ kernel to simulate a diffusion length of $1.7$~kpc, half the width of the convolving kernel. This is presented in 
Figure~\ref{fig:n6946-24}, which shows a slope of $0.99\pm0.02$, a linear RC--SFR relation within the uncertainty.
 Our results hence agree with those by
 \citet{tabatabaei_13a}, who also found a diffusion length of $1.7~\rm kpc$ in NGC~6946.
The explanation of sub-linear slopes with cosmic-ray transport is
corroborated by our 
 finding in Section~\ref{res:pix}, where the
pixel-by-pixel plots in about half of
 our sample show that the data points can
be characterized by their radio
 spectral index: data points at low $\Sigma_{\rm SFR}$
(radio and hybrid) have steeper
 non-thermal spectral indices, indicating
older CREs. The opposite is
 true for high $\Sigma_{\rm SFR}$, where, at sites of current SF,
the
 radio spectral index is $\alpha\approx-0.6$, similar to that measured in
supernova remnants
 \citep{green_09a}. As CREs lose their energy mainly by
synchrotron and IC losses, both processes having loss rates
$\propto E^2$, the CREs with the highest energies lose their energy
fastest. This results in a steepening of the electron spectral index with
time
 and hence also in the radio spectral index. In the pixel-by-pixel plots
we find that data points with
 steep spectral indices (old CREs) are relatively
radio bright and data
 points with flatter spectral indices (young CREs) are
relatively radio dim as
 expected for cosmic-ray transport. 

Our discussion would be incomplete without at least mentioning the influence
of the magnetic field structure on cosmic-ray diffusion. The galaxies in our
sample have spiral magnetic fields with pitch angles of typically $20\degr$
\citep{heald_09a}. In galaxies, the ordered component of the magnetic
  field has a strength that is comparable to the turbulent component, so
  that anisotropic diffusion along the ordered magnetic field is the most
  likely transport scenario \citep[e.g.,][]{tabatabaei_13b}. Thus, we can
  speculate that radial diffusion may be suppressed in comparison to azimuthal
  diffusion. This would explain why the ratio $\mathscr{R}$ has only quasi-periodic
fluctuations as a function of galactocentric radius (Section~\ref{res:rad}). However, a more
quantitative study of the influence of the magnetic field structure on the
cosmic-ray diffusion would be necessary in order to affirm this conclusion.

\subsubsection{Inverse Compton Radiation}
As we have seen in
Section~\ref{dis:global}, the variation of radiation energy density to
magnetic field energy density, $U_{\rm rad}/U_{\rm B}$, is globally not large
enough to explain the fluctuations in the integrated ratio $\mathscr{R}_{\rm int}$ of RC luminosity
to hybrid SFR. Locally, however, this may be different as the variation of
$U_{\rm rad}/U_{\rm B}$ within an individual galaxy can be much larger.

As a first-order approximation, the radiation energy density scales with
 the
hybrid $\Sigma_{\rm SFR}$ as $U_{\rm rad} \propto \rm (\Sigma_{\rm
  SFR})_{\rm hyb}$. The radiation energy density from dust is proportional to
the hybrid $\Sigma_{\rm SFR}$, and the stellar radiation field is assumed to
scale also with the hybrid $\Sigma_{\rm SFR}$ and neglecting the constant
background radiation field of CMB. On the other hand, the magnetic field
energy 
 density scales with the hybrid $\Sigma_{\rm SFR}$ as $U_{\rm B}
\propto \rm
 (\Sigma_{\rm SFR})_{\rm hyb}^{2\zeta/(3-\alpha_{\rm nt})}$. With
$\alpha_{\rm nt} = -0.85$ and
 $\zeta=1.11$ we have $U_{\rm B} \propto \rm
(\Sigma_{\rm SFR})_{\rm hyb}^{0.58}$. The ratio $U_{\rm
 rad} / U_{\rm B}$
hence scales approximately as $U_{\rm rad} / U_{\rm
 B}\propto \rm (\Sigma_{\rm
  SFR})_{\rm hyb}^{0.5}$. We note that this is an upper limit, as the stellar
radiation field probably does not depend as much on the hybrid $\Sigma_{\rm
  SFR}$ as the re-radiated FIR emission from dust. The typical range of the
hybrid $\Sigma_{\rm SFR}$ is a factor
 of 100, between the peak $\Sigma_{\rm
  SFR}$ and the lowest $\Sigma_{\rm SFR}$ in outskirts and inter-arm
 regions,
so that the ratio $U_{\rm rad}/U_{\rm B}$ may change by as much as a
 factor of
10 within a single galaxy.  As the RC luminosity scales with the
 inverse of
$1+U_{\rm rad}/U_{\rm B}$, we expect a variation of the RC--SFR
 relation. At
the peaks of the hybrid $\Sigma_{\rm SFR}$ the RC intensity will be slightly
suppressed, because some of the energy of the CREs is radiated
 via IC radiation.

In Figure~\ref{fig:ic-comb}(a) we show the
 expected RC
intensity as a function of the hybrid $\Sigma_{\rm SFR}$ for average ratio
across our sample $\langle U_{\rm rad}/U_{\rm B}\rangle =0.38$ and a typical $\Sigma_{\rm
  SFR}$ range. It
 can be seen that $U_{\rm rad}/U_{\rm B}$ varies between 10\%
at the lowest
 $\Sigma_{\rm SFR}$ and 100\% at the highest $\Sigma_{\rm SFR}$,
resulting in an RC decrease of a factor
 of two as compared to a linear
extrapolation. For higher ratios of $U_{\rm
 rad}/U_{\rm B}$ we expect an
increased effect on the pixel-by-pixel
 slopes. This is tested in
Figure~\ref{fig:ic-comb}(b), where we plot the
 pixel-by-pixel slopes as a function
of integrated ratio $U_{\rm rad}/U_{\rm B}$
 within individual galaxies. For
comparison, we also plot the expected slopes
 for our model including IC radiation. The model can explain a
 small fraction of the sub-linear
slopes but not all of it. For $U_{\rm
 rad}/U_{\rm B}=0.1$, as in NGC~6946, we
find a slope of 0.9, and only for
 ratios $U_{\rm rad}/U_{\rm B}> 0.5$ we do find
slopes of 0.8. IC
 losses alone therefore cannot explain the
sub-linear slopes, but they reduce the
 required cosmic-ray transport lengths.
\subsection{Cosmic-ray Transport}
\label{dis:cr_transport}
\subsubsection{Disk-parallel Cosmic-ray Transport}
We can now try to parameterize the
cosmic-ray transport as a function of the hybrid $\Sigma_{\rm SFR}$. The CRE lifetime
scales as $t_{\rm syn}\propto E^{2} B^{2}$, assuming that synchrotron losses
dominate as shown in Section~\ref{dis:global}. The synchrotron emission of a
single CRE has a broad peak at the critical frequency, so that
we can relate the CRe energy via $E\propto B^{-1/2}$ to the magnetic field strength,
resulting in $t_{\rm syn}\propto B^{-3/2}$. On the other hand, cosmic rays are
transported by diffusion as long as there is no galactic wind, which is
important for transport in the vertical direction only. For cosmic-ray diffusion one has
$L_{\rm diff}^2/t_{\rm syn}=D$, where $L_{\rm diff}$ is the transport length and $D$
is the diffusion coefficient. The energy dependence of the diffusion
coefficient is normally assumed to be $D\propto E^\kappa$ with $\kappa=0.5$
for Kolmogorov-type turbulence. The diffusion length $L_{\rm diff}=(D\cdot
t_{\rm syn})^{1/2}$ scales hence as $L_{\rm diff}\propto B^{-(3+\kappa)/4}$. In
Section~\ref{dis:mf-gas} we will find the relation $B\propto \rm SFR_{\rm hyb}^{0.30}$,
so that we find for dependence of the diffusion length $L_{\rm diff}\propto
\rm SFR_{\rm hyb}^{-0.26}$, where we take the global magnetic field strength for the
average value of the CRE lifetime.
This can be compared with observations such as of \citet{murphy_08a}, who 
measured the diffusion length in the WSRT SINGS galaxy sample. They studied
the
 dependence of $L_{\rm diff}$ on the radiation energy density, which they
derived from the FIR luminosity divided by the physical area of the galaxy
disk (similar to our approach in Section~\ref{dis:global}). Following for
this part of the discussion their assumption of a direct proportionality
between the hybrid SFR and FIR luminosity (although probably not strictly
true), their result is $L_{\rm
 diff}\propto \rm (\Sigma_{\rm SFR})_{\rm
  hyb}^{-0.5}$, significantly steeper than what our simple
 estimate above
indicates. However, \citet{tabatabaei_13b} find no such strong dependence
 on $\Sigma_{\rm SFR}$ and argue that the diffusion length is mainly determined by the magnetic
 field structure. This is a sensible assumption, observing that diffusion along
 magnetic field lines is
faster and measured diffusion coefficients can be large
 such as $2.0\times
10^{29}\rm~cm^2~s^{-1}$ \citep[in NGC~253,][]{heesen_09a}. On the other hand,
\citet{heesen_11a}
 found a much smaller diffusion coefficient of $3.0\times
10^{28}\rm~
 cm^2~s^{-1}$ in the same galaxy but for perpendicular
 diffusion,
which is in agreement with theoretical considerations
\citep{buffie_13a}. \citet{murphy_12a} found an even smaller diffusion
coefficient of
 $1.0 \times 10^{27}\rm~cm^2~s^{-1}$ in the vicinity of
30~Doradus, a 
 massive SF region in the Large Magellanic Cloud.
 This is
probably due to the high degree of turbulence in the magnetic field structure.
\subsubsection{Cosmic-ray Transport within Galaxy Halos}
\citet{krause_09a} notes
 that in her sample of edge-on galaxies the scale
height of the RC emission at
 $\lambda$6~cm is $\rm 1.8~kpc$, across an SFR span
in her sample of a factor of
 10. For the same sample, \citet{krause_11a} finds
$B=6~\mu\rm G$ and a
 scale height of $\rm 1.8~kpc$ for NGC~3628, whereas in
NGC~253 she finds $B=12~\mu\rm G$ and a scale height of $1.7$~kpc. Our
predicted diffusion length of $L_{\rm diff}\propto \rm B^{-0.88}$, using
$\kappa=0.5$
, would predict that the scale height of NGC~253 should be only
about half of that of NGC~3628. But the scale height is almost independent of
the magnetic field strength, in contrast to our expectation from cosmic-ray
diffusion. \citet{tabatabaei_13b} find that the diffusion lengths in the
  disks of galaxies do not depend on $\Sigma_{\rm SFR}$, which they interpret
  as support for the non-calorimetric model of \citet{niklas_97a}. If the diffusion length is not a function of
  $\Sigma_{\rm SFR}$ and hence the magnetic field strength, the timescale
  over which the CRE are transported within the disk is not given by the CRE
  lifetime $t_{\rm syn}$ as measured from synchrotron losses, but by the CRE
  escape time $t_{\rm esc}$ from the disk. In this case we have $t_{\rm
    esc}<t_{\rm syn}$ and the electrons escape freely into the halo.  The CRE
  scale heights in the halo are also constant and independent of the magnetic
  field strength \citep{krause_09a}, further strengthening this argument.
However, the alternative explanation of the larger than predicted
vertical scale height in NGC~253 is
a galactic
 wind, which transports cosmic rays convectively from the disk into
the halo. This
 can happen in addition to any transport by diffusion. A
distinction whether
 transport happens via diffusion or convection is the
predicted dependence of
 the scale height on the magnetic field strength. The
transport length for convection is $L_{\rm conv} \propto
 {\rm v_{wind}}\cdot
B^{-3/2}$, because the transport
 length depends in this case only on the
electron cooling time here assumed to be dominated by synchrotron radiation
losses. So that the global scale height of the RC emission does not depend on
the magnetic field strength and thus the SFR, for convective
 transport the
wind speed $\rm v_{wind}$ thus has to increase with $B^{3/2}$ in order to
compensate for the shorter loss timescale. To distinguish between convective
and diffusive cosmic-ray transport from the disk into the halo,
\citet{heesen_09a} studied the
 local correlation between the CRE scale height in
NGC~253 and the CRE lifetime. In the northeastern halo they favored
convection and in the
 southwestern part diffusion as the dominant mode of
transport. Most importantly, they found that the CRE lifetime indeed
  regulates the local scale height as is expected when synchrotron and IC losses
  are dominating over the escape of CRE, so that one has in this case $t_{\rm
    syn}<t_{\rm esc}$, which would mean that NGC~253 is a CRE calorimeter.
  But the non-thermal spectral index of the diffuse RC emission, excluding the
  nuclear starburst region, is $-1.0$ between $\lambda\lambda$ 90 and
  $3.6$~cm, indicative that some fraction of the CREs still do escape.
NGC~253 seems
 to be close to the hybrid $\Sigma_{\rm SFR}$ value
where convection in a wind becomes an important
 transport mode. At even higher
hybrid $\Sigma_{\rm SFR}$ in the nuclear starburst of NGC~253 and in the starburst galaxy M82, the RC scale height is smaller with
 $\approx 150~\rm pc$ at
$\lambda 6~\rm cm$, indicating that the wind speed
 does not increase
sufficiently to compensate for the shorter electron cooling
 times
\citep{heesen_11a,adebahr_13a}.
 These galaxies are supposed to be
  electron calorimeters as inferred from a comparison of RC and gamma-ray
  observations \citep{yoast-hull_13a}. In summary, the cosmic-ray
  diffusion lengths in the disk and scale heights in the halo argue in favor
  of a non-calorimetric model. However, it needs more investigation whether
  the vertical scale heights are at least partially explained by a galactic
  wind, which may explain the constant vertical scale heights by an increased
  wind speed due to a higher SFR, compensating the shorter CRE lifetimes.
\subsection{Magnetic Field--SFR Relation}
\label{dis:mf-gas}
As we have seen, the relation between RC emission and the hybrid SFR is
markedly different for integrated measurements, where it is a slightly
super-linear relation, and resolved measurements on a 1~kpc scale, where it
is a significantly sub-linear relation. We attributed this finding largely to
the diffusion of CREs within galaxies. We now attempt to reconcile these two
findings by proposing a relation between the magnetic field strength and the
SFR. Our model can be seen as a variety of the non-calorimetric model by
\citet{niklas_97a}, which is based on a magnetic field--gas relation. We assume (1) that the total cosmic-ray energy content of
a galaxy relates to the magnetic field strength, so that there is energy
equipartition between the cosmic rays and the magnetic field and (2)
that the magnetic field strength is proportional to the SFR. The
second assumption is supported by the theory that magnetic fields are
amplified by a turbulent dynamo, and turbulence is mostly stemming from
supernovae and hence the SFR \citep[e.g.,][]{cho_09a}. Our model is a close corollary of the models of \citet{dumas_11a} and \citet{tabatabaei_13a}, who assume a local magnetic
  field--gas relation.

Using the assumption of a global energy equipartition between the cosmic rays
and the magnetic field, we can express the magnetic field strength by the hybrid SFR
 as $B \propto {\rm
  SFR_{\rm hyb}}^{\zeta/(3-\alpha_{\rm nt})}$, where
 $\zeta_{\rm nt}=1.16\pm0.08$ (for
the non-thermal RC emission; Section~\ref{res:int}) and $\alpha_{\rm
  nt}=-0.85$. Our
 result is thus $B\propto \rm SFR_{\rm hyb}^{0.30\pm 0.02}$. This is
within the
 errors consistent with the results of \citet{niklas_97a}, who found
$B\propto \rm SFR^{0.34\pm 0.08}$, and those of \citet{chyzy_11a} who found
$B\propto \rm SFR^{0.25\pm 0.06}$, for dwarf irregular galaxies. We now assume
that this relation holds also locally on a 1~kpc scale, so that we have
$B\propto \rm (\Sigma_{\rm SFR})_{\rm hyb}^{0.30\pm 0.02}$. But we have
  to drop the assumption of local energy equipartition: because of cosmic-ray diffusion, the
cosmic-ray energy density is smoothed to a scale of several kpc, depending on
the diffusion length scale, which presumably varies in our sample
galaxies. In the following, we use the extreme assumption of a constant
  cosmic-ray energy density on a 1~kpc scale, in order to estimate the minimum
  expected slope of the non-thermal RC--$\Sigma_{\rm
  SFR}$ correlation.
We
hence expect a non-thermal RC--$\Sigma_{\rm SFR}$ correlation as $I_{\rm syn}\propto
B^{1-\alpha_{\rm nt}}\propto \rm (\Sigma_{\rm SFR})_{\rm hyb}^{0.56\pm 0.04}$. This indeed agrees
within the errors with the observed value of $\delta_{\rm nt}=0.61\pm
0.22$. We notice that the slope of $0.56$ is a lower limit, because the
cosmic-ray energy density probably does still depend slightly on the
$(\Sigma_{\rm SFR})_{\rm hyb}$.


We can
 now also relate the magnetic field to the local gas surface density,
$\Sigma_{\rm gas}$, if we use the known relations between $(\Sigma_{\rm
  SFR})_{\rm hyb}$ and
$\Sigma_{\rm gas}$ \citep{leroy_08a, bigiel_08a}. They have shown that if the
gas is predominantly molecular, $(\Sigma_{\rm SFR})_{\rm hyb}$ is directly proportional to the mass
surface density $\Sigma_{\rm gas}$. If the gas is predominantly atomic,
$(\Sigma_{\rm SFR})_{\rm hyb}$ depends on $\Sigma_{\rm gas}$ in a super-linear relation with a slope of
$1.5$. Thus, we find
 $B\propto \rm \Sigma_{\rm gas}^{0.30 ... 0.45}$,
the lower slope for predominantly molecular gas and the higher slope for
predominantly atomic gas. This is indeed in good agreement with the result of
\citet{niklas_97a}, who found for atomic hydrogen $B \propto M_{\rm
  HI}^{0.48\pm
 0.05}$. We notice that a small difference in the slope of
either the magnetic field--gas or the magnetic field--SFR relation has a
large impact on the slope of the RC--SFR relation, because the non-thermal RC
emission depends on the equipartition magnetic field strength as $I_{\rm syn}
\propto B^{3-\alpha_{\rm nt}}$. \citet{schleicher_13a}
derive a magnetic field--SFR relation of $B\propto \rm SFR^{1/3}$. This
would result in a super-linear RC--SFR relation with $\rm SFR_{RC}\propto
SFR_{hyb}^{1.28}$. Our observed non-thermal slope of $\zeta_{\rm
  nt}=1.16\pm0.08$ is slightly super-linear, but not as much as expected from
the theory of turbulent magnetic field amplification. A small change, however, of
the magnetic field--SFR relation would be sufficient to observe an only
slightly super-linear RC--SFR relation as found for our sample, or possibly
even a strictly linear RC--SFR relation, which makes RC emission an
excellent SF tracer as we discuss below.

It is important to point out that dropping the assumption of local energy
  equipartition differentiates our results from those of, e.g.,
  \citet{chyzy_08a}, \citet{dumas_11a}, and \citet{tabatabaei_13a}. Using
  equipartition, they do find flatter slopes of the magnetic field--SFR and
  magnetic field--gas relation, even though the slope of the RC--SFR relation
  is similar to ours. However, firstly, it is instructive to study the
  extreme case of a constant cosmic-ray energy density on the above
  relations. The effect of cosmic-ray transport is a smoothing of the
  corresponding energy
  distribution  over a kpc scale. No such transport is postulated for the
  magnetic field, so that away from SF sites equipartition is probably not
  strictly true anymore. Secondly, the
  appeal of this model lies in a unique magnetic field--SFR and possibly also
  magnetic field--gas relation that holds both globally and locally. And
  thirdly, we notice that in our scenario the cosmic-ray energy
density is only constant locally, but not globally across our galaxy sample,
where it varies as a function of SFR. A constant cosmic-ray energy density, independent of SFR,
would require a calorimetric model, which our observations and that of others
disfavor (Section~\ref{dis:global}). 
\subsection{Radio continuum as an SF Tracer}
\label{dis:rc_sf}
We will now try to summarize what we have learned from our study on how we can
use RC emission to determine integrated SFRs or spatially resolved
$\Sigma_{\rm SFR}$. Firstly, we notice that the Condon relation holds for the integrated RC
emission in comparison to the measured SFR from the hybrid SF tracer, the
combination of \emph{GALEX} FUV and \emph{Spitzer} $24~\rm \mu m$ MIR emission
(Section~\ref{res:int}). This contains two important sub-results: (1)
the radio derived SFR agrees within one standard deviation with the SFR as
derived from the state-of-the-art hybrid tracer, (2) the RC luminosity
of a galaxy is almost linearly dependent on the SFR as predicted by the Condon
relation. We did not find any galaxy parameter, which clearly influences the
ratio $\mathscr{R}_{\rm int}$ of radio to hybrid SFR (Section~\ref{dis:global} and
Appendix~\ref{app:par}). The integrated radio SFR
varies by $\pm 50\%$ with respect to the hybrid SFR. As the hybrid SFR has a
similar calibration error, we can conclude that for integrated galaxy-wide
measurements, the RC emission provides an SFR with similar accuracy as the
hybrid FUV/MIR tracer.
For resolved studies, the situation is more complicated. If averaging
azimuthally, the profiles of the radio $\Sigma_{\rm SFR}$ have almost exactly the same shape
as the radial profile of the hybrid $\Sigma_{\rm SFR}$, only offset by an almost constant
factor. We did not find any conclusive radial trends; the ratio $\mathscr{R}$ of radio
to hybrid $\Sigma_{\rm SFR}$ changes quasi-periodically. Furthermore, the relative amplitude of
the fluctuations is small, on average across our sample only $\pm 25\%$. This
means if one is interested in measuring the radial distribution of SF in
galaxies only, again the radio-based method provides with a similar accuracy
($\pm 50\%$) as using the hybrid FUV/MIR tracer.
For resolved studies on a 1~kpc scale, however, one has to be somewhat more
cautious. Individual data points lie typically a factor of 2--3 off from the
best fit to the data points. This is comparable with the small-scale scatter of
the hybrid FUV/MIR tracer \citep{leroy_12a}. Hence, on a 1~kpc scale, the
scatter of the RC--SFR correlation could be stemming entirely from the
calibration uncertainty of the hybrid SF tracer. However, the sub-linear slopes in our
pixel-by-pixel plot mean that the local RC--SFR correlation is not linear. Using
just one local measurement of the RC emission will hence not provide an
accurate estimate of the $\Sigma_{\rm SFR}$. As we have discussed in
Section~\ref{dis:crs}, this is almost entirely due to cosmic-ray transport
within the disk, whereas the influence of IC radiation can be
neglected. A high-$z$ study of star-forming galaxies may reveal, if spatial
resolution is high enough, only the peaks of the $\Sigma_{\rm SFR}$ within a galaxy, whereas
the fainter RC emission is missed. From our small sample of galaxies, it
appears that if one restricts measurements to high radio and hybrid $\Sigma_{\rm SFR}$,
$(\Sigma_{\rm SFR})_{\rm RC}\gtrapprox 10^{-2}~ M_\odot\,\rm
yr^{-1}\,kpc^{-2}$ and $(\Sigma_{\rm SFR})_{\rm hyb}\gtrapprox 10^{-2}~
M_\odot\,\rm yr^{-1}\,kpc^{-2}$, the spread around the Condon
relation is much reduced, approximately to a factor of 2--3
(Figure~\ref{fig:grand}). In case of only having
radio measurements available, one can restrict to areas with ``young'' non-thermal
spectral indices ($\alpha<-0.75$) in order to reduce the spread around the
Condon relation. However, we have to caution that this is
based only on the measurements of four galaxies (NGC~4736, NGC~5055,
NGC~5194, and NGC~6946), which have such high $\Sigma_{\rm SFR}$. Therefore,
it is not clear, whether this is only a property of our galaxy sample or will indeed hold true
for a much larger galaxy sample.
\section{Conclusions}
\label{sec:conclusions}
We have studied the resolved RC--SFR relation in a selection drawn from the
THINGS sample of galaxies. We have used WSRT maps at $\lambda\lambda$ 22 and
18~cm as RC maps and a combination of \emph{GALEX} FUV and \emph{Spitzer}
24~$\mu$m maps as hybrid SF tracers. To estimate the fraction of thermal RC
emission, we used maps of Balmer H$\alpha$ emission. We have compared our
results with the well-established Condon relation. These are our main
conclusions:
\begin{enumerate}
\item The integrated RC luminosity is almost directly proportional to the
  integrated
 hybrid SFR as expected from the Condon relation. We find $\rm
  SFR_{RC} \propto SFR_{hyb}^{1.11\pm0.08}$, an only very moderately
  super-linear relation. The averaged ratio of integrated radio to
 hybrid SFR is
  $\langle \mathscr{R}_{\rm int}\rangle =0.8\pm 0.4$, where the error reflects the
 standard deviation in our
  sample.
 The Condon relation hence predicts also the numerical value of the
  SFR correctly, within the uncertainty of our measurements. Because the
  integrated hybrid SFR has an intrinsic uncertainty of $\pm 50\%$, we
  conclude that the radio SFR has a similar accuracy if used for integrated
  measurements.
\item We have averaged the RC emission and hybrid SF tracers azimuthally in radial
  annuli to determine the radial profiles of both $\Sigma_{\rm SFR}$. The ratio $\mathscr{R}$ of
  radio to hybrid $\Sigma_{\rm SFR}$ stays remarkably constant as a function of galactocentric radius and
  undergoes only quasi-periodic variations. The relative variation of
  $\mathscr{R}$ as a
  function of galactocentric radius is only $25\%$. We fitted a constant function to
  the profile of the ratio to measure the radial average $\langle \mathscr{R}\rangle $ within a
  galaxy. We find that $\langle \mathscr{R}\rangle $ agrees within the uncertainty with the
  integrated ratio $\mathscr{R}_{\rm int}$.
 Again, we can conclude that the radio SFR
  has a similar accuracy as the hybrid SFR if used on azimuthally averaged
  data, i.e., radial profiles.
\item We have locally averaged the RC emission and hybrid SF tracers in pixels
  of
 $1.2$ and $\rm 0.7~kpc$ resolution. We find a clear correlation between the
  $\Sigma_{\rm SFR}$ as derived from the RC emission and hybrid SF tracers. Averaged over
  our sample, the result is $\rm (\Sigma_{SFR})_{RC} \propto (\Sigma_{SFR})_{hyb}^{0.63\pm0.25}$,
  where the uncertainty is the standard deviation of the sample
  galaxies. Results for the two resolutions are almost identical compared to
  the variation across our sample.
 The spread around the least-squares fit is
  a factor of 2--3, comparable with the uncertainty of the local hybrid
  $\Sigma_{\rm SFR}$. However, because the relation of radio to hybrid $\Sigma_{\rm SFR}$ is sub-linear,
  one cannot derive the local $\Sigma_{\rm SFR}$ from the radio emission alone.
\item Diffusion of CRE is responsible for flattening the local
  RC--SFR relation, resulting in a sub-linear relation. This is supported
  by our finding that at low radio and hybrid $\Sigma_{\rm SFR}$, the non-thermal spectral
  index is predominantly steeper than $-1.2$, indicative of an aged CRE population. At high $\Sigma_{\rm SFR}$, the non-thermal radio spectral index is
  flatter than $-0.75$, as expected from young CRE. At low
  $\Sigma_{\rm SFR}$ we hence have an excess of CRE, resulting in an excess
  of RC emission, which is dominated by non-thermal synchrotron emission
  around $\lambda$20~cm.
\item We have compared the ratio $\mathscr{R}_{\rm int}$ of integrated radio to hybrid
  SFR with various
 fundamental galaxy parameters. We did not find any obvious
  dependence on any
 parameter we have tested for. The same result was found
  for the slope in the
 pixel-by-pixel plots. We find no indication of the
  escape of cosmic rays in
 a galactic wind as traced by the radio spectral
  index or influence of IC radiation dependent on the ratio of
  radiation to magnetic field energy density $U_{\rm rad}/U_{\rm B}$.
\item We have shown the importance of FUV as an SF tracer. In the outer parts
  of galaxies, where the amount of dust is low, \emph{Spitzer} $24~\rm\mu m$
  emission alone is not a reliable SF tracer as it is not sensitive to the
  unobscured SF as measured by \emph{GALEX} FUV emission. Thus, the ratio
  $\mathscr{R}$ is constant when using the hybrid SF tracer from a combination of FUV
  and $24~\rm\mu m$, but rises significantly at larger radii when using
  $24~\rm\mu m$ emission alone.
\item When restricting to the highest $\Sigma_{\rm SFR}$, both $(\Sigma_{\rm
    SFR})_{\rm RC}\gtrapprox 10^{-2}~M_\odot\,\rm yr^{-1}\,kpc^{-2}$ and
  $(\Sigma_{\rm SFR})_{\rm hyb}\gtrapprox 10^{-2}~M_\odot\,\rm yr^{-1}\,kpc^{-2}$, the spread around the Condon
  relation reduces to a factor of 3, similar to the local uncertainty of the
  hybrid tracers, allowing the sole use of RC emission as an SF tracer. However,
  this result is based only on a sub-set of four galaxies that fulfill the
  above requirement. It hence has to be taken with caution.

\item We can explain the almost linear RC--SFR relation and the sub-linear
  resolved (on a 1~kpc scale) RC--$\Sigma_{\rm SFR}$ relation by proposing a
  non-linear magnetic field--SFR relation, which holds both globally and locally (1~kpc
  scale). Assuming energy equipartition between the cosmic rays and the
  magnetic field, we find $B\propto \rm SFR_{hyb}^{0.30\pm 0.02}$. Magnetic field amplification by a
  small-scale dynamo, powered by SF-driven turbulence, can explain our result.

\end{enumerate}

\begin{acknowledgements}
 V.H.\ is funded as a Postdoctoral Research Assistant
  by the UK's Science and
 Technology Facilities Council (STFC). This
  research has made use of the NASA/IPAC Extragalactic Database (NED), which is
  operated by the Jet Propulsion Laboratory, California Institute of
  Technology, under contract with the National Aeronautics and Space
  Administration. The authors thank the anonymous referee for a detailed
  reading of the manuscript and for comments that have led to improvements in
  its clarity and presentation. Elly Berkhuijsen is thanked for carefully
  reading the almost finished manuscript and providing further comments and suggestions.
\end{acknowledgements}
\appendix

\section{Further parameter studies}
\label{app:par}
\begin{figure}[tbhp]
\centering
\resizebox{1.0\hsize}{!}{ \includegraphics{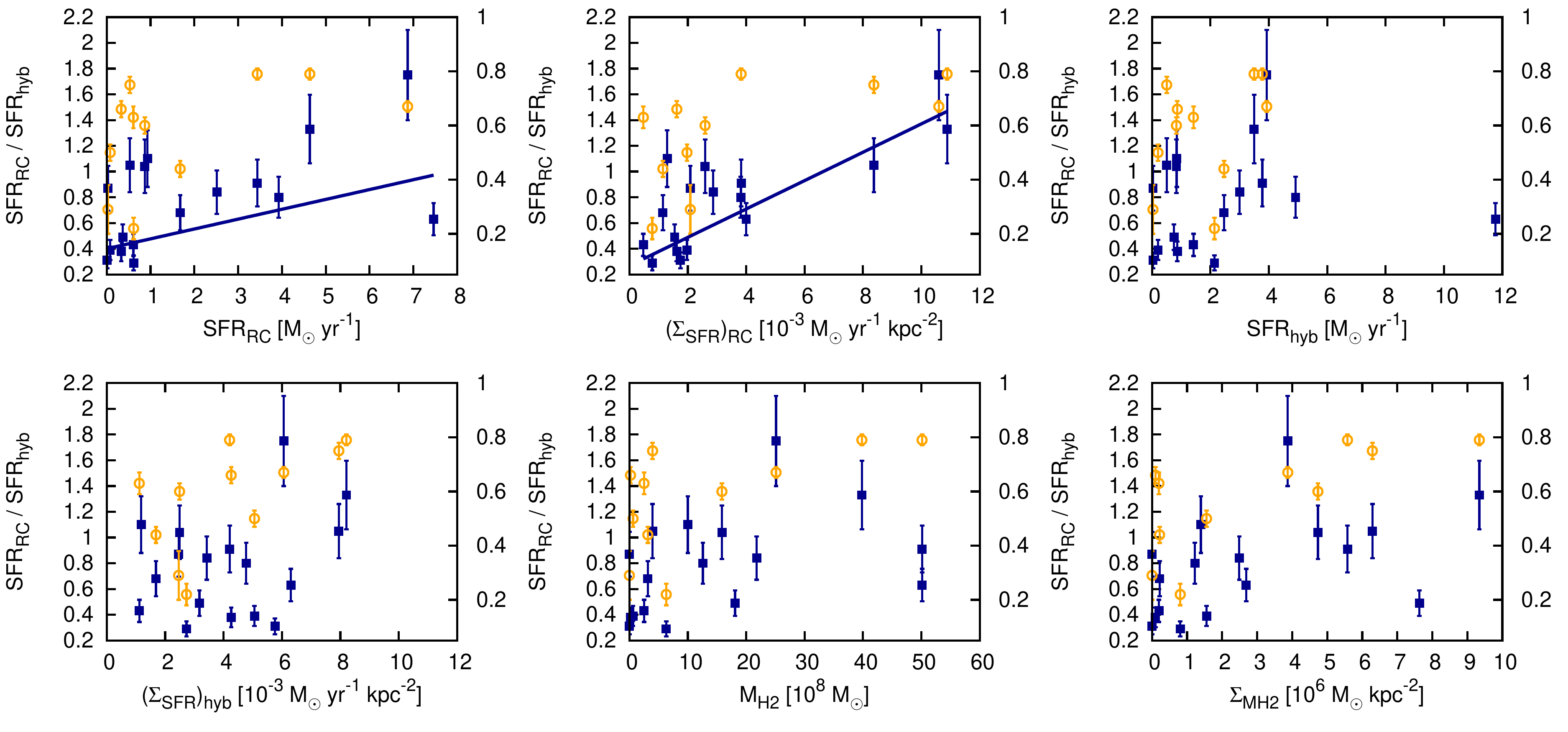}}
\caption{Parameter studies --- I. We compare the ratio of the SFRs as derived
  by the RC emission and hybrid SF tracers as a function of various galaxy
  parameters (filled squares). Solid lines represent least-squares fits to the
  filled squares. We also plot the pixel-by-pixel slopes as measured at
  $\rm 1.2~kpc$ resolution (open circles), where the scale is on the right
  axis. (a) SFR from RC luminosity. (b) $\Sigma_{\rm SFR}$ as calculated
 from integrated
  radio SFR divided by physical area. (c) Integrated SFR from
 hybrid SF
  tracers. (d) $\Sigma_{\rm SFR}$ as calculated from integrated hybrid SF
 tracers divided by
  physical area. (e) Molecular hydrogen mass. (f) Molecular
 hydrogen mass
  surface density.}
\label{fig:sfr}
\end{figure}

\begin{figure}[tbhp]
\centering
\resizebox{1.0\hsize}{!}{ \includegraphics{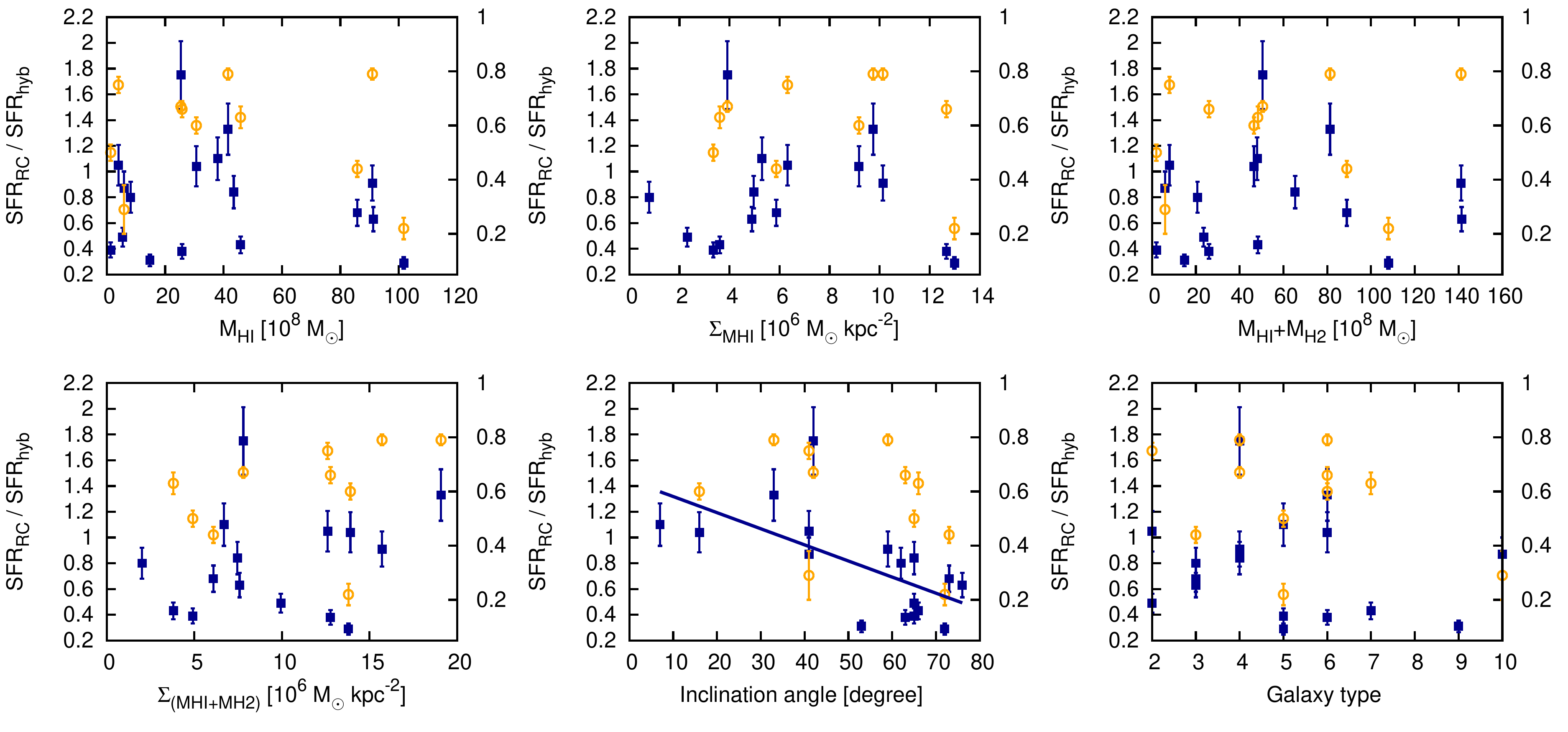}}
\caption{Parameter studies --- II. We compare the ratio of the SFRs as derived by
  the RC emission and hybrid SF tracers as a function of various galaxy
  parameters (filled squares). Solid lines represent
  least-squares fits to the filled squares. We also plot the pixel-by-pixel slopes as
  measured at $\rm 1.2~kpc$ resolution (open circles), where the scale is on the
  right axis. (a) Atomic hydrogen
  mass. (b) Atomic hydrogen mass divided by physical area. (c) Total hydrogen
  mass. (d) Total hydrogen mass divided by physical area. (e)
  Inclination angle. (f) Galaxy type.}
\label{fig:sfr2}
\end{figure}

\begin{figure}[tbhp]
\centering
\resizebox{1.0\hsize}{!}{ \includegraphics{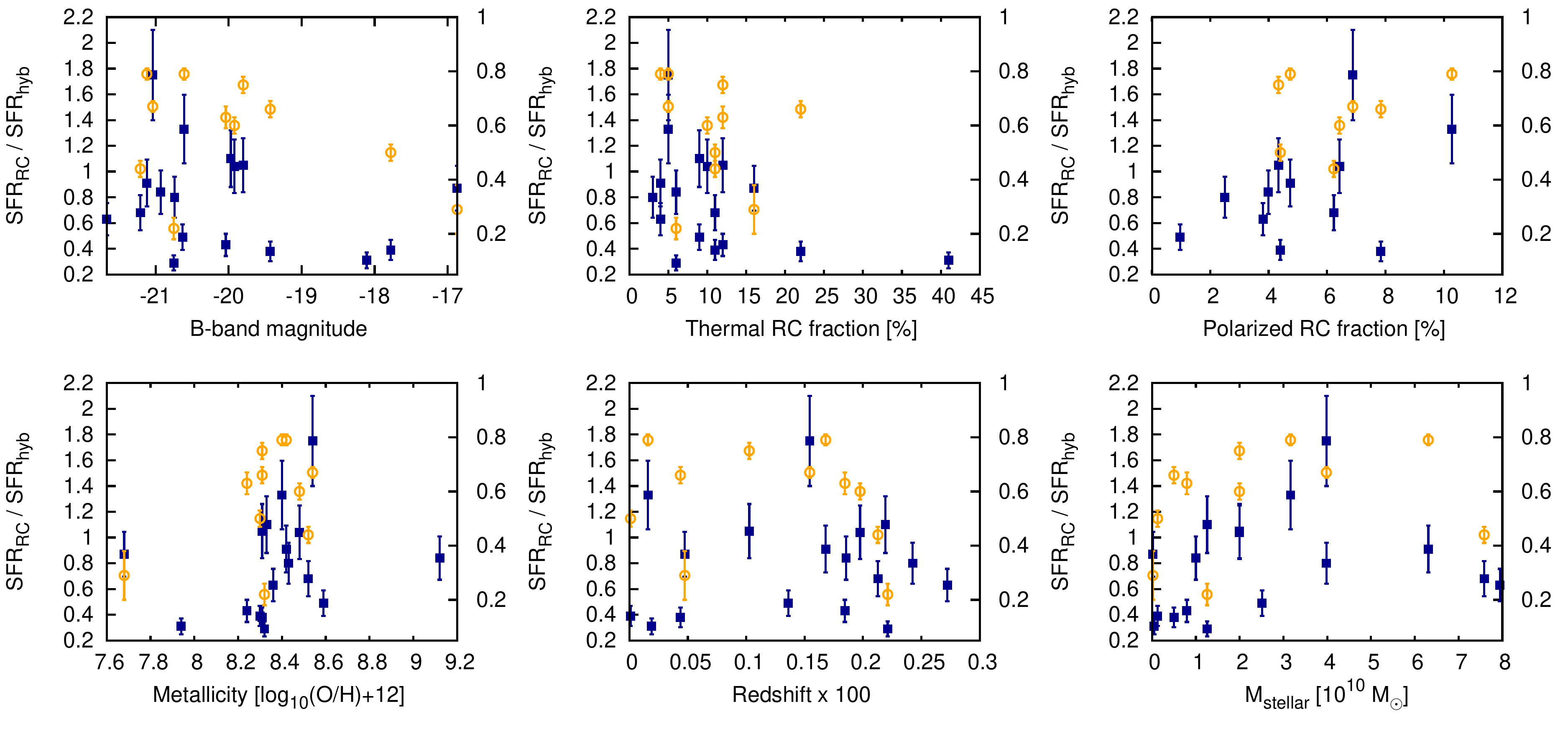}}
\caption{Parameter studies --- III. We compare the ratio of the SFRs as
  derived
 by the RC emission and hybrid SF tracers as a function of various
  galaxy
 parameters (filled squares). We also plot the pixel-by-pixel slopes as
  measured at $\rm 1.2~kpc$ resolution (open circles), where the scale is on the
  right axis. (a) $B$-band magnitude. (b) Thermal
 RC fraction. (c) Linearly
  polarized RC fraction. (d) Metallicity. (e) Redshift. (f) Stellar mass.}
\label{fig:sfr3}
\end{figure}

As we have seen in Section~\ref{res:int} the RC luminosity is a good tracer for
the SF as measured from hybrid tracers. We have shown that the ratio
$\mathscr{R}_{\rm int}$ of the integrated SFRs as measured from the RC to that from
the hybrid SF tracers has an average close to one, as expected for the Condon
relation, and a standard deviation of $0.4$. It would be very important to know
whether this ratio can be related to any global parameter of the galaxies in
our sample. If we cannot find such a parameter, then the ratio would be a
\emph{universal} number and can be assumed to be equal for all galaxies,
albeit with some uncertainty. Tables~\ref{tab:data_sample} and
\ref{tab:sample} contain the literature and measured data we plot the
ratio $\mathscr{R}_{\rm int}$ against.
We now compare the ratio $\mathscr{R}_{\rm int}$ with various galaxy parameters and
look for
 possible trends. We applied a standard least-squares fit with error
weighting of $\mathscr{R}_{\rm int}$ as a function
 of all parameters and show them for
those, where hints of a correlation with $\mathscr{R}_{\rm int}$ can be seen. The
results are shown in
 Figures~\ref{fig:sfr}--\ref{fig:sfr3}. We will discuss them
in the following.
 We note that we have discussed the correlation between the
ratio $\mathscr{R}_{\rm int}$ and both the 
radiation energy density and non-thermal
radio spectral index already in Section~\ref{dis:global}.
{\it SFR.} As we have discussed in Section~\ref{res:int} the
ratio $\mathscr{R}_{\rm int}$ of integrated radio to hybrid SFRs is rising slightly
with increasing
 SFRs. This can be seen by the fact that the power-law index
in
 Equation~(\ref{eq:int_corr}), $\zeta$, is larger than one. In
Figure~\ref{fig:sfr}(a) we plot the ratio $\mathscr{R}_{\rm int}$ as a function of the
SFR as derived from
 the RC luminosity. There is no clear correlation with a
significant spread in
 the data points. But the ratio appears to be indeed
rising with increasing
 radio SFR as expected. We also have tested the relation
with the hybrid
 SFR (Figure~\ref{fig:sfr}(c)), where the correlation is, if anything,
weaker than for the radio SFR. This is of course expected, because the radio
SFR is in the numerator of the ratio $\mathscr{R}_{\rm int} = {\rm SFR_{RC} /
SFR_{hyb}}$, whereas the hybrid SFR is in the denominator.

{\it SFR density.}  We have made the same plots for the SFR
averaged over the physical area of the galaxies. For this we chose the galactocentric radius
that we used for the integrations with {\sc iring} as described in
Section~\ref{subsec:radial}. The plots are presented in Figures~\ref{fig:sfr}(
b)
and (d). The data points show again a large scatter, but the ratio rises with
higher $\Sigma_{\rm SFR}$, particularly for the radio $\Sigma_{\rm SFR}$. For the hybrid $\Sigma_{\rm SFR}$ the
correlation is less pronounced than for the  radio $\Sigma_{\rm SFR}$, as it is for the
integrated SFRs.

{\it Molecular hydrogen mass.} In galaxies, there is a well-pronounced
correlation between the mass of the molecular hydrogen and the integrated
SFR,
 known as the ``molecular Schmidt law''
\citep{bigiel_08a,kennicutt_07a}. For
 higher masses of molecular hydrogen we
have higher SFRs and thus as shown
 before higher ratios $\mathscr{R}_{\rm
  int}$. Therefore, we expect the ratio $\mathscr{R}_{\rm int}$ rising not only
 with
the SFR but also with the mass of the molecular hydrogen. We plot the
correlation in Figures~\ref{fig:sfr}(e) and (f) for the integrated molecular
hydrogen mass and mass surface density. We
 indeed find some weak
correlation, where the ratio $\mathscr{R}_{\rm int}$ rises with both the molecular hydrogen
mass and mass surface density.

{\it Atomic hydrogen mass.} We have compared the ratio $\mathscr{R}_{\rm int}$ with
the atomic
 hydrogen mass, but found no obvious correlation either for the
mass or for the mass surface density as shown in
 Figures~\ref{fig:sfr2}(a) and
(b).

{\it Total hydrogen mass.} The ratio $\mathscr{R}_{\rm int}$ is shown also as a function
of the total hydrogen mass in Figures~\ref{fig:sfr2}(c) and (d), the sum of atomic
and molecular hydrogen. No correlation is found for either the mass or the
mass surface density.

{\it Inclination angle.} We found a dependence of the ratio $\mathscr{R}_{\rm int}$
on the
 inclination angle of the galaxy as shown in
Figure~\ref{fig:sfr2}(e). With
 increasing inclination angle, i.e., the galaxy is
seen closer to an edge-on
 position, the ratio $\mathscr{R}_{\rm int}$ drops. This
dependence is not easily understood. We
 adjusted the size of the ellipses of
integration to include any RC halo
 emission (Section~\ref{subsec:radial}), so
deficiency in RC emission can be
 excluded. It could be that we overestimate
the hybrid $\Sigma_{\rm SFR}$, because in an
 edge-on position the diffuse ``cirrus''
emission at $24~\rm \mu m$ may add up
 to levels such that it exceeds the
detection threshold due to longer
 lines of sight.

{\it Galaxy type.} We tested whether the ratio $\mathscr{R}_{\rm int}$ depends on the type of galaxy,
e.g.\ whether it is an early or late type galaxy. This plot is shown in
Figure~\ref{fig:sfr2}f with no correlation visible.

{\it B-band magnitude.} The $B$-band magnitude is a coarse tracer for
SF showing mainly the young, blue population of massive stars. We
find again a slight dependence of higher ratios $\mathscr{R}_{\rm int}$ toward higher SFRs as traced
by the $B$-band as shown in Figure~\ref{fig:sfr3}(a).

{\it Thermal RC fraction.} We compare the ratio $\mathscr{R}_{\rm int}$ with the
thermal fraction of the RC emission as measured from H$\alpha$ emission in
Figure~\ref{fig:sfr3}(b). We do not find any clear correlation. The two galaxies
with the largest thermal fraction, NGC~2403~ and IC~2574, have low ratios with
$\mathscr{R}_{\rm int}\approx 0.35$.

{\it Fractional polarization.} We used the values of \citet{heald_09a}, who
measured the fractional polarization of the WSRT SINGS galaxies at $\lambda$
22~cm. No correlation with the ratio $\mathscr{R}_{\rm int}$ was found as shown in
Figure~\ref{fig:sfr3}(c). The
 fractional polarization is a measure of how
ordered the magnetic field
 orientation is within a resolution
element. However, one has to caution that
 at this long wavelength of
$\lambda$22~cm the polarized emission mostly comes
 from the near side of the
halo, because of Faraday depolarization by the disk
 magnetic field
\citep{braun_10a}.

{\it Metallicity.} Metallicity values of \citet{walter_08a} were used to study
the dependence of the ratio $\mathscr{R}_{\rm int}$ on metallicity. Our sample
galaxies do not show any obvious correlation (Figure~\ref{fig:sfr3}(d)).

{\it Redshift.} The galaxies in our sample are all nearby within 25~Mpc and
thus have only small redshifts. The ratio $\mathscr{R}_{\rm int}$ as a function of redshift is shown
in Figure~\ref{fig:sfr3}(e). No obvious correlation is visible.

{\it Stellar mass.} Figure~\ref{fig:sfr3}(f) shows the plot of the ratio as a
function of stellar mass. A weak correlation appears that larger ratios
$\mathscr{R}_{\rm int}$ are found at higher stellar masses.

We have also tested the above correlation with the slope of the pixel-by-pixel
plots as measured in Section~\ref{res:pix}. These plots are shown in
Figures~\ref{fig:sfr}--\ref{fig:sfr3}, overlaid as open circles. We found no obvious correlation for any of the
parameters. In conclusion to this section we find that the ratio $\mathscr{R}_{\rm
  int}$ appears to be indeed a universal parameter, as we do not find any
obvious correlation with any galaxy parameter. There appears to be a weak
correlation of a rising ratio $\mathscr{R}_{\rm int}$ with SFR, $\Sigma_{\rm SFR}$, or SF tracer. But the scatter is
substantial and we cannot be confident at this stage if any relation would
become evident when dealing with a larger sample. We particularly note that we
do not find any dependence of the integrated ratio $\mathscr{R}_{\rm int}$ on either the ratio of radiation energy density to
magnetic field energy density or the non-thermal radio spectral index. This points to the
fact that neither high IC radiation losses nor CRE escape are responsible for radio dim
galaxies, as we have discussed in Section~\ref{dis:global}.

\clearpage

\section{Atlas of Maps and Profile Plots}
\label{app:sample}
\begin{figure}[tbhp]
\centering
\resizebox{0.9\hsize}{!}{ \includegraphics{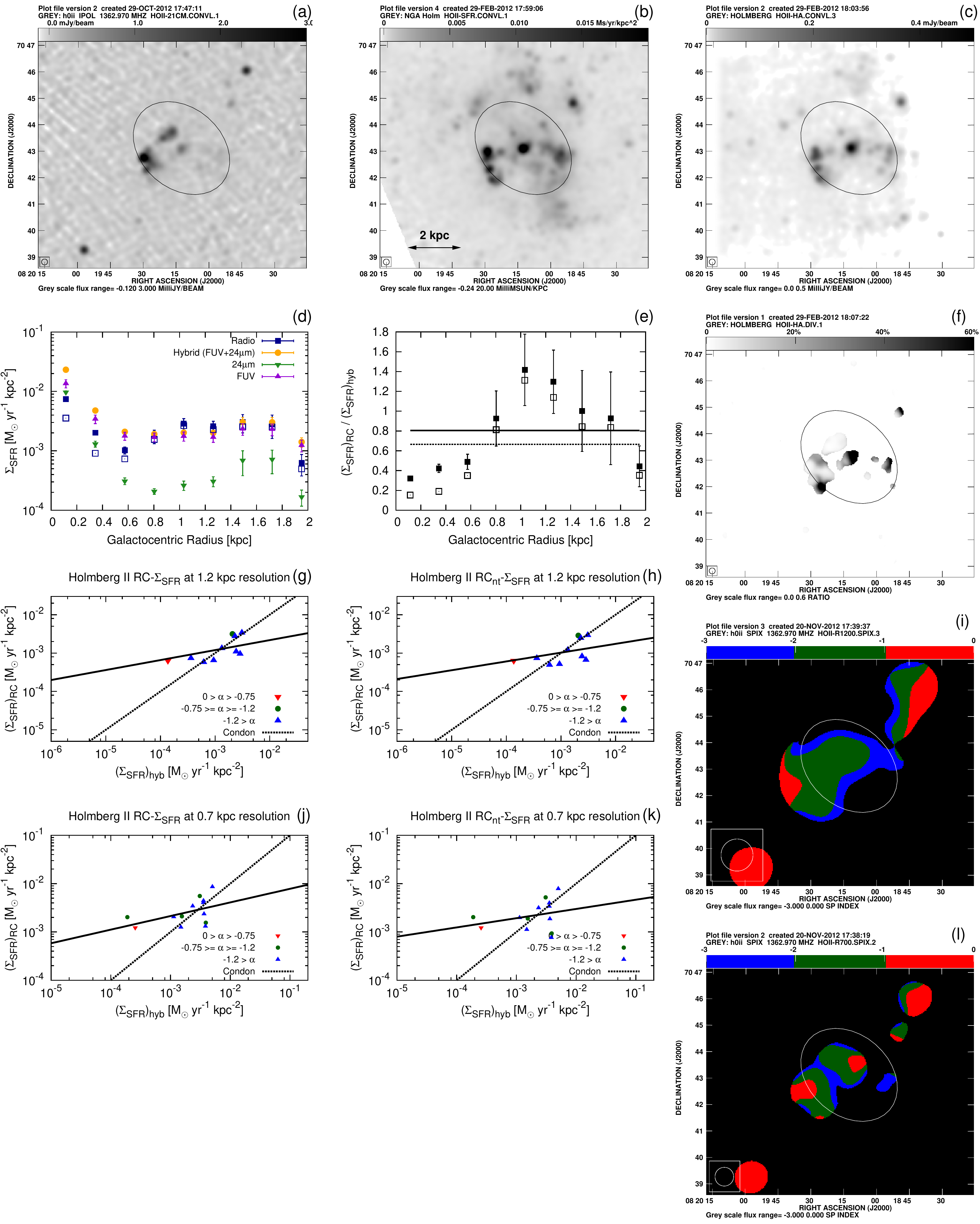}}
\caption{Holmberg~II. (a) RC emission at $\lambda 22~\rm
  cm$. (b) Hybrid SFR surface density $(\Sigma_{\rm SFR})_{\rm
    hyb}$. (c) Thermal RC emission as derived from H$\alpha$. (d) Radio
  $\Sigma_{\rm SFR}$ (dark blue) and
 hybrid $\Sigma_{\rm SFR}$ (orange) as a function
  of galactocentric radius. (e) Ratio $\mathscr{R}$ of radio to hybrid
 $\Sigma_{\rm SFR}$. Open
  symbols represent the non-thermal RC emission alone. (f) Predicted thermal
  RC fraction. (g) Radio $\Sigma_{\rm SFR}$ as a function of hybrid $\Sigma_{\rm
    SFR}$ at $\rm 1.2~kpc$
 resolution. (h) Same as (g) but with thermal emission
  subtracted. (i) Radio spectral
 index at $\rm 1.2~kpc$ resolution. (j--l)
  Same as
  (g--i) but at $\rm 0.7~kpc$ resolution. The
 typical uncertainty of the radio
  spectral index is $0.6$.}
\label{fig:hoii}
\end{figure}

\begin{figure}[tbhp]
\centering
\resizebox{1.0\hsize}{!}{ \includegraphics{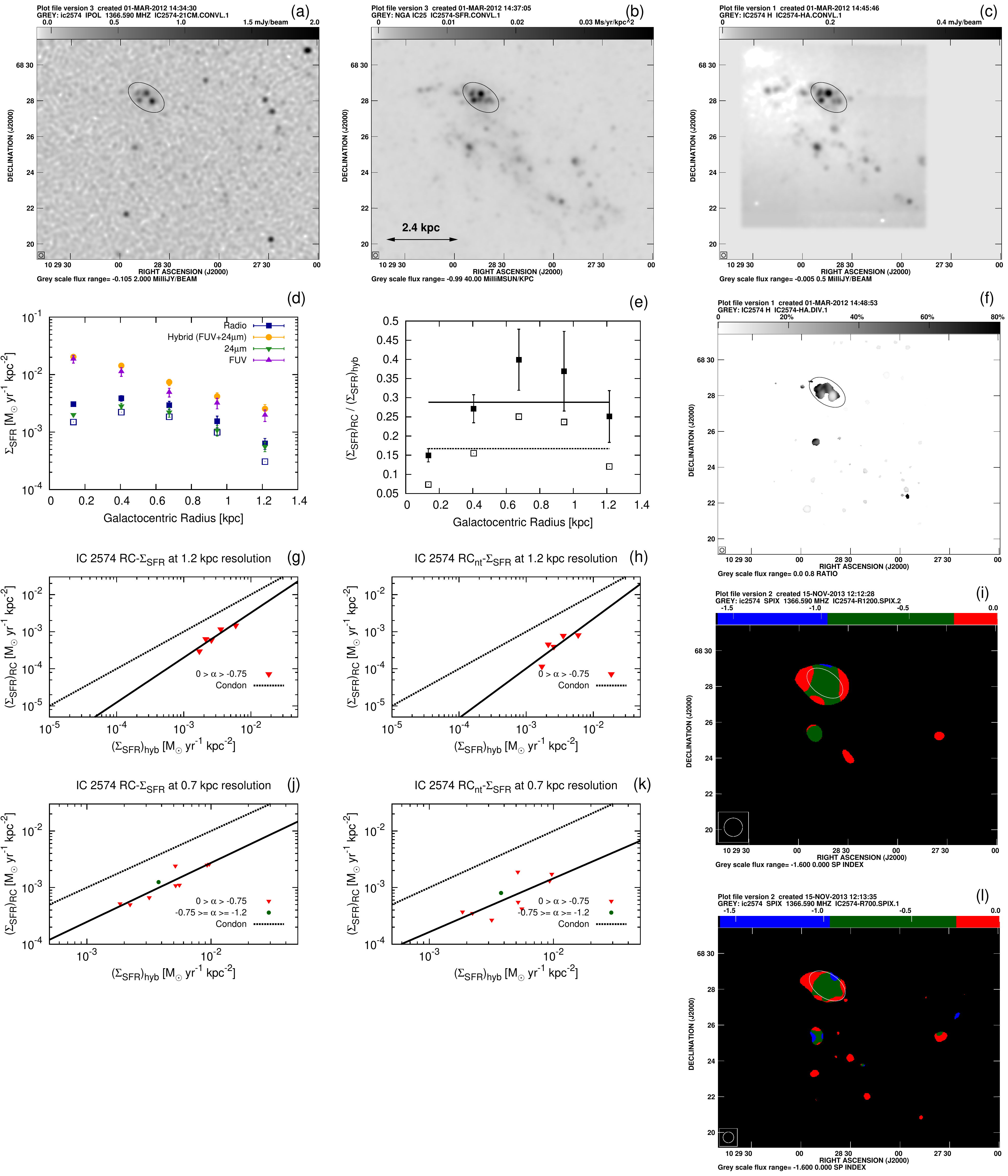}}
\caption{IC~2574. (a) RC emission at $\lambda 22~\rm
  cm$. (b) Hybrid SFR surface density $(\Sigma_{\rm SFR})_{\rm hyb}$. (c) Thermal RC
  emission as derived from H$\alpha$. (d) Radio $\Sigma_{\rm SFR}$ (dark blue) and
  hybrid $\Sigma_{\rm SFR}$ (orange) as a function of galactocentric radius. (e) Ratio $\mathscr{R}$ of radio to hybrid
  $\Sigma_{\rm SFR}$. Open symbols represent the non-thermal RC emission
  alone. (f) Predicted thermal RC fraction. (g) Radio $\Sigma_{\rm SFR}$ as a
  function of hybrid $\Sigma_{\rm SFR}$ at $\rm 1.2~kpc$ resolution. (h) Same
  as (g) but with thermal
  emission subtracted. (i) Radio spectral index at $\rm 1.2~kpc$ resolution. (j--l)
  Same as (g--i) but at $\rm 0.7~kpc$ resolution. The typical uncertainty of the radio
  spectral index is $0.6$.}
\label{fig:ic2574}
\end{figure}

\begin{figure}[tbhp]
\centering
\resizebox{1.0\hsize}{!}{ \includegraphics{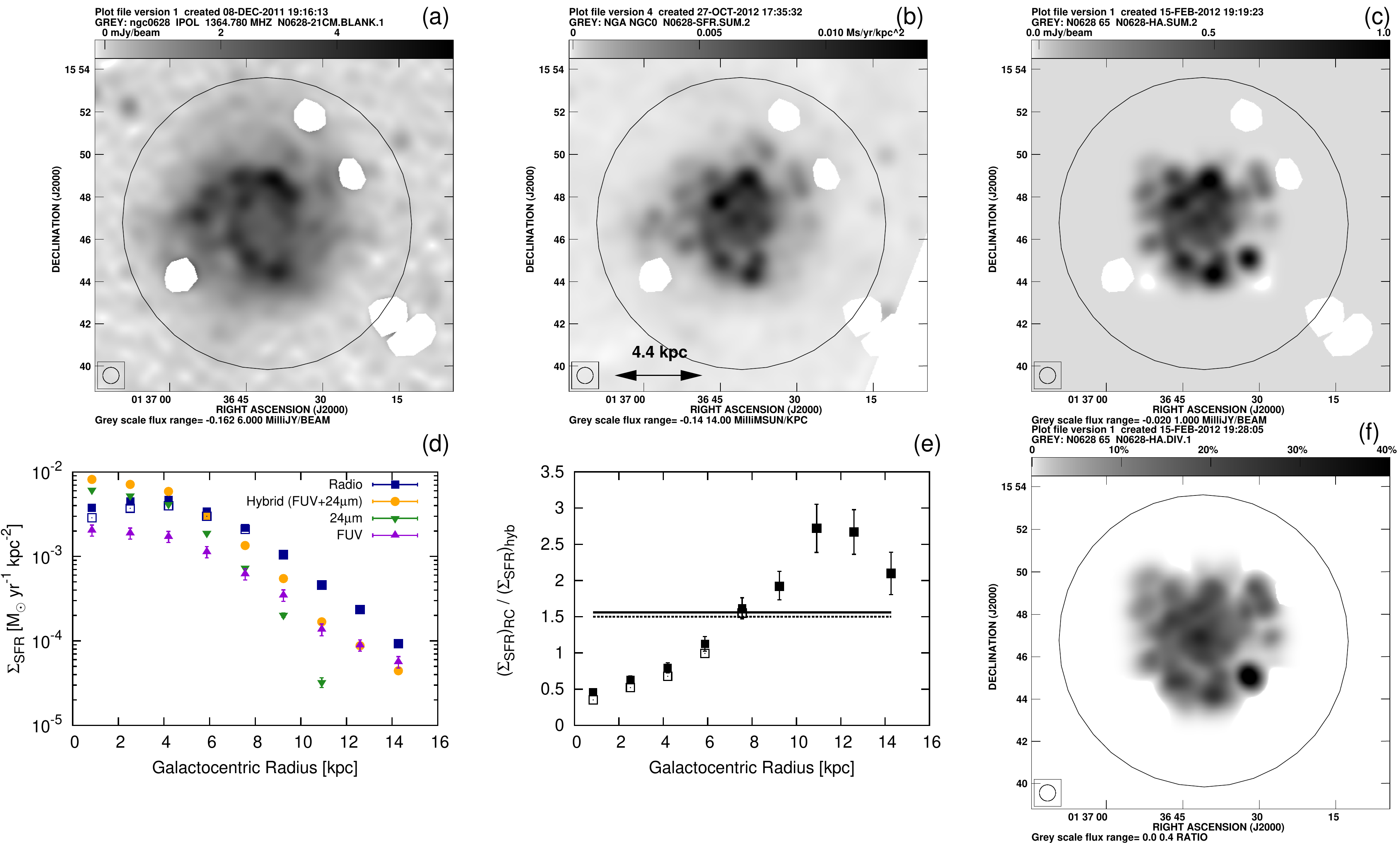}}
\caption{NGC~628.  (a) RC emission at $\lambda 22~\rm
  cm$. (b) Hybrid SFR surface density $(\Sigma_{\rm SFR})_{\rm hyb}$. (c) Thermal RC
  emission as derived from H$\alpha$. (d) Radio $\Sigma_{\rm SFR}$ (dark blue) and
  hybrid $\Sigma_{\rm SFR}$ (orange) as a function of galactocentric radius. (e) Ratio $\mathscr{R}$ of radio to hybrid
  $\Sigma_{\rm SFR}$. Open symbols represent the non-thermal RC emission
  alone. (f) Predicted thermal RC fraction.}
\label{fig:n0628}
\end{figure}

\begin{figure}[tbhp]
\centering
\resizebox{1.0\hsize}{!}{ \includegraphics{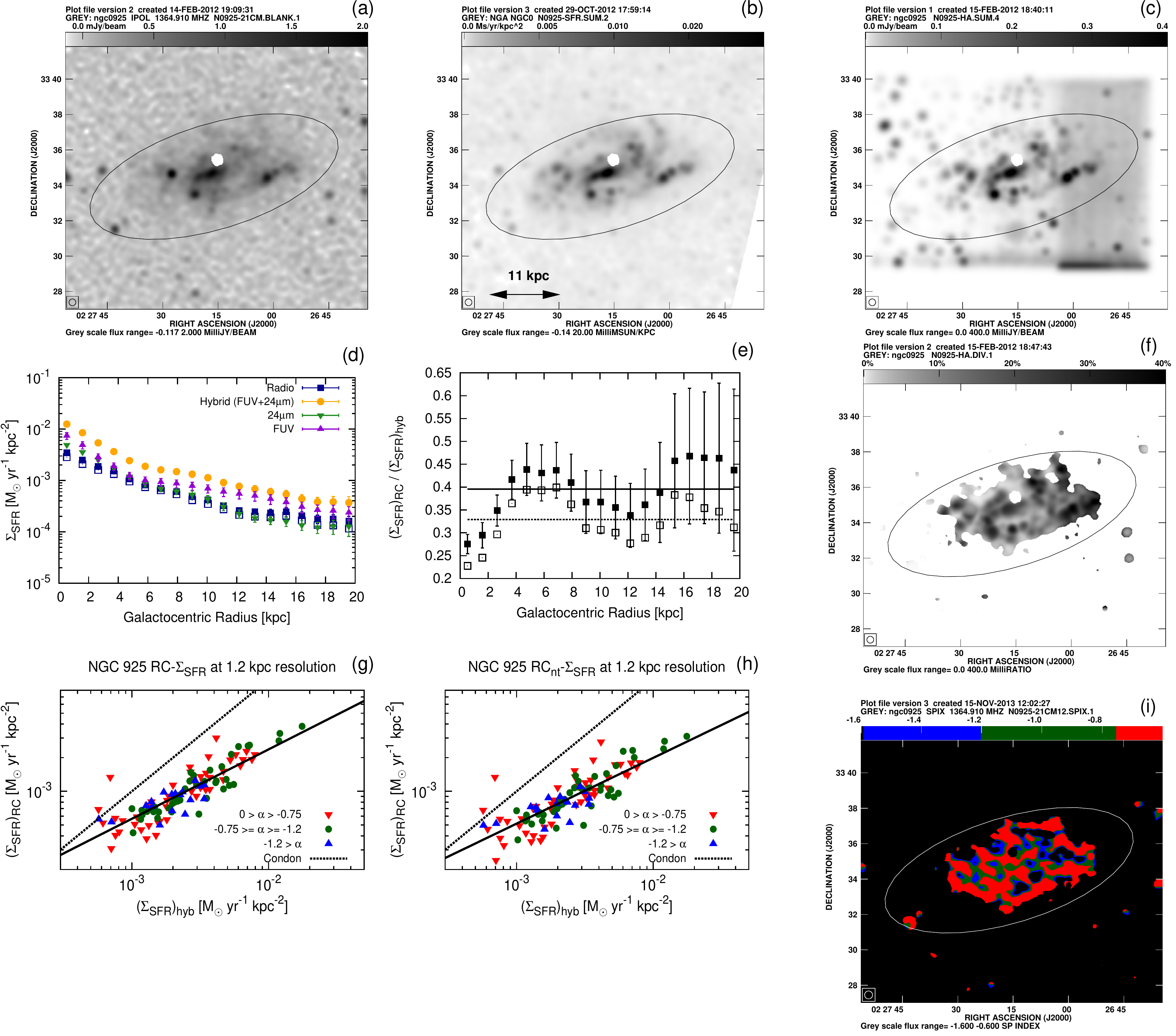}}
\caption{NGC~925. (a) RC emission at $\lambda
 22~\rm
  cm$. (b) Hybrid SFR surface density $(\Sigma_{\rm SFR})_{\rm
    hyb}$. (c) Thermal
 RC emission as derived from H$\alpha$. (d) Radio
  $\Sigma_{\rm SFR}$ (dark blue) and hybrid $\Sigma_{\rm SFR}$ (orange) as a function
  of galactocentric radius. (e) Ratio $\mathscr{R}$ of
 radio to hybrid $\Sigma_{\rm SFR}$. Open
  symbols represent the non-thermal RC emission
 alone. (f) Predicted thermal
  RC fraction. (g) Radio $\Sigma_{\rm SFR}$ as a function of
 hybrid $\Sigma_{\rm
    SFR}$ at $\rm 1.2~kpc$ resolution. (h) Same as (g) but with thermal emission
  subtracted. (i) Radio spectral index at $\rm 1.2~kpc$ resolution. The
  typical
 uncertainty of the radio spectral index is $1.5$ and $0.6$ for radio
  $\Sigma_{\rm SFR}$ of
 $10^{-3}$ and $3\times 10^{-3}~M_\odot~ \rm yr^{-1}~
  kpc^{-2}$,
 respectively.}
\label{fig:n0925}
\end{figure}

\begin{figure}[tbhp]
\centering
\resizebox{1.0\hsize}{!}{ \includegraphics{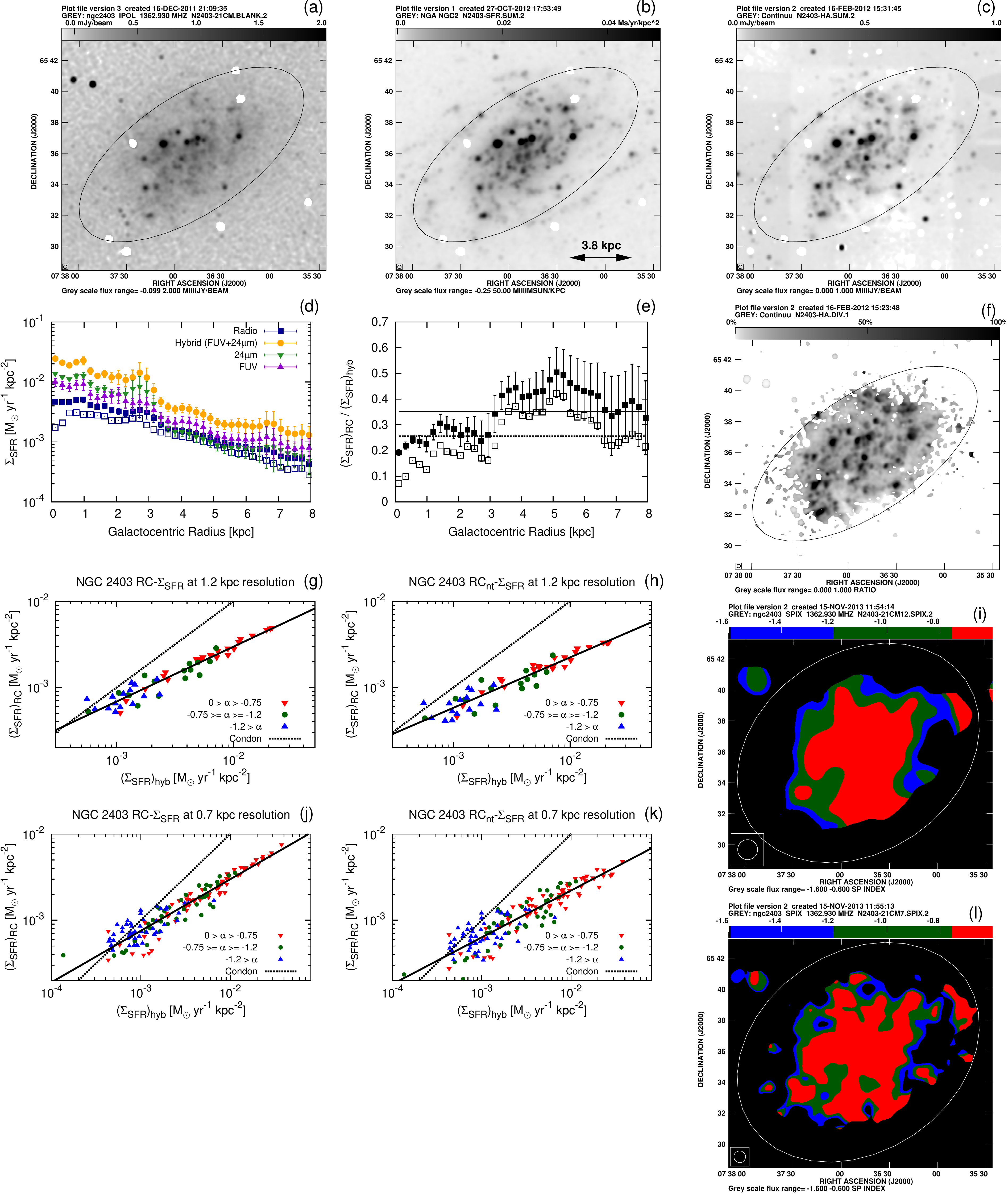}}
\caption{NGC~2403. (a) RC emission at
 $\lambda 22~\rm
  cm$. (b) Hybrid SFR surface density $(\Sigma_{\rm SFR})_{\rm
    hyb}$. (c)
 Thermal RC emission as derived from H$\alpha$. (d)
 Radio
  $\Sigma_{\rm SFR}$ (dark blue) and hybrid $\Sigma_{\rm SFR}$ (orange) as a function
  of galactocentric radius. (e) Ratio $\mathscr{R}$
 of radio to hybrid $\Sigma_{\rm SFR}$. Open
  symbols represent the non-thermal RC emission
 alone. (f) Predicted thermal
  RC fraction. (g) Radio $\Sigma_{\rm SFR}$ as a function of
 hybrid $\Sigma_{\rm
    SFR}$ at $\rm 1.2~kpc$ resolution. (h) Same as (g) but with thermal emission
  subtracted. (i) Radio spectral index at $\rm 1.2~kpc$ resolution. (j--l))
  Same as (g--i)
 but at $\rm 0.7~kpc$ resolution. The typical uncertainty of the radio
  spectral index is $1.2$, $0.6$, and $0.3$ for the blue, green, and red data
  points,
 respectively.}
\label{fig:n2403}
\end{figure}

\begin{figure}[tbhp]
\centering
\resizebox{1.0\hsize}{!}{ \includegraphics{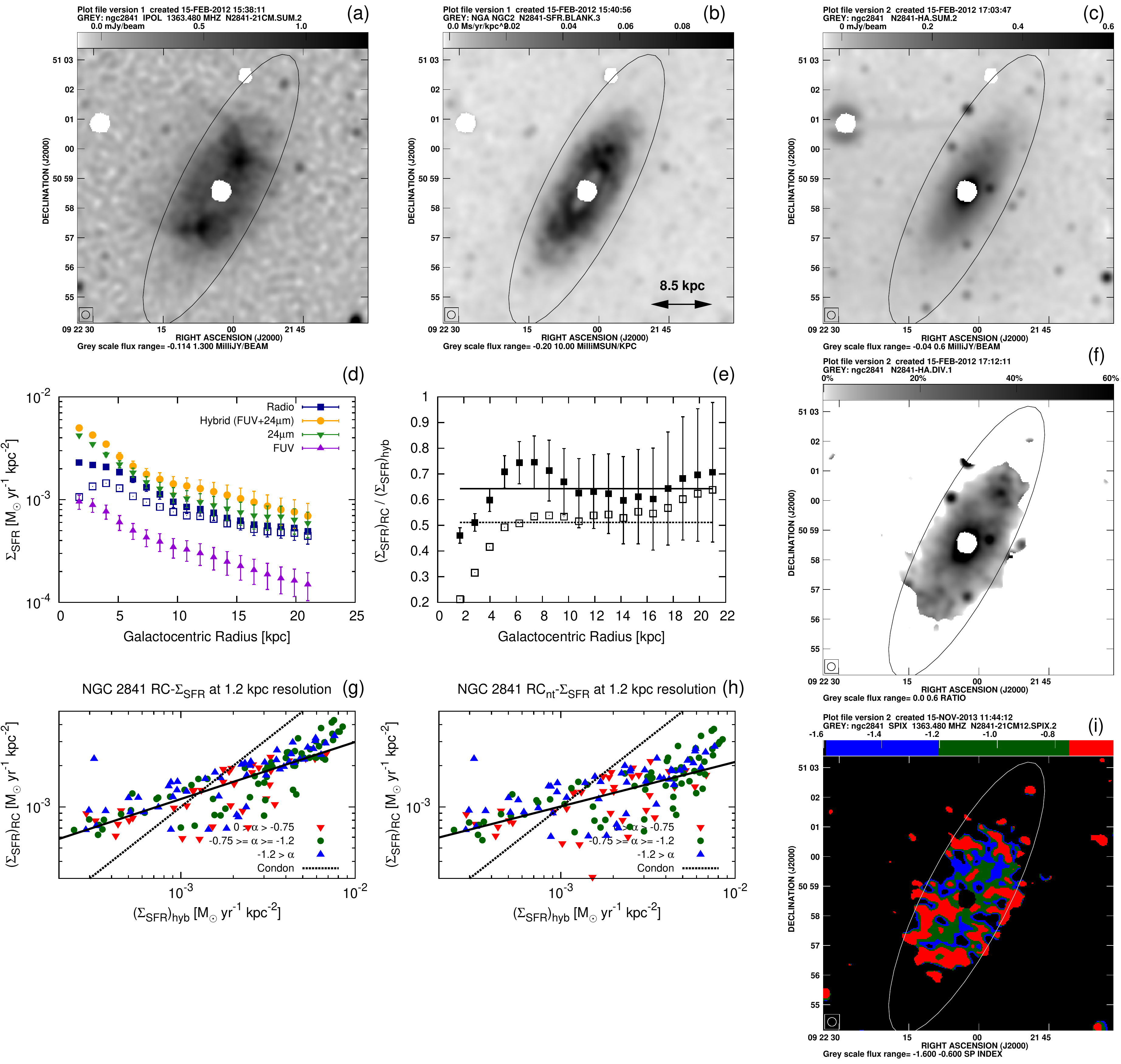}}
\caption{NGC~2841. (a) RC emission at
 $\lambda 22~\rm
  cm$. (b) Hybrid SFR surface density $(\Sigma_{\rm SFR})_{\rm
    hyb}$. (c)
 Thermal RC emission as derived from H$\alpha$. (d)
 Radio
  $\Sigma_{\rm SFR}$ (dark blue) and hybrid $\Sigma_{\rm SFR}$ (orange) as a function
  of galactocentric radius. (e) Ratio $\mathscr{R}$
 of radio to hybrid $\Sigma_{\rm SFR}$. Open
  symbols represent the non-thermal RC emission
 alone. (f) Predicted thermal
  RC fraction. (g) Radio $\Sigma_{\rm SFR}$ as a function of
 hybrid $\Sigma_{\rm
    SFR}$ at $\rm 1.2~kpc$ resolution. (h) Same as (g) but with thermal emission
  subtracted. (i) Radio spectral index at $\rm 1.2~kpc$ resolution. The
  typical
 uncertainty of the radio spectral index is $1.2$ and $0.6$ for radio
  $\Sigma_{\rm SFR}$ of
 $2\times10^{-3}$ and $4\times 10^{-3}~M_\odot~\rm
  yr^{-1}~ kpc^{-2}$,
 respectively.}
\label{fig:n2841}
\end{figure}

\begin{figure}[tbhp]
\centering
\resizebox{1.0\hsize}{!}{ \includegraphics{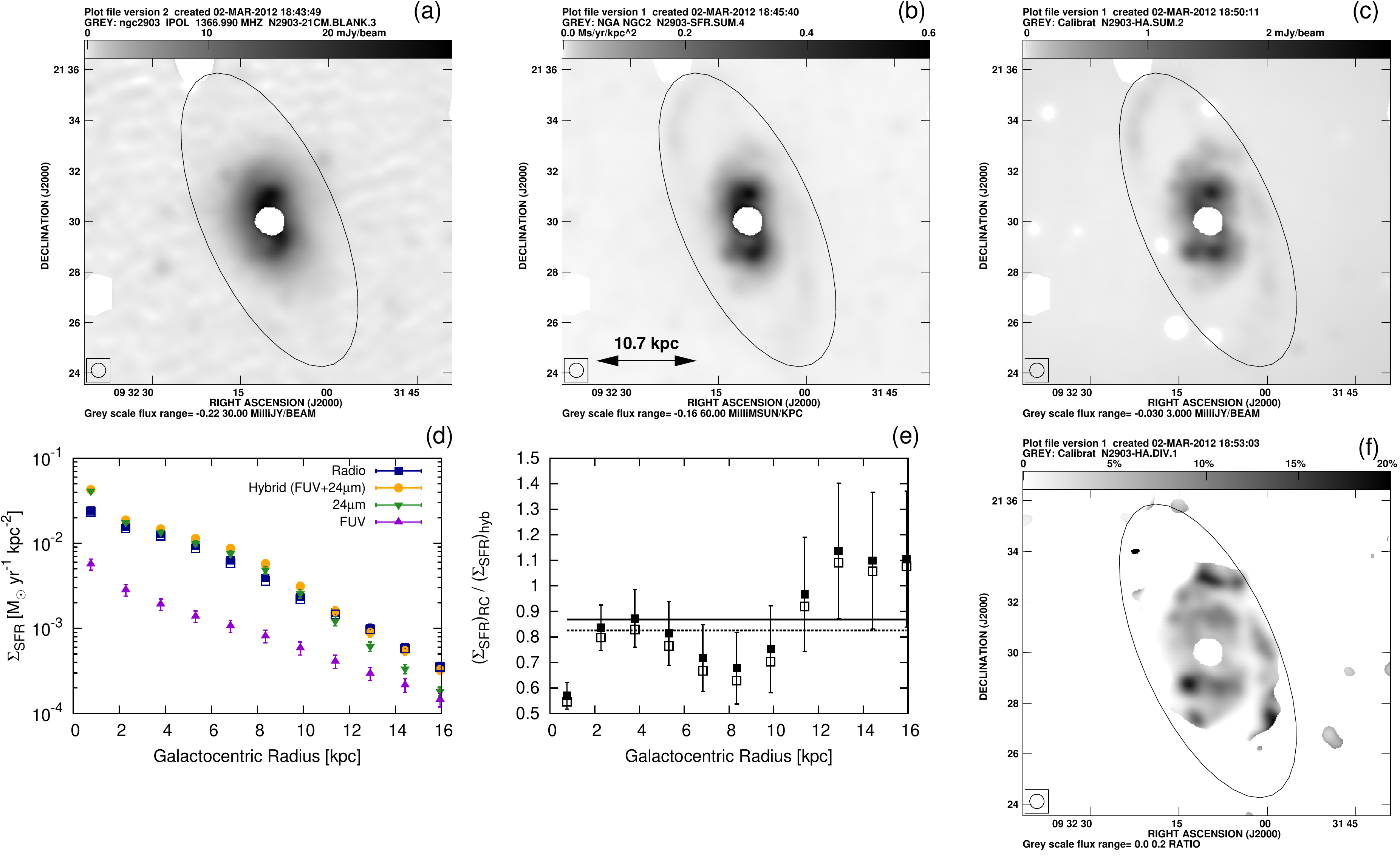}}
\caption{NGC~2903. (a) RC emission at $\lambda 22~\rm
  cm$. (b) Hybrid SFR surface density $(\Sigma_{\rm SFR})_{\rm
    hyb}$. (c) Thermal RC
 emission as derived from H$\alpha$. (d) Radio
  $\Sigma_{\rm SFR}$ (dark blue) and
 hybrid $\Sigma_{\rm SFR}$ (orange) as a function
  of galactocentric radius. (e) Ratio $\mathscr{R}$ of radio to hybrid
 $\Sigma_{\rm SFR}$. Open
  symbols represent the non-thermal RC emission
 alone. (f) Predicted thermal
  RC fraction.}
\label{fig:n2903}
\end{figure}

\begin{figure}[tbhp]
  \centering
  \resizebox{1.0\hsize}{!}{ \includegraphics{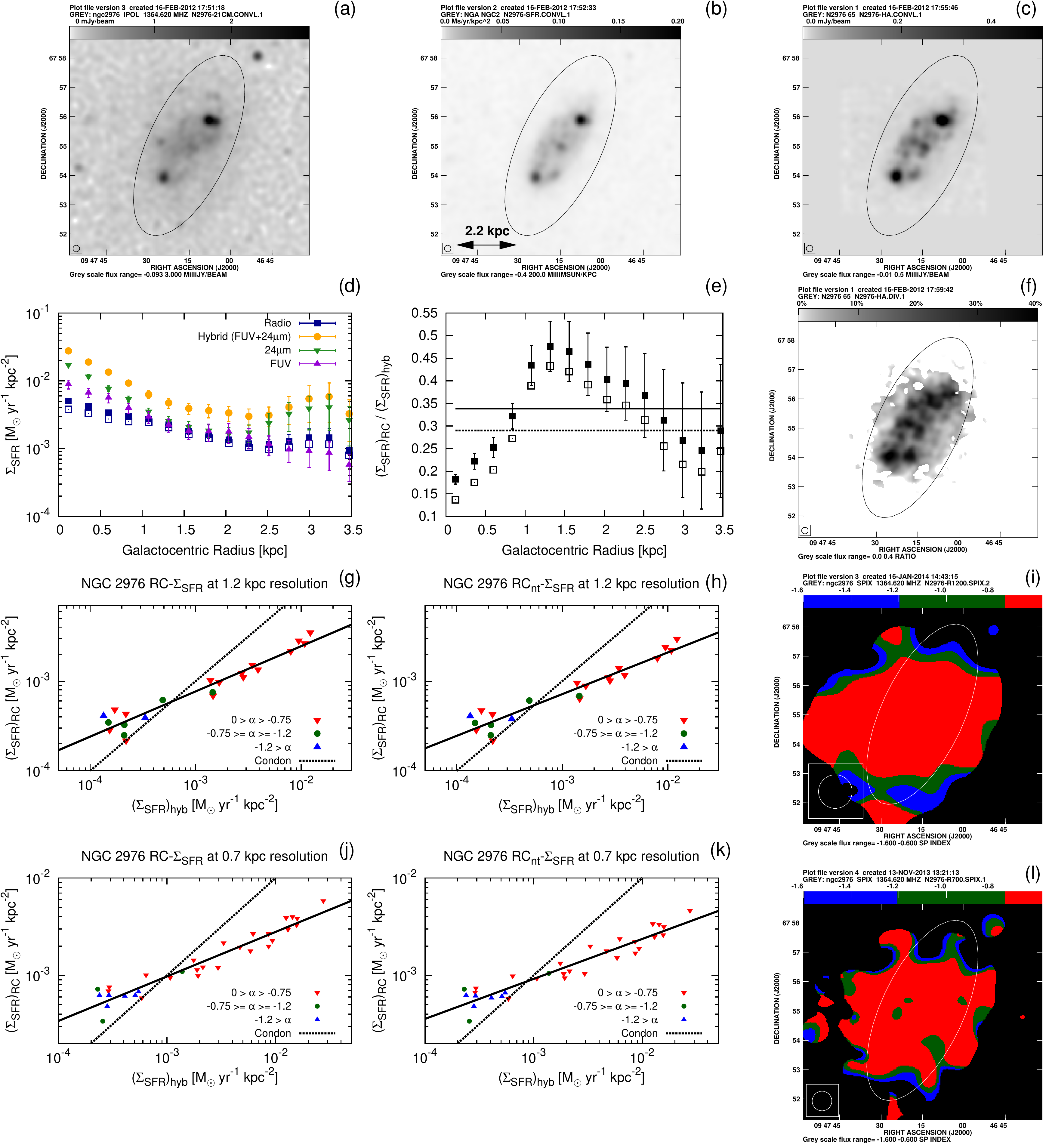}}
  \caption{NGC~2976. (a) RC emission at $\lambda 22~\rm
    cm$. (b) Hybrid SFR surface density $(\Sigma_{\rm SFR})_{\rm
      hyb}$. (c) Thermal RC emission
 as derived from H$\alpha$. (d) Radio
    $\Sigma_{\rm SFR}$ (dark blue) and hybrid $\Sigma_{\rm SFR}$ (orange) as
 a function
    of galactocentric radius. (e) Ratio $\mathscr{R}$ of radio to hybrid $\Sigma_{\rm SFR}$. Open
    symbols
 represent the non-thermal RC emission alone. (f) Predicted thermal
    RC fraction. (g)
 Radio $\Sigma_{\rm SFR}$ as a function of hybrid
    $\Sigma_{\rm SFR}$ at $\rm 1.2~kpc$ resolution. (h) Same as (g)
 but with thermal
    emission subtracted. (i) Radio spectral index at $\rm 1.2~kpc$
    resolution. (j--l) Same as (g--i) but at $\rm 0.7~kpc$ resolution. The typical
    uncertainty
 of the radio spectral index is $0.8$ and $0.3$ for radio
    $\Sigma_{\rm SFR}$ of
 $7\times10^{-4}$ and $2\times 10^{-3}~M_\odot~\rm
    yr^{-1}~ kpc^{-2}$,
 respectively.}
  \label{fig:n2976}
\end{figure}

\begin{figure}[tbhp]
\centering
\resizebox{1.0\hsize}{!}{ \includegraphics{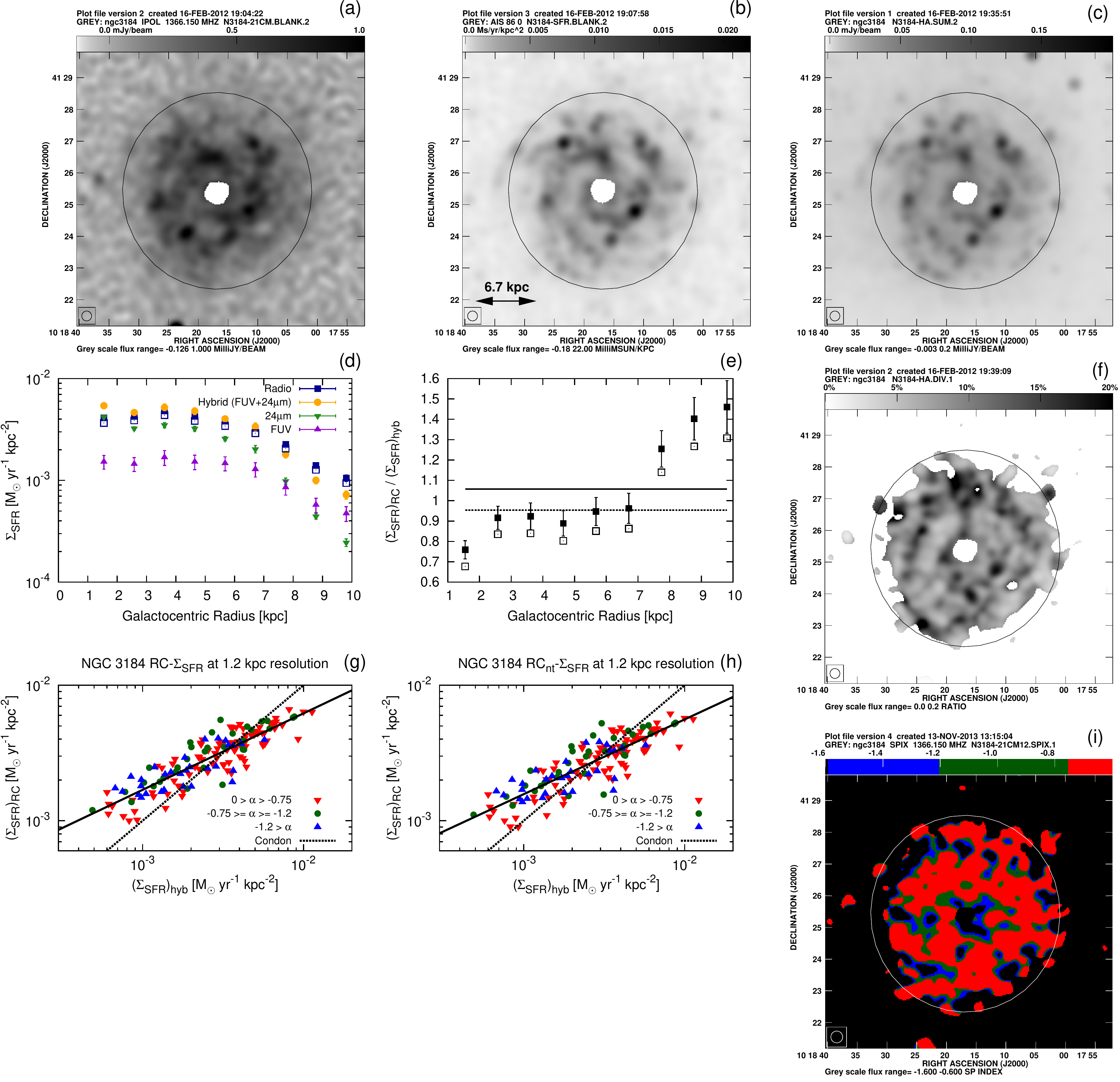}}
\caption{NGC~3184. (a) RC emission at
 $\lambda 22~\rm
  cm$. (b) Hybrid SFR surface density $(\Sigma_{\rm SFR})_{\rm
    hyb}$. (c)
 Thermal RC emission as derived from H$\alpha$. (d)
 Radio
  $\Sigma_{\rm SFR}$ (dark blue) and hybrid $\Sigma_{\rm SFR}$ (orange) as a function
  of galactocentric radius. (e) Ratio $\mathscr{R}$
 of radio to hybrid $\Sigma_{\rm SFR}$. Open
  symbols represent the non-thermal RC emission
 alone. (f) Predicted thermal
  RC fraction. (g) Radio $\Sigma_{\rm SFR}$ as a function of
 hybrid $\Sigma_{\rm
    SFR}$ at $\rm 1.2~kpc$ resolution. (h) Same as (g) but with thermal emission
  subtracted. (i) Radio spectral index at $\rm 1.2~kpc$ resolution. The
  typical
 uncertainty of the radio spectral index is $1.2$ and $0.5$ for radio
  $\Sigma_{\rm SFR}$ of
 $3\times10^{-3}$ and $8\times 10^{-3}~M_\odot~\rm
  yr^{-1}~ kpc^{-2}$,
 respectively.}
\label{fig:n3184}
\end{figure}

\begin{figure}[tbhp]
\centering
\resizebox{1.0\hsize}{!}{ \includegraphics{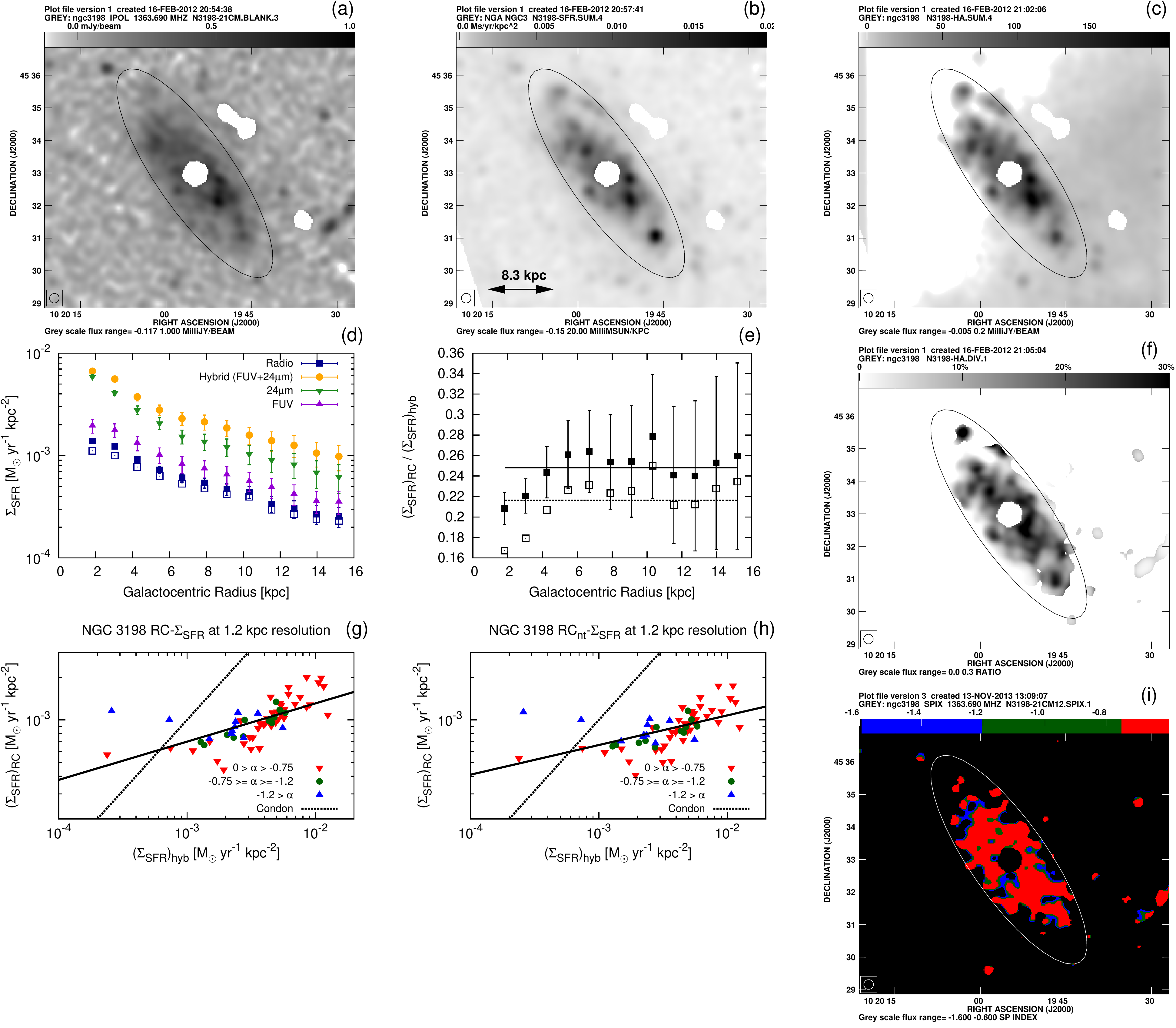}}
\caption{NGC~3198. (a) RC emission at
 $\lambda 22~\rm
  cm$. (b) Hybrid SFR surface density $(\Sigma_{\rm SFR})_{\rm
    hyb}$. (c)
 Thermal RC emission as derived from H$\alpha$. (d)
 Radio
  $\Sigma_{\rm SFR}$ (dark blue) and hybrid $\Sigma_{\rm SFR}$ (orange) as a function
  of galactocentric radius. (e) Ratio $\mathscr{R}$
 of radio to hybrid $\Sigma_{\rm SFR}$. Open
  symbols represent the non-thermal RC emission
 alone. (f) Predicted thermal
  RC fraction. (g) Radio $\Sigma_{\rm SFR}$ as a function of
 hybrid $\Sigma_{\rm
    SFR}$ at $\rm 1.2~kpc$ resolution. (h) Same as (g) but with thermal emission
  subtracted. (i) Radio spectral index at $\rm 1.2~kpc$ resolution. The
  typical
 uncertainty of the radio spectral index is $1.2$ for radio
  $\Sigma_{\rm SFR}$ of
 $1\times10^{-3}~M_\odot~\rm
  yr^{-1}~ kpc^{-2}$.}
\label{fig:n3198}
\end{figure}

\begin{figure}[tbhp]
\centering
\resizebox{1.0\hsize}{!}{ \includegraphics{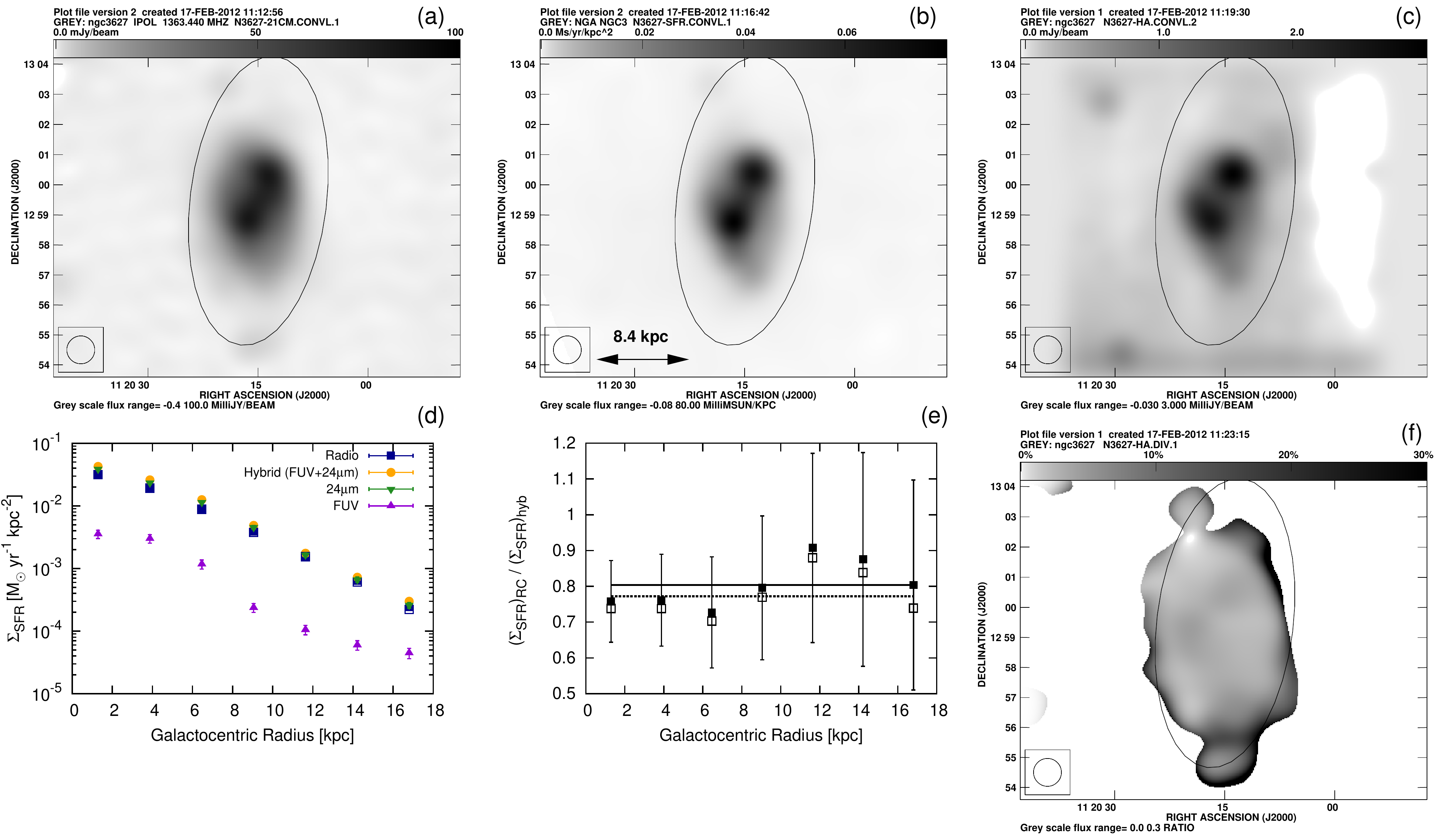}}
\caption{NGC~3627. (a) RC emission at $\lambda 22~\rm
  cm$. (b) Hybrid SFR surface density $(\Sigma_{\rm SFR})_{\rm hyb}$. (c) Thermal RC
  emission as derived from H$\alpha$. (d) Radio $\Sigma_{\rm SFR}$ (dark blue) and
  hybrid $\Sigma_{\rm SFR}$ (orange) as a function of galactocentric radius. (e) Ratio $\mathscr{R}$ of radio to hybrid
  $\Sigma_{\rm SFR}$. Open symbols represent the non-thermal RC emission
  alone. (f) Predicted thermal RC fraction.}
\label{fig:n3627}
\end{figure}

\begin{figure}[tbhp]
\centering
\resizebox{1.0\hsize}{!}{ \includegraphics{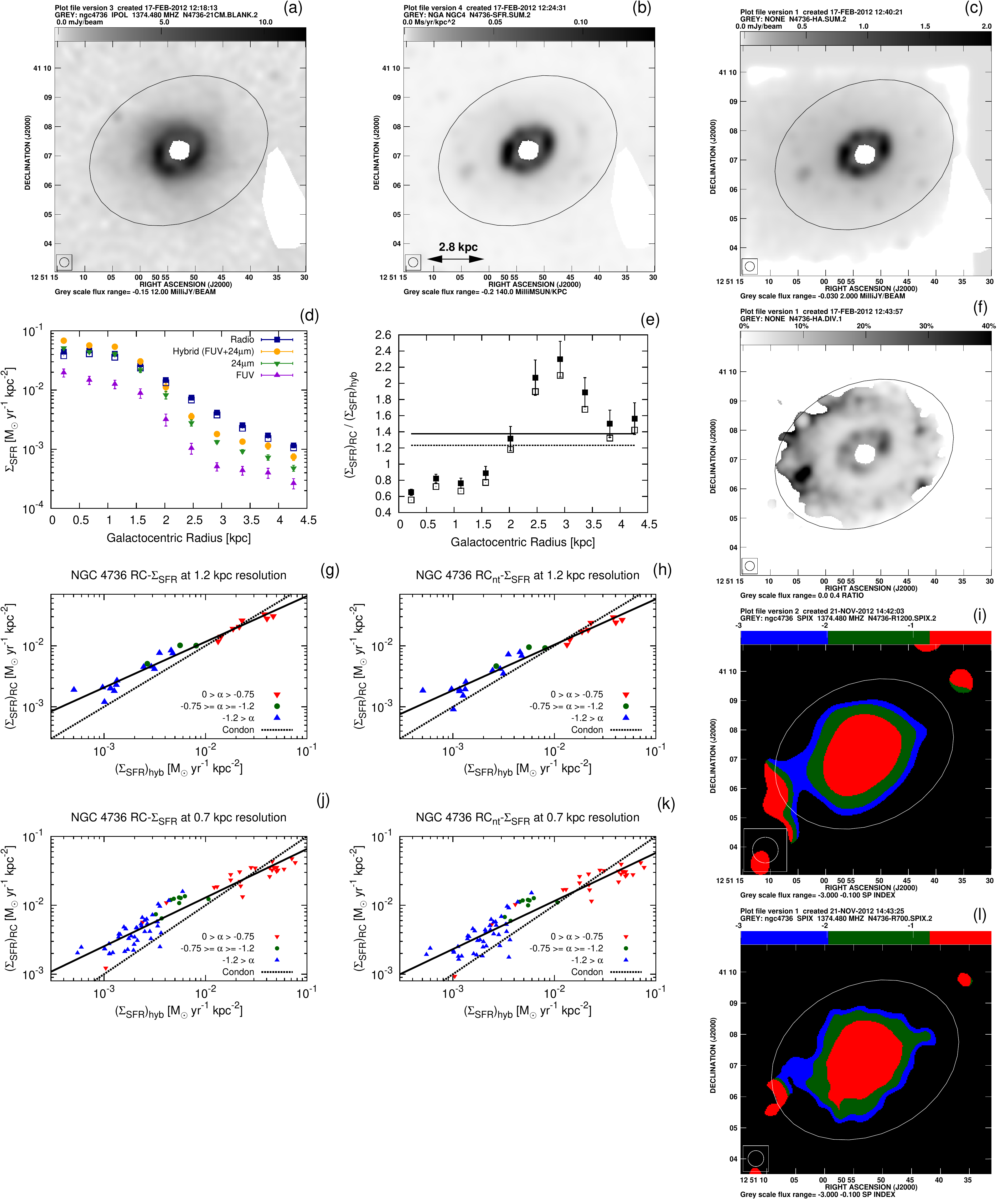}}
\caption{NGC~4736. (a) RC emission at $\lambda 22~\rm
  cm$. (b) Hybrid SFR surface density $(\Sigma_{\rm SFR})_{\rm
    hyb}$. (c) Thermal RC
 emission as derived from H$\alpha$. (d) Radio
  $\Sigma_{\rm SFR}$ (dark blue) and
 hybrid $\Sigma_{\rm SFR}$ (orange) as a function
  of galactocentric radius. (e) Ratio $\mathscr{R}$ of radio to hybrid
 $\Sigma_{\rm SFR}$. Open
  symbols represent the non-thermal RC emission
 alone. (f) Predicted thermal
  RC fraction. (g) Radio $\Sigma_{\rm SFR}$ as a
 function of hybrid $\Sigma_{\rm
    SFR}$ at $\rm 1.2~kpc$ resolution. (h) Same as (g) but with thermal
 emission
  subtracted. (i) Radio spectral index at $\rm 1.2~kpc$ resolution. (j--l)
 as
  (g--i) but at $\rm 0.7~kpc$ resolution. The typical uncertainty of the radio
  spectral index is $0.5$, $0.1$, and $0.05$ for the blue, green, and red data
  points,
 respectively.}
\label{fig:n4736}
\end{figure}

\begin{figure}[tbhp]
\centering
\resizebox{1.0\hsize}{!}{ \includegraphics{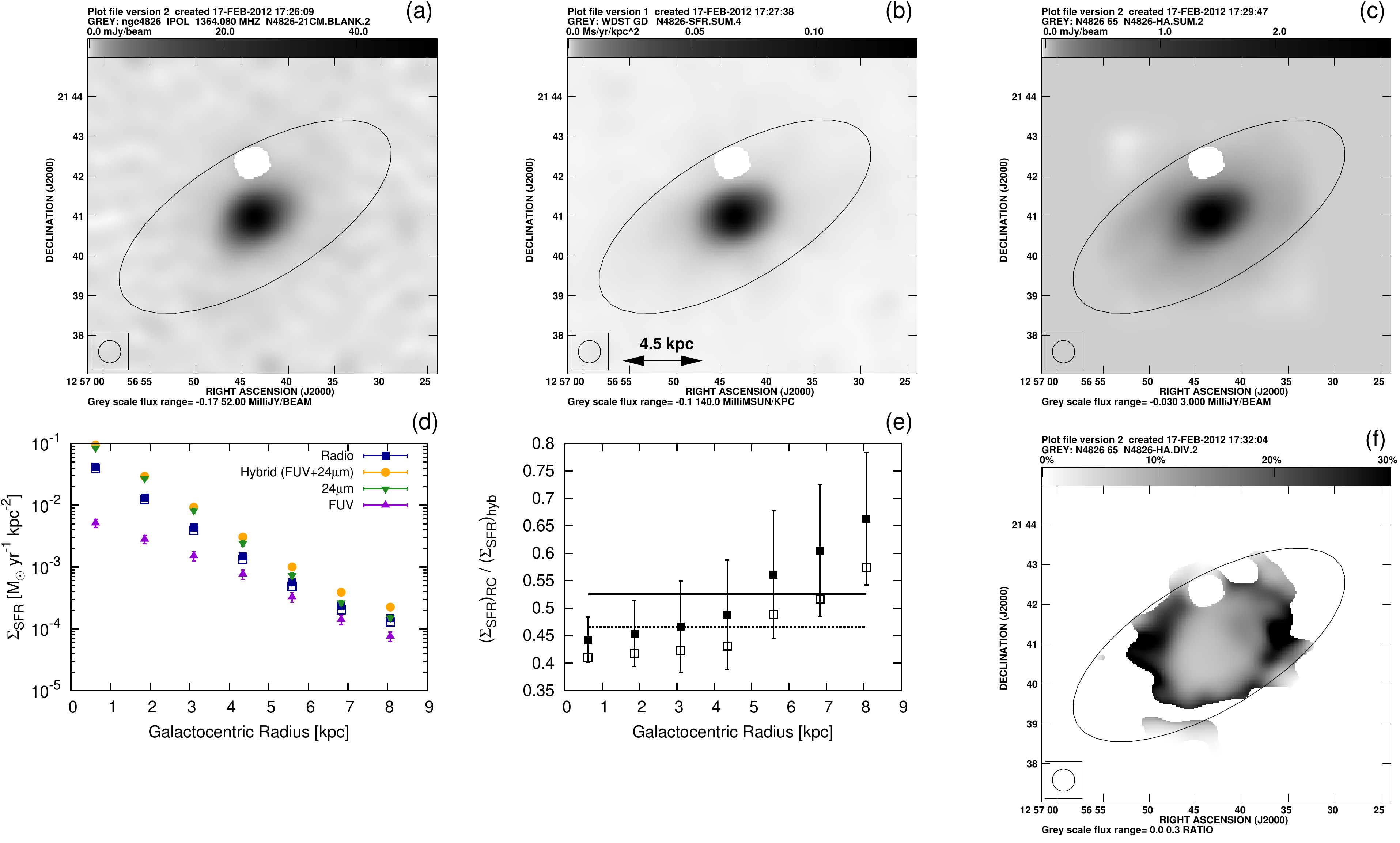}}
\caption{NGC~4826. (a) RC emission at $\lambda 22~\rm
  cm$. (b) Hybrid SFR surface density $(\Sigma_{\rm SFR})_{\rm
    hyb}$. (c) Thermal RC
 emission as derived from H$\alpha$. (d) Radio
  $\Sigma_{\rm SFR}$ (dark blue) and
 hybrid $\Sigma_{\rm SFR}$ (orange) as a function
  of galactocentric radius. (e) Ratio $\mathscr{R}$ of radio to hybrid
 $\Sigma_{\rm SFR}$. Open
  symbols represent the non-thermal RC emission
 alone. (f) Predicted thermal
  RC fraction.}
\label{fig:n4826}
\end{figure}

\begin{figure}[tbhp]
\centering
\resizebox{1.0\hsize}{!}{ \includegraphics{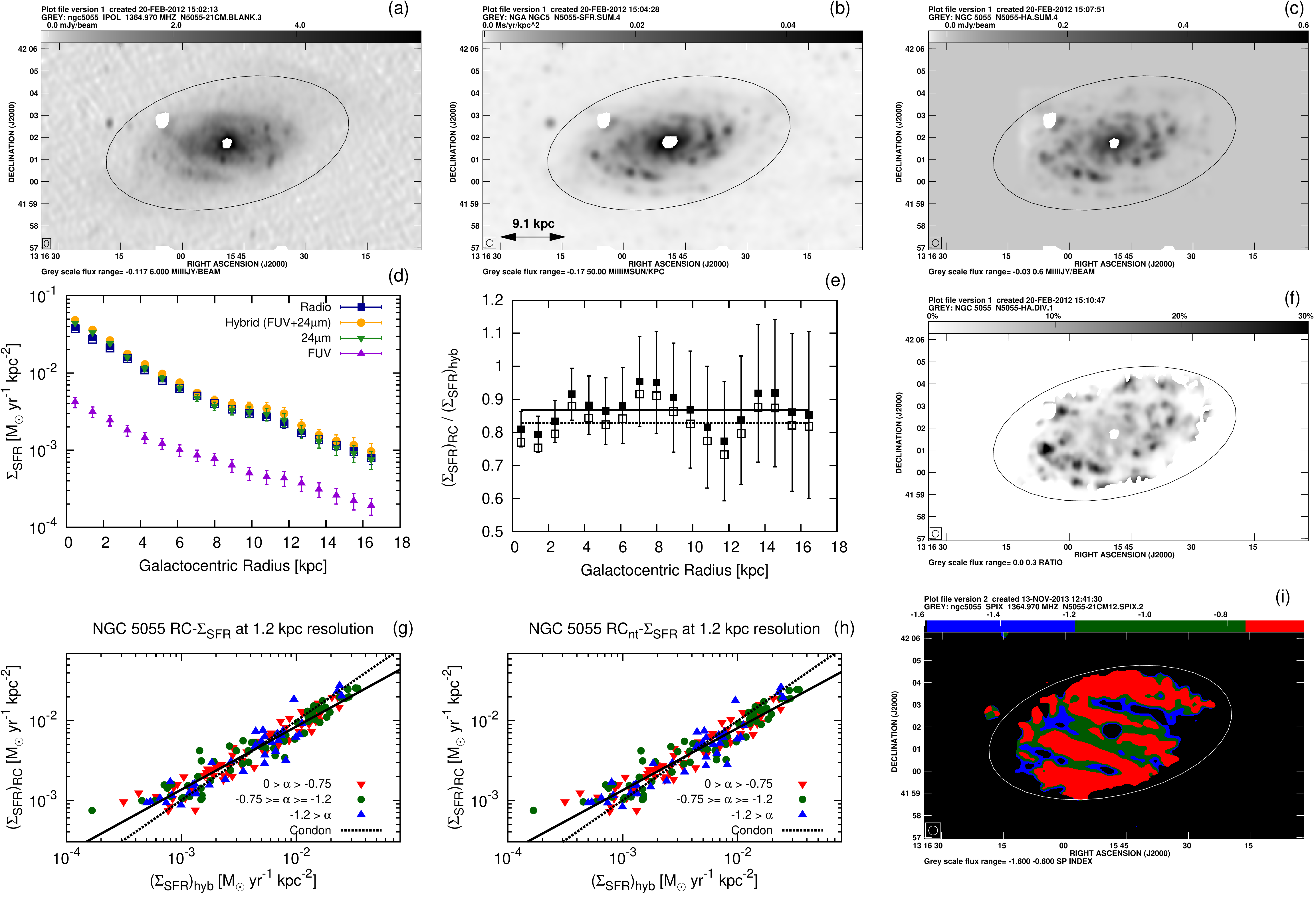}}
\caption{NGC~5055. (a) RC emission at $\lambda 22~\rm
  cm$. (b) Hybrid SFR surface density $(\Sigma_{\rm SFR})_{\rm
    hyb}$. (c) Thermal RC
 emission as derived from H$\alpha$. (d) Radio
  $\Sigma_{\rm SFR}$ (dark blue) and
 hybrid $\Sigma_{\rm SFR}$ (orange) as a function
  of galactocentric radius. (e) Ratio $\mathscr{R}$ of radio to hybrid
 $\Sigma_{\rm SFR}$. Open
  symbols represent the non-thermal RC emission
 alone. (f) Predicted thermal
  RC fraction. (g) Radio $\Sigma_{\rm SFR}$ as a
 function of hybrid $\Sigma_{\rm
    SFR}$ at $\rm 1.2~kpc$ resolution. (h) Same as (g) but with thermal
 emission
  subtracted. (i) Radio spectral index at $\rm 1.2~kpc$ resolution. The
  typical
 uncertainty of the radio spectral index is $1.2$, $0.3$, and $0.1$
  for radio $\Sigma_{\rm SFR}$ of
 $2\times10^{-3}$, $8\times10^{-3}$, and $3\times
  10^{-2}~M_\odot~\rm  yr^{-1}~ kpc^{-2}$,
 respectively.}
\label{fig:n5055}
\end{figure}

\begin{figure}[tbhp]
\centering
\resizebox{\hsize}{!}{ \includegraphics{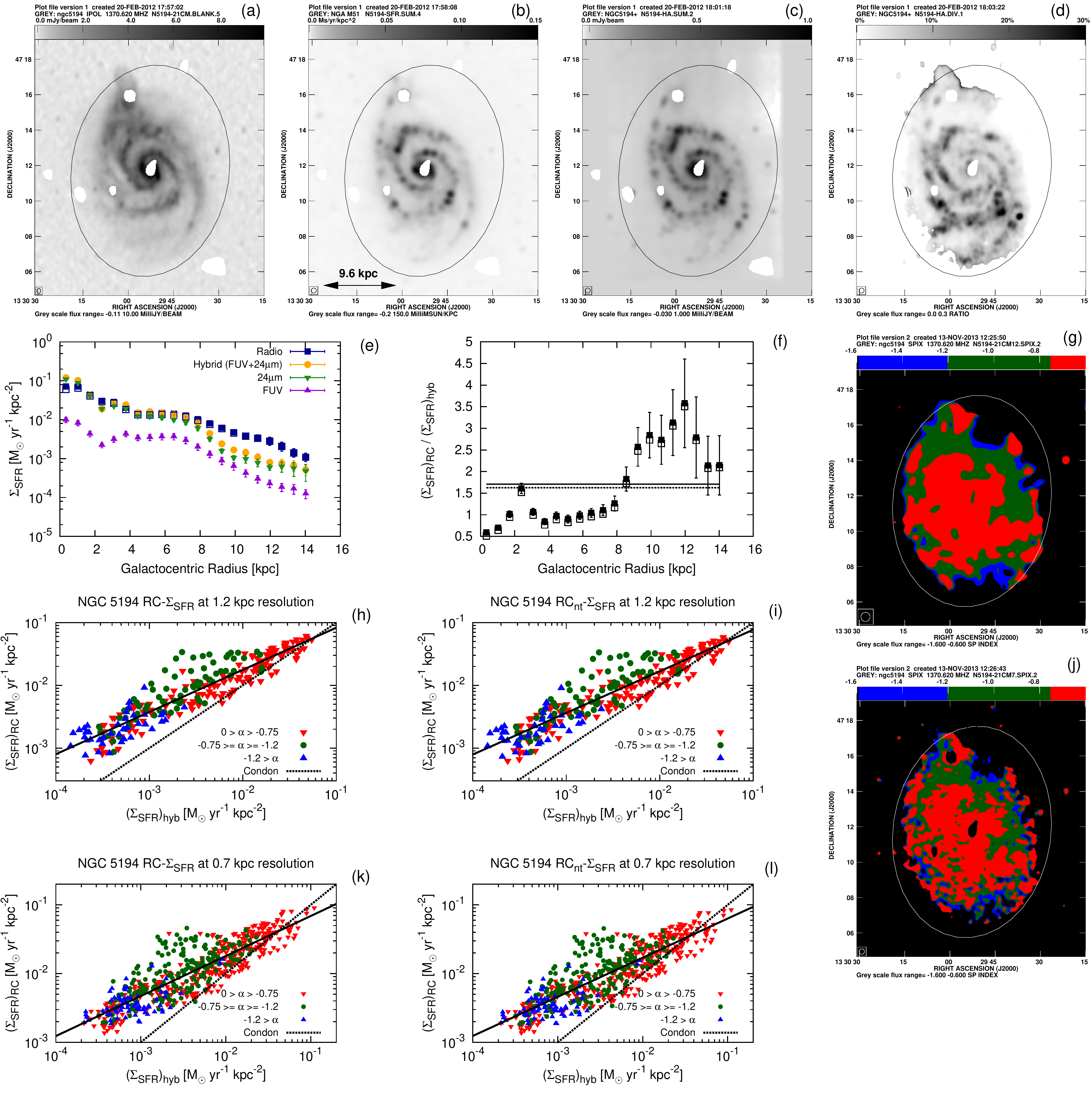}}
\caption{NGC~5194. (a) RC emission at $\lambda 22~\rm
  cm$. (b) Hybrid SFR surface density $(\Sigma_{\rm SFR})_{\rm
    hyb}$. (c) Thermal RC emission
 as derived from H$\alpha$. (d) Predicted
  thermal RC fraction.  (e) Radio $\Sigma_{\rm SFR}$ (dark blue)
 and hybrid
  $\Sigma_{\rm SFR}$ (orange) as a function of galactocentric radius. (f) Ratio $\mathscr{R}$ of radio to
  hybrid $\Sigma_{\rm SFR}$. Open symbols represent the non-thermal RC
  emission alone. (g)
 Radio spectral index at $\rm 1.2~kpc$ resolution. (h)
  Radio $\Sigma_{\rm SFR}$ as a function of
 hybrid $\Sigma_{\rm SFR}$ at $\rm
  1.2~kpc$ resolution. (i) Same as (h) but with thermal emission
  subtracted. (j--l) Same as
  (g--i) but at $\rm 0.7~kpc$ resolution. The typical uncertainty
 of the radio
  spectral index is $0.8$, $0.2$, and $0.1$ for the blue, green, and red
 data
  points, respectively.}
\label{fig:n5194}
\end{figure}

\begin{figure}
\centering
\resizebox{\hsize}{!}{ \includegraphics{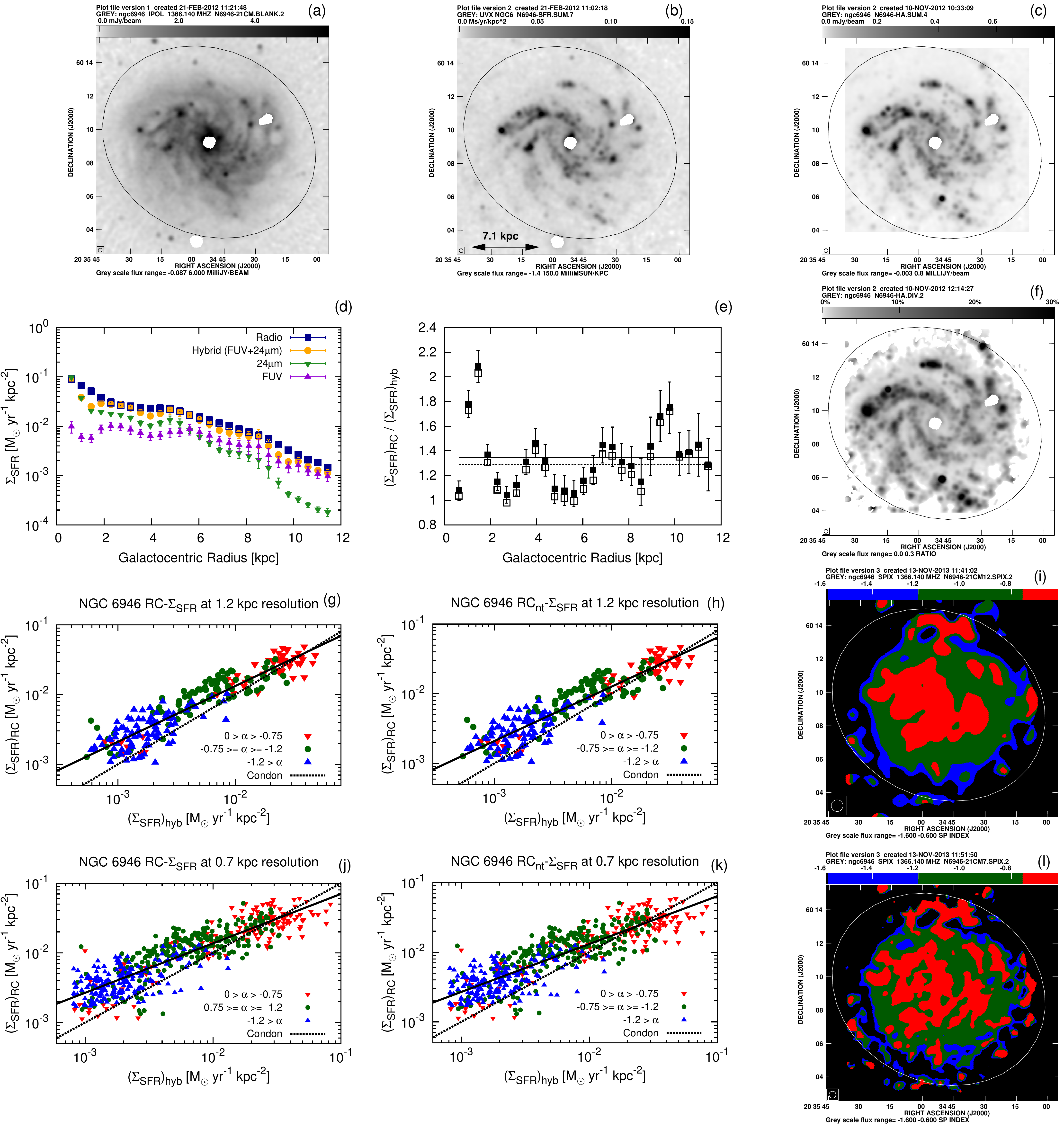}}
\caption{NGC~6946. (a) RC emission at
 $\lambda 22~\rm
  cm$. (b) Hybrid SFR surface density $(\Sigma_{\rm SFR})_{\rm
    hyb}$. (c)
 Thermal RC emission as derived from H$\alpha$. (d) Radio
  $\Sigma_{\rm SFR}$
 (dark blue) and hybrid $\Sigma_{\rm SFR}$ (orange) as a function
  of galactocentric radius. (e) Ratio $\mathscr{R}$ of radio to
 hybrid $\Sigma_{\rm SFR}$. Open
  symbols represent the non-thermal RC emission alone. (f)
 Predicted thermal
  RC fraction. (g) Radio $\Sigma_{\rm SFR}$ plotted against hybrid
  $\Sigma_{\rm SFR}$ at $\rm 1.2~kpc$ resolution. (h) Same as (g) but with thermal
  emission subtracted. (i)
 radio spectral index at $\rm 1.2~kpc$
  resolution. (j--l) Same as (g--i) but at
 $\rm 0.7~kpc$ resolution. The typical
  uncertainties for the radio spectral index
 are $0.4$, $0.2$, and $0.1$ for
  the blue, green, and red data points respectively.}
\label{fig:n6946}
\end{figure}

\begin{figure}[tbhp]
\centering
\resizebox{1.0\hsize}{!}{ \includegraphics{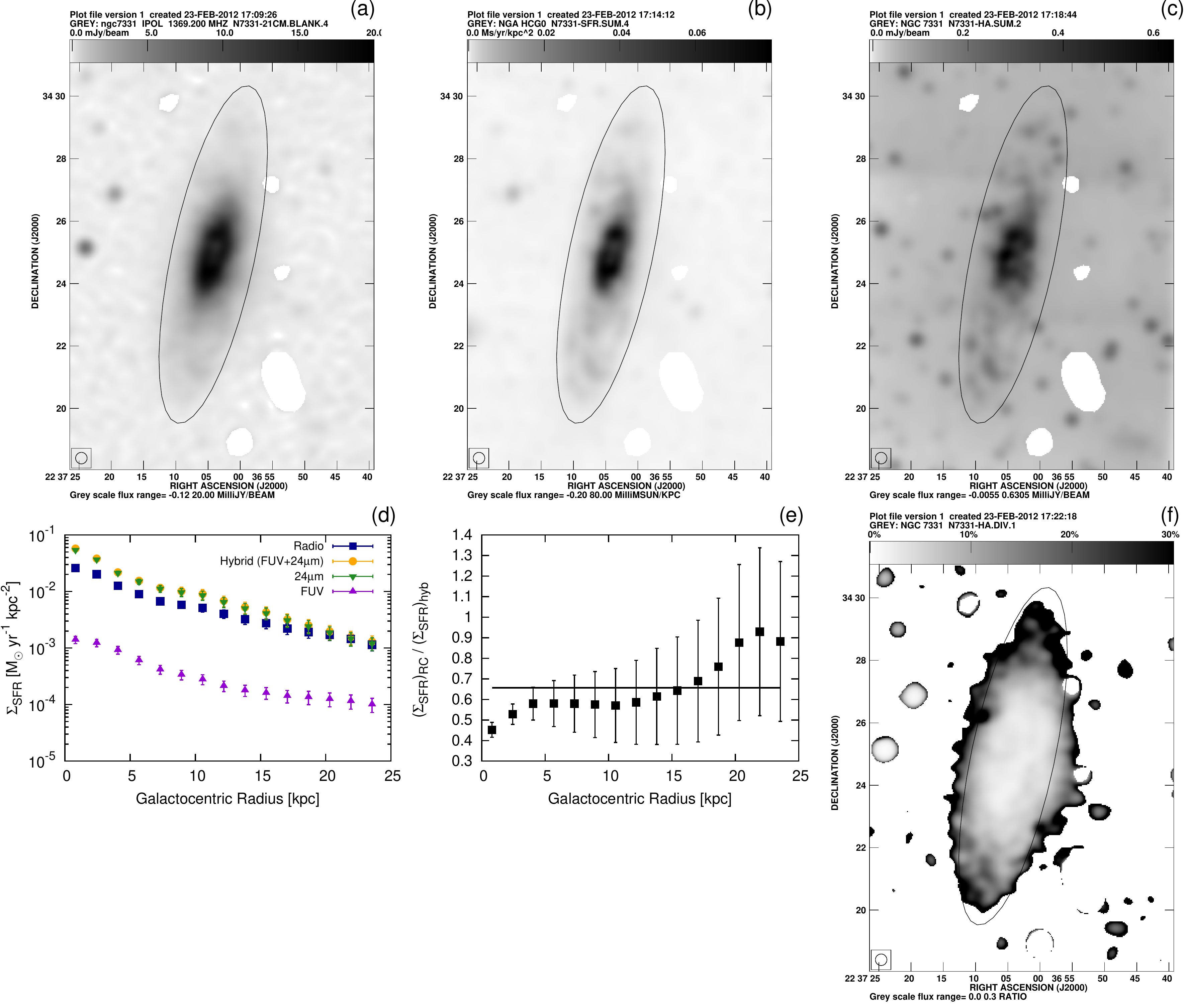}}
\caption{NGC~7331. (a) RC emission at $\lambda 22~\rm
  cm$. (b) Hybrid SFR surface density $(\Sigma_{\rm SFR})_{\rm hyb}$. (c) Thermal RC
  emission as derived from H$\alpha$. (d) Radio $\Sigma_{\rm SFR}$ (dark blue) and
  hybrid $\Sigma_{\rm SFR}$ (orange) as function of galactocentric radius. (e) Ratio $\mathscr{R}$ of radio to hybrid
  $\Sigma_{\rm SFR}$. Open symbols represent the non-thermal RC emission
  alone. (f) Predicted thermal RC fraction.}
\label{fig:n7331}
\end{figure}

In Figures \ref{fig:hoii}--\ref{fig:n7331}, we present maps and radial profiles for individual
galaxies. Each page shows results for one galaxy. The top row shows maps of RC
emission at $\lambda$22~cm, hybrid SFR density $\rm (\Sigma_{SFR})_{hyb}$, and thermal RC emission at $\lambda$22~cm. In the second row at the
right, we show the thermal RC fraction at $\lambda$22~cm. These maps use a gray-scale
linear transfer function, where a bar at the top illustrates the gray tones
used. The ellipse indicates the area we have used for integrating the RC
emission and studying radial profiles. A bar in the middle panel of the top
row indicates the spatial scale in each galaxy. The size of the FWHM, the
angular resolution of our maps, is shown in the left bottom corner by the open
circle. Please note that each galaxy has a different angular and spatial
resolution.

In the second row at the left, we show profiles
of the azimuthally averaged star formation rate surface density based on RC
emission at  $\lambda$22~cm (dark blue), based on the hybrid prescription of \emph{GALEX} FUV
and \emph{Spitzer} $24~\rm\mu m$ (orange), based on \emph{Spitzer} $24~\rm\mu
m$ only (green), and based on \emph{GALEX} FUV only (violet). Open dark blue
symbols show the SFR surface density based on the non-thermal RC emission at $\lambda$22~cm only. The
latter one is shown for comparison in order to highlight the different
behavior of thermal and non-thermal RC emission. In the second row, the middle
panel shows the profile of the ratio of RC to hybrid SFR surface
density. Open symbols show this ratio for the non-thermal RC emission
only. The solid line shows the constant ratio of a linear least-squares fit to
the data and the dashed line shows the same for the non-thermal RC only.

The
following plots are only shown if our maps of an individual galaxy has
sufficient spatial resolution. In the left panel of the third row, we show the
pixel-by-pixel plots at $1.2$~kpc spatial resolution. The SFR
surface density based on the RC emission at $\lambda$22~cm is plotted as a function of
the hybrid SFR surface density. The solid line shows a
least-squares fit to the data and the dashed line shows the relation as
predicted by \citet{condon_92a}. The color of the data points depends on the radio
spectral index: spectral indices $\alpha < -0.75$ are color coded in red, spectral
indices $-1.2 \leq \alpha \leq -0.75$ are color coded in green, and spectral
indices $\alpha < -1.2$ are color coded in blue. The middle panel in the third row shows the
pixel-by-pixel plots for the non-thermal RC emission only. The right panel in
the third row shows the map of the radio spectral index between $\lambda\lambda$ 22 and 18~cm, where we have color coded the data as in the pixel-by-pixel plots (with
the exception of Holmberg~II and NGC~4736). The fourth row shows the same plots as
the third row but for a spatial resolution of $0.7~\rm kpc$.
%
\clearpage
\bibliographystyle{apj}
\bibliography{sfr_bib}

\end{document}